\documentclass[fleqn,usenatbib]{mnras}

\usepackage{newtxtext,newtxmath}

\usepackage[T1]{fontenc}
\usepackage{ae,aecompl}

\usepackage{graphicx}	
\usepackage{blindtext}  
\usepackage{natbib}
\usepackage{rotating,times,graphicx,latexsym}
\usepackage{color}
\usepackage{longtable}
\usepackage{lscape}
\usepackage{lipsum} 
\usepackage{array}
\usepackage{flafter}
\usepackage{color, colortbl}

\usepackage{soul} 
\usepackage{xargs} 
\usepackage[pdftex,dvipsnames]{xcolor}  
\usepackage[colorinlistoftodos,prependcaption]{todonotes}
\newcommandx{\tobedone}[2][1=]{\todo[linecolor=red,backgroundcolor=red!25,bordercolor=red,inline,#1]{#2}}
\newcommandx{\changed}[2][1=]{\todo[linecolor=blue,backgroundcolor=blue!25,bordercolor=blue,inline,#1]{#2}\noindent}
\newcommandx{\thiswillnotshow}[2][1=]{\todo[disable,#1]{#2}}
\newcommandx{\alex}[2][1=]{\todo[linecolor=Orchid,backgroundcolor=Orchid!25,bordercolor=Orchid,inline,#1]{#2}\noindent}
\newcommandx{\dirk}[2][1=]{\todo[linecolor=BurntOrange,backgroundcolor=BurntOrange!25,bordercolor=BurntOrange,inline,#1]{#2}\noindent}
\newcommandx{\fenia}[2][1=]{\todo[linecolor=SpringGreen,backgroundcolor=SpringGreen!40,bordercolor=SpringGreen,inline,#1]{#2}\noindent}
\newcommandx{\jochen}[2][1=]{\todo[linecolor=BurntOrange,backgroundcolor=Purple!60,bordercolor=SpringGreen,inline,#1]{#2}\noindent}
\newcommandx{\fred}[2][1=]{\todo[linecolor=BurntOrange,backgroundcolor=BurntOrange!40,bordercolor=BurntOrange,inline,#1]{#2}\noindent}
\newcommandx{\niall}[2][1=]{\todo[linecolor=BurntOrange,backgroundcolor=BurntOrange!40,bordercolor=SpringGreen,inline,#1]{#2}\noindent}
\newcommandx{\jcw}[2][1=]{\todo[linecolor=SpringGreen,backgroundcolor=SpringGreen!40,bordercolor=Orchid,inline,#1]{#2}\noindent}
\newcommandx{\bst}[2][1=]{\todo[linecolor=red,backgroundcolor=Green!40,bordercolor=BurntOrange,inline,#1]{#2}\noindent}
\newcommandx{\siegfried}[2][1=]{\todo[linecolor=Green,backgroundcolor=Red!40,bordercolor=BurntOrange,inline,#1]{#2}\noindent}
\newcommandx{\svm}[2][1=]{\todo[linecolor=Orchid,backgroundcolor=Orchid!40,bordercolor=Orchid,inline,#1]{#2}\noindent}

\newcommand{\hc}{{HOYS}}

\graphicspath{ {images/} }
\newcommand{\hii}{H{\sc ii}~}

\title[Periods in IC\,5070]{A survey for variable young stars with small telescopes: IV - Rotation Periods of YSOs in IC\,5070}

\author[Dirk Froebrich et al.]{Dirk Froebrich$^{1}$\thanks{E-mail: df@star.kent.ac.uk},
Efthymia Derezea$^{2}$,
Aleks Scholz$^{3}$, 
Jochen Eisl\"offel$^{4}$, 
\newauthor
Siegfried Vanaverbeke$^{5,6,7}$,
Alfred Kume$^{2}$,
Carys Herbert$^{8}$,
Justyn Campbell-White$^{8,9}$, 
\newauthor
Niall Miller$^{8,10}\ddagger$, 
Bringfried Stecklum$^{4}$,
Sally V. Makin$^{8}$, 
Thomas Urtly$^{11}$\thanks{\hc\ Observer}, 
\newauthor
Francisco C. Sold\'{a}n Alfaro$^{12,13}\dagger$, 
Erik Schwendeman$^{13}\dagger$, 
Geoffrey Stone$^{14}\dagger$, 
\newauthor
Mark\,Phillips$^{15}\dagger$, 
George Fleming$^{11}\dagger$, 
Rafael Gonzalez Farf\'{a}n$^{12}\dagger$, 
Tonny Vanmunster$^{16,17}\dagger$, 
\newauthor
Michael A. Heald$^{13}\dagger$, 
Esteban Fern\'{a}ndez Ma\~{n}anes$^{12}\dagger$, 
Tim Nelson$^{18}\dagger$, 
\newauthor
Heinz-Bernd Eggenstein$^{19}\dagger$,
Franky Dubois$^{5,6}\dagger$, 
Ludwig Logie$^{5,6}\dagger$, 
Steve Rau$^{5,6}\dagger$, 
\newauthor
Klaas Wiersema$^{20,21}\dagger$, 
Nick Quinn$^{11}\dagger$, 
Diego Rodriguez$^{13}\dagger$, 
Rafael Castillo Garc\'{i}a$^{12,13,22}\dagger$, 
\newauthor
Thomas Killestein$^{11,20}\dagger$, 
Tony Vale$^{11,23,24}\dagger$,
Domenico Licchelli$^{25}\dagger$, 
Marc Deldem$^{13}\dagger$, 
\newauthor
Georg Piehler$^{26}\dagger$, 
Dawid Mo\'{z}dzierski$^{27}\dagger$, 
Krzysztof Kotysz$^{27}\dagger$, 
\newauthor
Katarzyna Kowalska$^{27}\dagger$, 
Przemys{\l}aw Miko{\l}ajczyk$^{27}\dagger$, 
Stephen R.L. Futcher$^{11,18,28}\dagger$, 
\newauthor
Timothy P. Long$^{29}\dagger$, 
Mario Morales Aimar$^{12,13}\dagger$, 
Barry Merrikin$^{18}\dagger$, 
\newauthor
Stephen Johnstone$^{11,13}\dagger$, 
Pavol A. Dubovsk\'{y}$^{30}\dagger$,
Igor Kudzej$^{30}\dagger$, 
Roger Pickard$^{11}\dagger$,
\newauthor
Samuel J. Billington$^{8}\thanks{Observer Beacon Observatory}$, 
Lord Dover$^{8}\ddagger$, 
Tarik Zegmott$^{8}\ddagger$, 
Jack J. Evitts$^{8}\ddagger$, 
\newauthor
Alejandra Traspas Munia$^{8}\ddagger$, 
Mark C. Price$^{8}$
\\
$^{1}$School of Physical Sciences, University of Kent, Canterbury CT2 7NH, UK\\
$^{2}$School of Mathematics, Statistics and Actuarial Sciences, The University of Kent, Canterbury, CT2 7FS, U.K.\\
$^{3}$SUPA, School of Physics \& Astronomy, University of St Andrews, North Haugh, St Andrews KY16 9SS, UK\\
$^{4}$Th\"{u}ringer Landessternwarte, Sternwarte 5, 07778 Tautenburg, Germany\\
$^{5}$Public Observatory AstroLAB IRIS, Provinciaal Domein De Palingbeek, Verbrandemolenstraat 5, B-8902 Zillebeke, Ieper, Belgium\\
$^{6}$Vereniging voor Sterrenkunde, werkgroep veranderlijke sterren, Oostmeers 122 C, 8000 Brugge, Belgium\\
$^{7}$Center for Mathematical Plasma Astrophysics, University of Leuven, Belgium\\
$^{8}$Centre for Astrophysics and Planetary Science, School of Physical Sciences, University of Kent, Canterbury CT2 7NH, UK\\
$^{9}$SUPA, School of Science and Engineering, University of Dundee, Nethergate, Dundee DD1 4HN, UK\\
$^{10}$Centre for Astrophysics Research, University of Hertfordshire, Hatfield AL10 9AB, UK\\
$^{11}$The British Astronomical Association, Variable Star Section, Burlington House Piccadilly, London W1J 0DU, UK\\
$^{12}$Observadores de Supernovas$^{\thanks{\tt \href{https://sites.google.com/view/sn2017eaw/}{Observadores de Supernovas}}}$, Spain\\
$^{13}$AAVSO, 49 Bay State Road, Cambridge, MA 02138, USA\\
$^{14}$First Light Observatory Systems, 9 Wildflower Way, Santa Fe, NM 87506, USA\\
$^{15}$Astronomical Society of Edinburgh, Edinburgh, UK\\
$^{16}$Center for Backyard Astrophysics Extremadura, 06340 Fregenal de la Sierra, Spain\\
$^{17}$Vereniging voor Sterrenkunde VVS, 3401 Landen, Belgium\\
$^{18}$Hampshire Astronomical Group, Clanfield, UK\\
$^{19}$Volkssternwarte Paderborn, 33041 Paderborn, Germany\\
$^{20}$Department of Physics, University of Warwick, Coventry CV4 7AL, UK\\
$^{21}$University of Leicester, University Road, Leicester LE1 7RH, UK\\
$^{22}$Asociacion Astronomica Cruz del Norte, Calle Caceres 18, 28100 Alcobendas, Madrid, Spain\\
$^{23}$Wiltshire Astronomical Society, 2 Oathills, Corsham, SN13 9NL, UK\\
$^{24}$The Herschel Society, The Herschel Museum of Astronomy, 19 New King Street, Bath BA1 2BL, UK \\
$^{25}$Center for Backyard Astrophysics, Piazzetta del Ges\`{u} 3, 73034, Gagliano del Capo, Italy\\
$^{26}$Selztal Observatory, D-55278 Friesenheim, Bechtolsheimer Weg 26, Germany\\
$^{27}$Instytut Astronomiczny, Uniwersytet Wroc{\l}awski, Kopernika 11, 51-622 Wroc{\l}aw, Poland\\
$^{28}$Royal Astronomical Society, Burlington House, Piccadilly, London W1J 0BQ, UK\\
$^{29}$Tigra Astronomy, 16 Laxton Way, Canterbury, Kent, CT1 1FT, UK\\
$^{30}$Vihorlat Observatory, Mierov\'{a} 4, 06601 Humenn\'{e}, Slovakia\\
}

\date{Accepted XXX. Received YYY; in original form ZZZ}

\pubyear{2021}

\begin{document}
\label{firstpage}
\pagerange{\pageref{firstpage}--\pageref{lastpage}}
\maketitle

\begin{abstract}

Studying rotational variability of young stars is enabling us to investigate a multitude of properties of young star-disk systems. We utilise high cadence, multi-wavelength optical time series data from the Hunting Outbursting Young Stars citizen science project to identify periodic variables in the Pelican Nebula (IC\,5070). A double blind study using nine different period-finding algorithms was conducted and a sample of 59 periodic variables was identified. We find that a combination of four period finding algorithms can achieve a completeness of 85\,\% and a contamination of 30\,\% in identifying periods in inhomogeneous data sets. The best performing methods are periodograms that rely on fitting a sine curve. Utilising Gaia\,EDR3 data, we have identified an unbiased sample of 40 periodic YSOs, without using any colour or magnitude selections. With a 98.9\,\% probability we can exclude a homogeneous YSO period distribution. Instead we find a bi-modal distribution with peaks at three and eight days. The sample has a disk fraction of 50\,\%, and its statistical properties are in agreement with other similarly aged YSOs populations. In particular, we confirm that the presence of the disk is linked to predominantly slow rotation and find a probability of 4.8\,$\times$\,10$^{-3}$ that the observed relation between period and presence of a disk has occurred by chance. In our sample of periodic variables, we also find pulsating giants, an eclipsing binary, and potential YSOs in the foreground of IC\,5070.

\end{abstract}





\section{Introduction}

Variability is one of the key characteristics of Young Stellar Objects (YSOs). Time-domain observations of star forming regions provide reliable information about the formation and early evolution of stars. Rotational flux modulation has been used to measure rotation periods ranging from hours to weeks \citep{2007prpl.conf..297H,  2014prpl.conf..433B}. Optical fluxes of YSOs are also affected by variable excess emission from accretion shocks, variable emission from the inner disk, and variable extinction along the line of sight \citep{2001AJ....121.3160C}. Thus, they give insights into the structure and evolution of the environment of YSOs. 

While the interplay of these variability causes can lead to very complicated light curves and render the interpretation difficult, several prototypical phenomena have been successfully attributed to a single physical cause. AA\,Tau is the prototype for one category of dippers; a contingent of YSOs temporarily eclipsed by portions of the inner disks warped by the star's magnetic field \citep{2014prpl.conf..433B, 2015A&A...577A..11M}. Other dippers are possibly caused by companions or protoplanets in the inner disk \citep{2020MNRAS.493..184E}. FU\,Ori and EX\,Lupi are prototypes for stars with sharp increases in their mass accretion rates \citep{2014prpl.conf..387A}, a phenomenon that is now known to occur on a wide range of timescales (months to tens of years; \citet{2016AJ....151...60S}) and amplitudes (1--5\,mag; \citet{2017MNRAS.465.3011C, 2017MNRAS.465.3039C, 2017MNRAS.472.2990L}). This includes objects with continuous accretion rate changes and hence stochastic light curves \citep{2014AJ....147...83S, 2016AJ....151...60S} or periodic bursters \citep{2020AJ....160..278D}. 

The gold standard for optical studies of YSO variability are space-based observing campaigns with the COROT and Kepler/K2 satellite missions. Their combined monitoring of NGC\,2264 is unprecedented in cadence and photometric precision. Complemented by ground-based observations, it has led to a new comprehensive overview of the phenomenology of variable YSOs and the underlying causes \citep{2014AJ....147...82C}. Kepler/K2 has observed large numbers of YSOs continuously over campaigns of 70\,days. Its archive is a treasure trove for detailed studies of rotation periods, dippers, bursters, and related phenomena \citep{2016ApJ...816...69A}. These and other numerous studies of YSO variability have often focused on shorter term variability (weeks to months) with high cadence (hours to days). They often exclusively investigate periodic behaviour, and focus on outbursts and the study of accretion rate changes. Many of the past large-scale optical and infrared time-domain surveys are restricted to a single (or only two) wavelength/filter (e.g., UGPS, VVV(X), Pan-STARRS, (i)PTF, ASAS-SN, ZTF). 

Hence, there is a definite need for long-term, quasi-simultaneous monitoring in multiple bands, similar to the pioneering studies by \citet{2007A&A...461..183G}. The Hunting Outbursting Young Stars (HOYS) citizen science project \citep{2018MNRAS.478.5091F} has been initiated as such a survey. This project is performing long term, multi-wavelength, high cadence photometric monitoring of a number of nearby star forming regions and young clusters. It uses a mix of amateur, university, and professional telescopes. Participants submit reduced and stacked images to our database where an astrometric and a basic photometric calibration are performed. In this paper we aim to investigate how such a diverse data set can be used to reliably identify periodic variables with as little bias and contamination as possible. This study will focus on one of the HOYS target regions, IC\,5070 - The Pelican Nebula. Our goal here is to measure the rotation period distribution of YSOs in this star forming region.

Our paper is structured as follows: In Sect.\,\ref{target} we give an overview of the IC\,5070 star forming region. We briefly introduce the photometry data we use in Sect.\,\ref{data}. We then describe in detail our methodology to identify a sample of periodic variables in Sect.\,\ref{make_sample}. A discussion of our period-finding methods is given in Sect.\,\ref{res_methods}, while the properties of the sample of periodic variables, with particular focus on the YSOs in the region, are detailed in Sect.\,\ref{res_variables}.   

\section{IC\,5070 in the literature}\label{target}

The North American Nebula (NGC\,7000) and Pelican Nebula (IC\,5070) are part of the \hii\ region W\,80, separated by the foreground molecular dust cloud L\,985, with IC\,5070 in the west of the region. W\,80 has a measured distance of 795\,pc \citep{2020ApJ...899..128K} and a diameter of 3\,deg \citep{2014AJ....148..120B}. NGC\,7000 and IC\,5070 are associated with T\,Tauri stars first identified in \citet{1958ApJ...128..259H}. 

Using Spitzer data, \citet{2011ApJS..193...25R} identified over 2000 YSO candidates within a 7 square degree region toward NGC\,7000 and IC\,5070, 256 of which lie in IC\,5070. \citet{2020ApJ...899..128K} used Gaia parallax and proper motion data to confirm 395 young stars belonging to 6 groups within the region. The majority of them are aged $\sim$1\,Myr, and almost all are younger than 3\,Myr. The \citet{2020ApJ...904..146F} spectroscopic study identified sequential star formation between the groups laid out in \citet{2020ApJ...899..128K}, with those in IC\,5070 (groups C and D in their paper) in the second wave of star formation.

The photometric variability of YSOs in IC\,5070 has been the subject of limited study. Most recently \citet{2019A&A...627A.135B} identified 95 variable stars using BVRI observations of a 16$^\prime$ square taken in 90 nights over one year (2012-2013), and identified periods for 56 objects. \citet{2018RAA....18..137I} utilised BVRI observations of 15 pre-main-sequence stars in a 16$^{\prime}$ radius field in IC\,5070. One periodic variable star was identified. \citet{2014A&A...568A..49P} used archive photographic plates and data collected from seven observatories to create a data set covering 60 years. They investigated 17 previously detected pre-main sequence stars and 3 periodic sources were identified, all of which are outside our survey field. There are a number of other smaller studies of variable stars in the area, such as e.g., \citet{2011A&A...527A.133K} and \citet{2013ApJ...768...93F}.

\section{HOYS Observational Data}\label{data}

All photometry data for this project has been obtained as part of HOYS. The astrometric solution for all HOYS images has been obtained using the Astrometry.net\footnote{\href{http://astrometry.net/}{\tt Astrometry.net}} software \citep{2008ASPC..394...27H}. Source extraction for photometry is conducted with the Source Extractor software\footnote{\href{https://www.astromatic.net/software/sextractor}{\tt The Source Extractor}} \citep{1996A&AS..117..393B}. Relative photometry is performed against reference images taken under photometric conditions in Johnson U, B, V and Cousins Rc and Ic filters (R and I, hereafter). The calibration offsets into apparent magnitudes for those reference images are obtained using the Cambridge Photometric Calibration Server\footnote{\tt \href{http://gsaweb.ast.cam.ac.uk/followup}{Cambridge Photometric Calibration Server}}. In \citet{2020MNRAS.493..184E} we have developed an internal calibration procedure for the HOYS data to refine the photometric accuracy of the heterogeneous data set, i.e., for images taken in slightly different filters than our reference frames, or under non-photometric conditions. We identify non-variable stars in the data and utilise their magnitudes and colours to correct systematic colour terms in the relative photometry of all sources. The corrected data (used in the analysis for this project) typically achieves a relative photometric accuracy of a few percent for stars between 10$^{\rm th}$ and 16$^{\rm th}$ magnitude.

\section{Rotation Period Sample}\label{make_sample}

Our aim is to establish as unbiased a sample as possible of rotation periods of YSOs in the IC\,5070 field. We do not simply aim to identify periodic photometric changes in light curves of previously known members (e.g., \citet{2011ApJS..193...25R,2020ApJ...899..128K}). Instead our strategy is to first identify periodic signals in the light curves of all stars in the field, verify their periodic nature (in a double blind way), and only then select cluster members from the sample with periods to study in detail. This section details the entire process of establishing our sample of YSO rotation periods in IC\,5070.

\subsection{Photometry Data Selection}

Our target field in IC\,5070 is centred at RA\,=\,20:51:00 and DEC\,=\,+44:22:00 (J2000) and is about 1$^\circ \times 1^\circ$ in size. In order to establish a source list for the field, we extracted all sources from the HOYS I-band reference frame, a $10 \times 2$\,min stack taken with the University of Kent Beacon Observatory \citep{2018MNRAS.478.5091F}. Only sources with photometric errors less than 0.1\,mag in the reference frame are included. For all these stars we extracted the HOYS light curves from the database on  23$^{\rm rd}$ June 2020, at 1\,am UTC+1. We only select photometric data points in the individual images with errors of less than 0.2\,mag, Source Extractor flags \citep{1996A&AS..117..393B} less than five and a full width half maximum of less than seven arcseconds. All stars with fewer than 100 data points per HOYS light curve are removed from the list. There are 8548 stars in total. This contains stars in the range 9.5\,mag\,$<$\,I\,$<$\,17.5\,mag. There is, however, no universal magnitude cut-off as these limits do slightly depend on the stellar colour and position in the field. This occurs as not all parts of the survey area are covered with the same number of images to the same depth, due to the varying field of view and aperture size of the contributing telescopes. 

We only use the high cadence part of the long term HOYS light curves for the period determination. These data were taken during 80 days in the summer of 2018, i.e., between JD\,=\,2458330 and 2458410, as part of an AAVSO campaign\footnote{https://www.aavso.org/aavso-alert-notice-684} to monitor the YSO V\,1491\,Cyg \citep{2020MNRAS.493..184E}. Thus, for this period we extracted the data for each star in all broad-band filters (U, B, V, R and I) for period analysis. The data were only analysed if there were more than 50 data points in a particular filter during that time period. A total of 6063 stars had sufficient data in at least one filter to be analysed. 

To ensure an unbiased analysis, this data preparation was done by only one of us (DF). The resulting photometry data (only listing date, magnitude, magnitude error) was then given an ID number based on the position in our original source list, and an indication of the filter it was taken in. At no time, until the final source list had been made, did any other member of the team know the association of the ID numbers and the object names or coordinates. 

\subsection{Period Determination}

All the photometric time series data sets for all sources and filters were then handed to two other members of the team (ED, AK). The only instructions given were to return (for each light curve and filter) a list of the most likely real periods, as well as the powers in the utilised periodogram. No background information regarding the scientific aims of the project was given, or what light curve shapes in phase space were to be expected. It was only specified that the period search should be done for periods between 0.5\,d and 50\,d. These conditions are, however, imposed by the nature of the data, given the typical sampling and length of our light curves.

In total we applied nine different periodogram methods to our data. We utilise the two widely used {\it standard} methods Lomb-Scargle (LS) and Generalised Lomb-Scargle (GLS), based on \citet{1982ApJ...263..835S} and \citet{2009A&A...496..577Z}. A further seven additional methods have been used. These are based on sine/cosine or spline function fitting and employ different ways of obtaining the coefficients of the assumed models, e.g., least squares (L2) see Eq.\,\ref{L2} or absolute deviation optimisation (L1) see Eq.\,\ref{L1}. As a general comment, methods based on splines make little assumptions on the shape of the light curves and should be more flexible for data that depart from the sinusoidal shape. Methods based on L1 should be more robust in the presence of outliers. Appropriate critical values for each method were used to determine which periodogram peaks represent "valid" periods.

A sinusoidal wave was fitted to the phase-folded data according to Eq.\,\ref{siphmod} for the regression methods L2Beta, L2Boot, L1Beta and L1Boot. $m_{i}$ and $\epsilon_{i}$ are the magnitudes and their uncertainties for data point $i$, respectively. With $f_{ip}$ we denote the phase folded data point $i$ at a given period $p$ calculated as in Eq.\,\ref{phase}. The coefficients $\beta_{0}$, $\beta_{1}$, and $\beta_{2}$ where estimated using L2 or L1 regression. Methods L2spB and L1spB are based on fitting B-splines with 4 knots to the phase-folded data (instead of sine/cosine). Similarly, the coefficients of these models were estimated using L2 and L1 regression respectively.

\begin{equation} \label{siphmod}
m_{i}=\beta_{0}+\beta_{1} \sin(2\pi f_{pi}) +\beta_{2}\cos(2 \pi f_{pi})+\epsilon_{i}
\end{equation}

The periodogram values for these six regression methods were based on the coefficient of determination ($R^{2}$), see for example the corresponding expression for the L1 and L2 types denoted as $Perp(p)$ in Eqs.\,\ref{R2} or \ref{robR}. These calculations were performed with the R package {\tt RobPer} \citep{thieler2016robper}. The potentially valid periods were initially determined by a hypothesis test based on a Beta distribution \citep{thieler2013periodicity} or by bootstrapping of the periodogram. Finally, method L2persp is based both on splines and sine/cosine fitting in time (and not phase space). The coefficients of the model were estimated using L2 regression and the potential valid periods were determined by bootstrapping the periodogram. Table\,\ref{met_tab} gives a brief overview of the basic characteristics of all nine methods. The full details of all methods can be found in Appendix\,\ref{per_meth_details} in the online supplementary material.

\begin{table}
\caption{\label{met_tab} A summary of the characteristics of the period finding methods used. The check-mark denotes the presence of the relevant attribute. With L2 we denote least squares and with L1 absolute deviation determination. In the column ($\sigma$) we indicate where magnitude uncertainties have been taken into account. }
\centering
\begin{tabular}{|l|c|c|c|c|c|}
\hline
Name & sine & splines & L2 & L1 & $\sigma$ \\ \hline
L2Beta & \checkmark &            & \checkmark &            & \checkmark \\
L2Boot & \checkmark &            & \checkmark &            & \checkmark \\
L1Beta & \checkmark &            &            & \checkmark & \checkmark \\
L1Boot & \checkmark &            &            & \checkmark & \checkmark \\
L2spB  &            & \checkmark & \checkmark &            & \checkmark \\
L1spB &            & \checkmark &            & \checkmark & \checkmark \\
L2persp  & \checkmark &   \checkmark         &  \checkmark          &            &            \\
LS       & \checkmark &            &            &            &            \\
GLS      & \checkmark &            &            &            &            \\
\hline
\end{tabular}
\end{table}

\subsection{Generating the period master sample}

After the most likely periods and periodogram powers were determined for each object and filter separately, they were merged back together by DF. We then selected for each method all objects as candidate periodic variables where the most likely period in at least two filters agreed within two percent, and where the periodogram power was above 0.2 in both filters. This value corresponds to a typical false alarm probability of less than 0.1\,\% for our data. The candidate period was set as the average of the periods from the different filters. We also excluded all periods that were within one percent of one day, as these may be caused by the observing cadence. We show the number of candidate periods for each method in Table\,\ref{methods}.

For each of these candidates, phase-folded light curves in all filters (if available) were made. The plots only contained the photometric data points and only one period of phase space was shown. They were then placed on a website and shown to two members of the team (AS, JE), without identification or reference to the method the periods had been determined from. Both team members then independently selected all objects for which they thought a reliable  periodic signal was visible in the phase plots. As the amplitudes of the variability and noise in the data vary with filter, objects were selected as real if they were detected in at least two of the phase folded diagrams and also had a similar behaviour (peaks and troughs) in phase space.

The results of this selection were then collated by DF. All objects for each periodogram method where both team members agreed in their assessment that the periodicity is correct were selected as real variables. The numbers of objects considered  periodic variables for each method are listed in Table\,\ref{methods}. These lists for each method were then cross-matched to generate a master list of periodic variables. This results in a master list of 59 unique sources. For all objects we then proceeded to re-determine the period in a consistent way. We determined the peak in a simple LS periodogram in all filters within 10\,\% around the original period found for the object during the initial period search, i.e., the average of the periods from the different methods. We then determine the final period as the median period from all filters with data for the respective source. The data were then phase-folded in all filters with this final period. The phase-folded data for all sources are shown in Appendix\,\ref{appb} in the online supplementary material. In those plots we show two periods of data for all filters with photometry measurements. We over-plot a running median and one sigma deviations in each of the filters. The object ID and period are also indicated in each plot. 

\begin{table}
\caption{\label{methods} In this table we list the number of periodic candidate objects ($N_C$), the number of detected objects from the master ($N_M$) list of periodic variables, the completeness ($F_{comp}$) and contamination ($F_{cont}$). The highlighted rows are the {\it best} methods as discussed in Sect.\,\ref{res_methods}. }
\centering
\begin{tabular}{|l|c|c|c|c|}
\hline
Name & $N_C$ & $N_M$ & $F_{comp}$ [\%] & $F_{cont}$ [\%] \\ \hline
\rowcolor{lightgray} L2Beta  & 45 & 32 & 54 & 29 \\
L2Boot & 50 & 29 & 49 & 42 \\
\rowcolor{lightgray} L1Beta & 42 & 38 & 64 & 10 \\
\rowcolor{lightgray} L1Boot & 41 & 31 & 53 & 24 \\
L2spB & 59 & 24 & 41 & 59 \\
L1spB & 41 & 23 & 39 & 44 \\
L2persp & 27 & 0  &  0 & 100\\
LS       & 55 & 22 & 37 & 60 \\
\rowcolor{lightgray} GLS      & 54 & 31 & 53 & 43 \\
\hline
\end{tabular}
\end{table}

\subsection{Selection of periodic YSOs}

\begin{figure*}
\centering
\includegraphics[width=1.03\columnwidth]{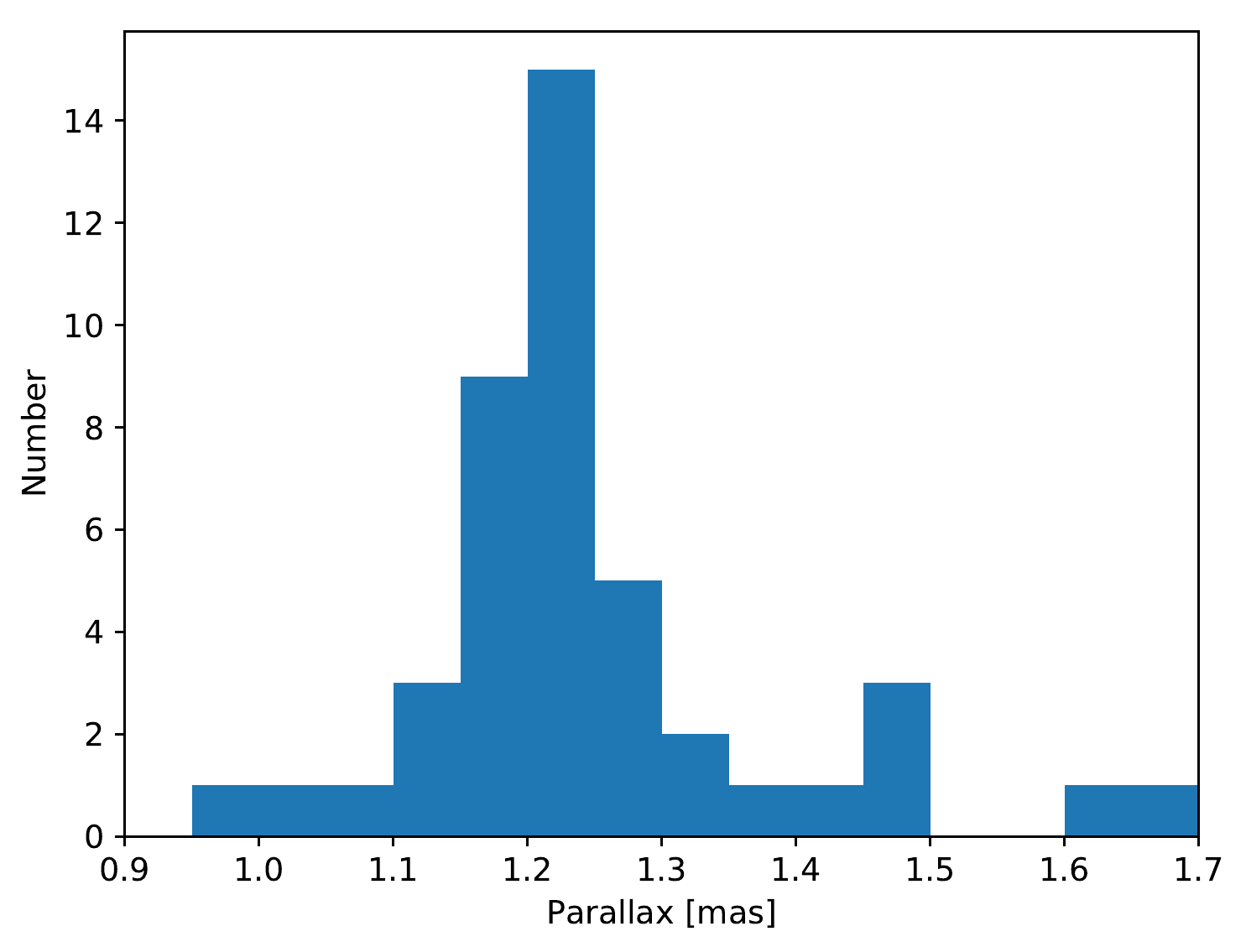} \hfill
\includegraphics[width=0.97\columnwidth]{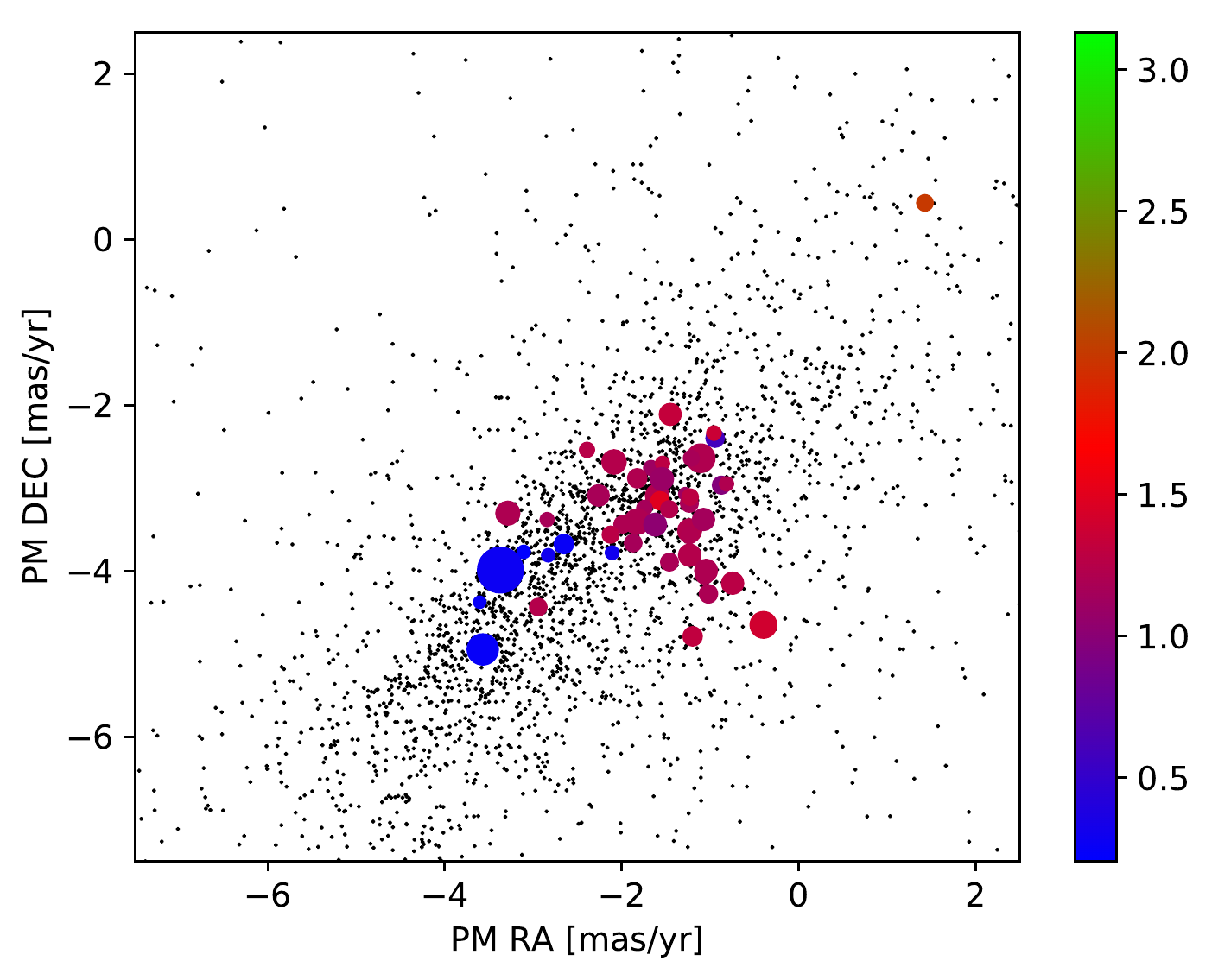}
\caption{{\bf Left:} Distribution of the parallaxes in our periodic object sample, limited to around the cluster parallax. {\bf Right:} Proper motion in RA and DEC for the all stars in the field (black dots) and the periodic stars as coloured symbols. The colour indicates the parallax in mas and the symbol size the period. All the foreground objects are outside the plotting area, which has been focused on the cluster proper motion. \label{yso_selection}}
\end{figure*}

All objects in our initial source list were cross-matched against Gaia-EDR3 \citep{2020arXiv201201533G} and ALLWISE/2MASS \citep{2014yCat.2328....0C}. Given the typically large FWHM of the stellar images in the HOYS data, we initially matched each of our sources to all sources within six arcseconds in the two catalogues. If more than one match was found, we selected the brightest source in G in Gaia and in J in ALLWISE/2MASS as match, as these are the most likely matches to the HOYS source.

Using Gaia parallaxes we created a sample of YSOs at the distance of IC\,5070 from our periodic objects based solely on their astrometric properties. In Fig.\,\ref{yso_selection} we show a histogram of the parallax values (left panel), for all periodic objects, centered around the cluster parallax. Fifteen of the 59 objects are not shown in the histogram as their parallax values show them to be foreground or background objects. In \citet{2020MNRAS.493..184E} the distance to the main YSO population in the field was determined as 870\,pc (parallax of 1.15\,mas, based on Gaia\,DR2). The dynamic groups C and D in \citet{2020ApJ...899..128K}, which form the population of YSOs in our field, have parallaxes of 1.23\,mas and 1.21\,mas, respectively. The peak in the histogram is in excellent agreement with our population of periodic variables. In the right panel of Fig.\,\ref{yso_selection} we show the proper motions of all stars in the field as black dots, again with focus on the main IC\,5070 cluster. Ten of the objects, the foreground population of periodic objects with higher proper motion (up to a few ten mas/yr), are outside the shown part of the parameter space. Over-plotted as coloured symbols are the periodic variables, with the parallax colour coded and sizes proportional to the periods. 

From these two plots we can clearly identify potential cluster members. We select all objects with 0.9\,mas\,$< p <$\,1.5\,mas, -4.0\,mas/yr\,$< PM_{RA} <$\,0.0\,mas/yr, and -5.0\,mas/yr\,$< PM_{DEC} <$\,-1.0\,mas/yr as YSOs in IC\,5070. This selection, especially in parallax, seems rather generous. However, note that \citet{2020ApJ...899..128K} estimate for their even slightly wider selection of sources a contamination with non-members of only 3\,\%. We hence estimate that at worst one of the selected YSOs is potentially a non-member. Our selection removes 19 of the periodic objects from the sample and we are left with 40 YSO cluster members of IC\,5070. Note that our sample of periodic YSOs has only been selected based on periodic variability, parallax and proper motion. At no stage has any selection based on colour or brightness been made. 

\section{Discussion of Period finding Methods}\label{res_methods}

In this section we discuss the results of the period finding methods. This refers to all 59 identified periodic variables. 

\subsection{Completeness and contamination}

We summarise the results of the identification of periodic variable objects in Table\,\ref{methods}. In this table we list the total number of candidates selected by each specific method, as well as the number of sources from the master list that each method selects. Based on the total number of 59 sources in the master list, we also determine the completeness and contamination for each method. We define the completeness of a method as the fraction of sources from the master list it identifies. Similarly, the contamination of a method is defined as the fraction of periodic candidates it identifies that are not part of the master list.

There is a wide range of success (or lack thereof) in finding periodic variables. The completeness ranges (with one exception discussed below) from 37\,\% to 64\,\%, while the contamination can be as low as just 10\,\% and as high as 60\,\%. Generally there is an anti-correlation between these two values. There is no single stand-out method that clearly outperforms all the others in either completeness or contamination. The best method in both metrics is L1Beta with a completeness of 64\,\% and a low contamination of just 10\,\%. There are four methods (L1Beta, L2Beat, L1Boot, GLS) with a more than 50\,\% completeness. They all perform similarly in terms of completeness but vary in contamination from 10 to 43\,\%. 

We have analysed how well these four best methods perform when used together. They only miss nine of the 59 periodic variables from the master list. This corresponds to a completeness of 85\,\% - a significant improvement over any of the individual methods. In total these four methods find 72 unique candidate objects. Thus, the contamination is 21 out of 72 sources, i.e., 29\,\%. With a completeness of 49\,\% the L2Boot methods comes very close to the four best methods. If one would combine these five best methods, then only one additional source would be found. This would slightly increase the completeness to 86\,\%, but the contamination would increase to 31 out of 83, i.e., 37\,\%. 

Most of the 59 stars in the master list are identified by multiple methods. Indeed only 12 of the objects are solely found by one single method. Five of those are found by one of the four best methods, the other seven are found by one of the other methods, without any single one prone to identify objects that all the others do not find. Thus, if one considers the sources found only by one of the methods as less reliable, then the four best methods identify 45 of the 47 objects, which is a completeness of 96\,\%. There is no tendency for these single-method sources to be of a certain type (YSO, fore/background star). 

In our analysis, one method (L2persp), which is based on fitting together splines and a sinusoidal wave, seems to fail completely at achieving the task. Indeed it does not find any period that was judged to be real. The problem with this model is that the overall behaviour of the light curve is captured by non periodic splines while the sinusoidal part is fitting periods along the spline lines, thus failing to capture the main underlying periods. Therefore, we do not recommend using it for the purpose of finding periodic variables in HOYS light curves or similar data sets.

\subsection{The ideal combination of period search routines}

Given that the completeness values for the four best methods are similar, it seems that none of these clearly outperforms any of the others. We investigated, however, whether there are significant differences (other than completeness and contamination) between the methods. These results are shown in Table\,\ref{methods_agree}. In the top right part of the table we list the fraction of all sources from the master list where two of the methods agree (for all combinations of the four best methods). The bottom left part lists the fraction of sources where the methods disagree. 

One can clearly see that GLS differs from the other methods, in that it typically agrees with them for only 56--58\,\% of the objects, while the other methods have agreements between 70\,\% and 80\,\%. Thus, if computing time is a limiting factor, the combination of GLS with one of the other three methods (L2Beta, L1Beta, L1Boot) would provide the highest completeness. Given the high completeness and very low contamination of L1Beta, the combination of this method with GLS should be the choice if only two period detection methods are used.

The four best methods, as described above, are L2Beta, L1Beta, L1Boot and GLS. A common attribute of these is that they are all based on fitting a sinusoidal wave to the light curves. The two methods L1Beta and L1Boot are based on a form of robust regression. They should be more resilient in the presence of outliers and therefore maybe more suited to heterogeneous data sets (like HOYS) than other methods. 

In summary, a combination of several period finding methods (L2Beta, L1Beta, L1Boot, GLS) provides the most robust way to identify periodic variables in our HOYS data. This combination maximises the completeness of the period sample (85\,\%) and achieves a contamination of lower than 30\,\%. Using more methods will generally slightly increase the completeness but comes at cost of increased computing time and contamination. At least two methods (preferably L1Beta and GLS) should be combined.

\begin{table}
\caption{\label{methods_agree} In this table we show, for the {\it best} 4 methods, the percentage of sources from the master list which each two methods find as periodic variables (top right), and the percentage that only one of them finds it (bottom left). }
\centering
\begin{tabular}{|c|c|c|c|c|}
\hline
 & L2Beta	& L1Boot	& L1Beta  & GLS \\ \hline
L2Beta & -	& 	71 & 	80 & 58 \\
L1Boot & 	29 & -	& 	76 & 	56 \\
L1Beta & 	20 & 	25 &   -& 	58 \\
GLS	   & 42	& 44 	& 42 & - \\ \hline
\end{tabular}
\end{table}

\subsection{Comments on eyeballing}

Visual examination of light curves (`eyeballing') was a crucial part of our period search routine. The value of eyeballing becomes apparent when comparing the sample of candidate periods with the final master sample. As can be gleaned from the completeness rates in Table\,\ref{methods}, even in the best case, fewer than two thirds of the robust and reliable objects are identified as candidate periodic variables by any method. Typically it is only a third to half. This is particularly relevant for `clumpy' light curves as they are typically obtained from ground-based long-term monitoring. 

We investigated the occasions where the two team members (AS, JE) who eyeballed light curves to check the candidate periods disagreed with each other. In total there are 19 stars for which this happens. However, 12 of these sources are in our master list. This is due to the fact that the periods measured by the individual methods slightly differ, i.e., the folded light curves for the same object will look slightly different between different methods. For these 12 objects, the two team members agreed in their assessment at least in one method, but disagreed in at least one other. The seven other candidate periods with disagreement between the two assessors come from a variety of methods, without preference for one particular method. 

We note that our approach to eyeballing yields a very robust sample of periods, but not necessarily a complete one. For example, we insisted on looking at the data itself, without an over plotted running median. Relaxing this constraint would have led to a larger sample of confirmed periods, but possibly slightly increasing the contamination in the master sample. In addition to the specific computational method, the exact design of the period search and the criteria adopted for accepting a period are relevant and need to be specific to enable meaningful comparisons between period searches. 

\subsection{Comparing with the literature}

To our knowledge, this study is one of the most comprehensive comparisons of period search algorithms and their various implementations, as commonly used in the literature. Typically, period searches in astrophysical data sets have focused on very few methods, and eyeballing is rarely carried out on blinded data sets. 

\citet{2011MNRAS.413.2595S} have run four different period searches, using entirely independent approaches, for ground-based light curves for low-mass stars in the open cluster Praesepe. The results are then combined and used to assess the reliability of the final period sample. Three of the four methods are based on periodograms and sine-fitting, but the implementation differs and uses different criteria for accepting a period. The fourth method used was the string-length method \citep{1983MNRAS.203..917D}, which stood out as being less complete. The techniques based on sine-fitting on the other hand yield comparable results, mirroring the results obtained in this current (more comprehensive) study.

For simulated light curves mimicking the data from the Kepler space telescope, \citet{2015MNRAS.450.3211A} carried out a blind period recovering test, using a variety of methods. Since these are light curves with uniform cadence and without the typical sampling issues in ground-based data, a wider range of period search algorithms are available, including auto-correlation. Contamination is less of a problem in this type of data, and many types of period searches perform similarly well.

\begin{figure*}
\centering
\includegraphics[width=1.02\columnwidth]{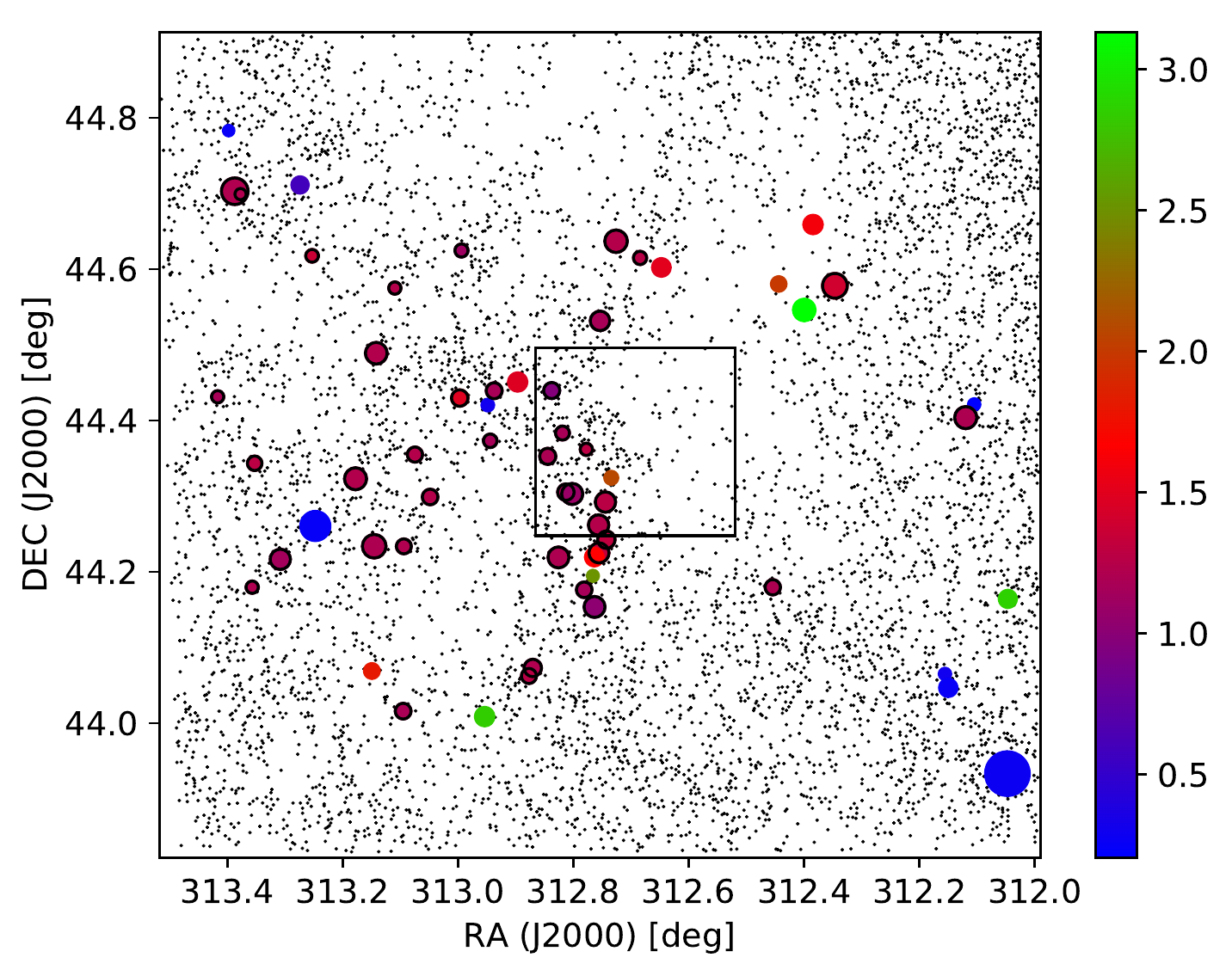} \hfill
\includegraphics[width=0.98\columnwidth]{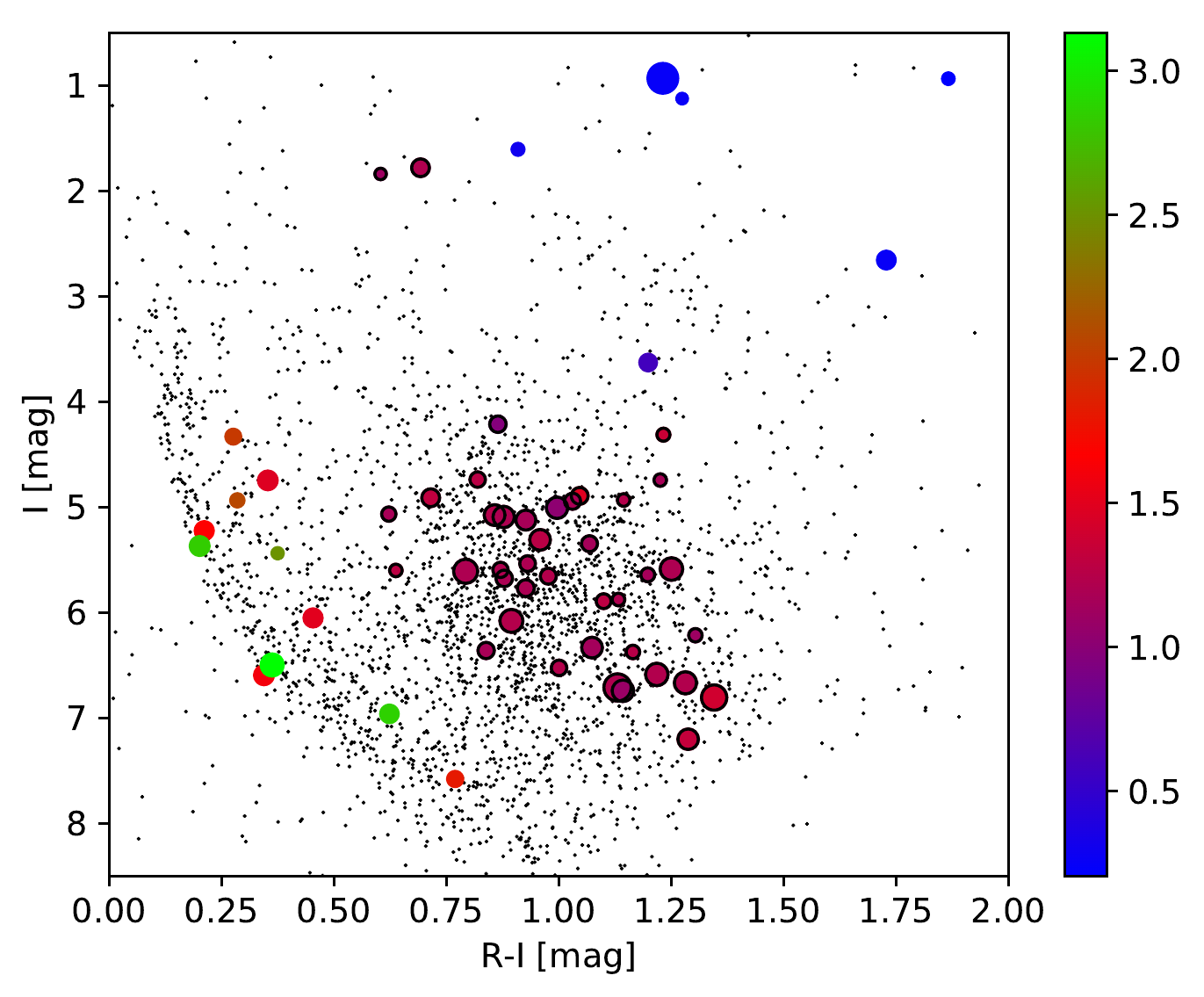}
\caption{{\bf Left:} Sky position of all stars (small dots) in our IC\,5070 field. The periodic variable YSOs are over plotted as coloured circles. The colours indicate the parallax in mas and the size of the circle is proportional to the period. Sources that are part of our final IC\,5070 YSO sample are surrounded by a black circle. The square field is the area investigated for periodic variables by \citet{2019A&A...627A.135B}. {\bf Right:} R-I vs. absolute I colour magnitude diagram of the HOYS data in the IC\,5070 field. No extinction correction has been applied to determine the absolute magnitudes. Median magnitudes along each light curve for all stars are shown. Symbol size and colours are the same as in the left panel. Several of the background giants have R-I colours outside the plot area and are not shown. \label{i_ri_cmd}}
\end{figure*}

\section{Discussion of Periodic Variables}\label{res_variables}

\subsection{Comparison with published periods}

We are only aware of one other systematic search for periodic variables in the IC\,5070 field, conducted by \citet{2019A&A...627A.135B}. They undertook deep imaging of a 16$\arcmin \times 16\arcmin$ field, which we have indicated as a square in the left panel of Fig.\,\ref{i_ri_cmd}. With their much deeper data, they find 56 periodic variables in this field. Based on the magnitudes of the stars, at the very best our HOYS data will only detect 30 of those. With our conservative selection of periodic variables, our sample contains only six of their 56 objects. However we also find three additional periodic variables that are not listed in \citet{2019A&A...627A.135B}. Two of those sources are identified as variable by them, but not as periodic. These six matching periodic sources are indicated in Tables\,\ref{source-list} and \ref{source-list-nonyso}. All but one source are part of our YSO list. The exception is object ID 6592, which we identify as part of the foreground population. From the periodic matches, only one (ID 6592) has a different period. 

We further matched our master list against the ASAS-SN list of variables \citep{2018MNRAS.477.3145J}. Only three of our objects have a counterpart with a period, and all three given periods are consistent with ours. We are only aware of two other known periods amongst our periodic variable sample. One is in the YSO list (7181, LkHA\,146, \citet{2018RAA....18..137I}) and one in the non-YSO list (7896, V\,1598\,Cyg, \citet{2020MNRAS.497.4602F}). With the one exception from \citet{2019A&A...627A.135B}, all published periods agree with the ones found in our analysis. All this information, as well as the commonly used designations for our sources obtained from SIMBAD\footnote{http://simbad.u-strasbg.fr/simbad/sim-fid} are also listed in Tables\,\ref{source-list} and \ref{source-list-nonyso}. Thus, to the best of our knowledge, we have measured periods for 50 of our periodic sources for the first time. This includes 34 of the periodic YSOs in the field.

\subsection{Spatial distribution}

In the left panel of Fig.\,\ref{i_ri_cmd} we show the position of all investigated stars in our roughly one square degree survey field as small black dots. As coloured symbols we over plot the periodic variables and indicate their parallax in a colour code and the period by the symbol size. The YSOs are shown as black edged symbols. The vast majority of YSOs are situated in the western half of the field. The foreground and background population of periodic objects however, seems to be homogeneously distributed. The distribution of our YSO population is very similar to the one found in \citet{2020ApJ...899..128K} (see their Fig.\,3). Indeed, based on their proper motions, the majority (about 35) of our YSOs seems to be associated with the expanding group D, while about five should be part of the compact group C. The fraction of sources belonging to the two groups is roughly identical to the number ratio of group members from the YSO sample established in \citet{2020ApJ...899..128K}. Hence, there is no preference of finding periodic variable stars in either of the two YSO groups. 

\begin{figure*}
\centering
\includegraphics[width=1.02\columnwidth]{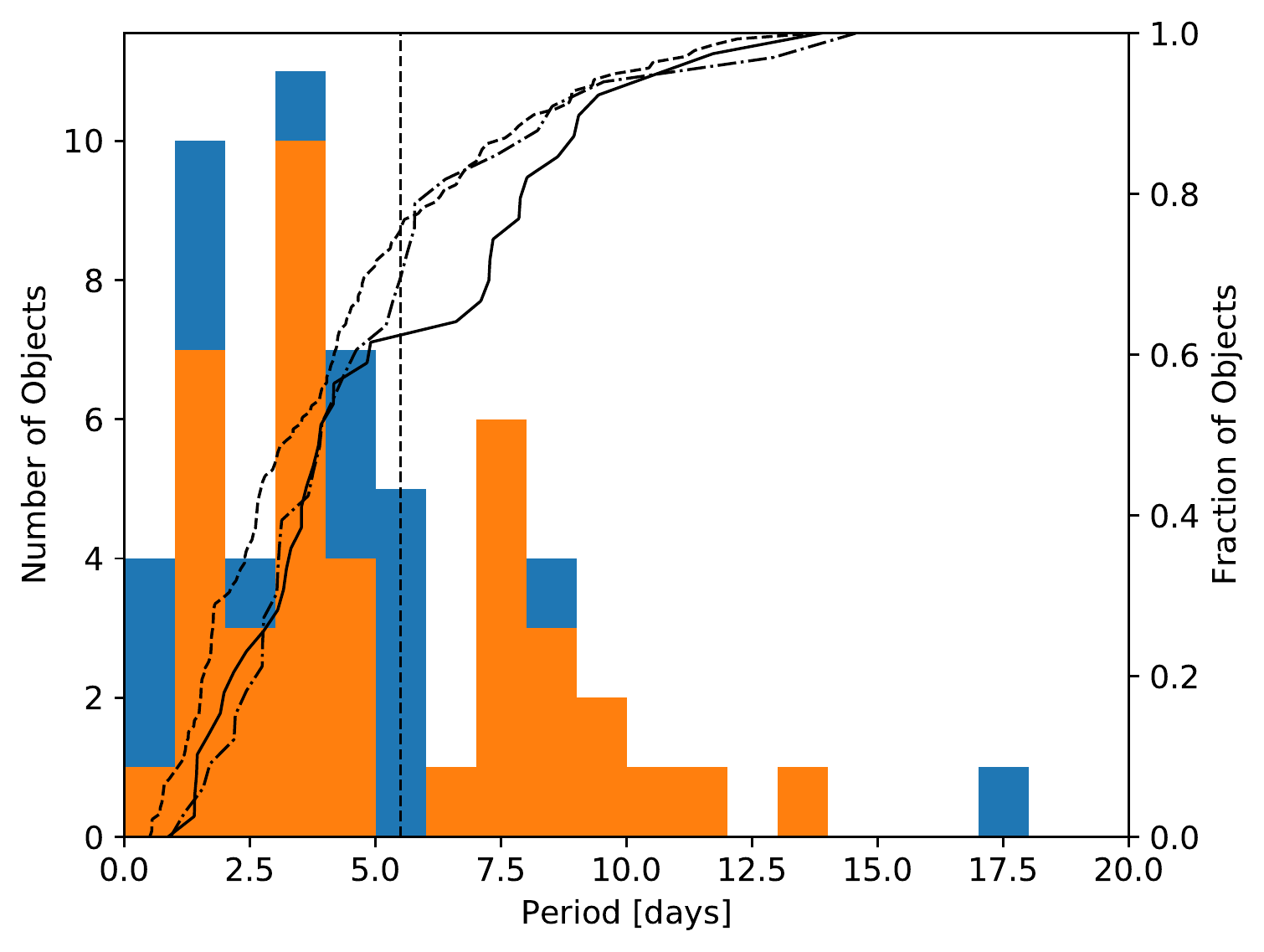} \hfill
\includegraphics[width=0.98\columnwidth]{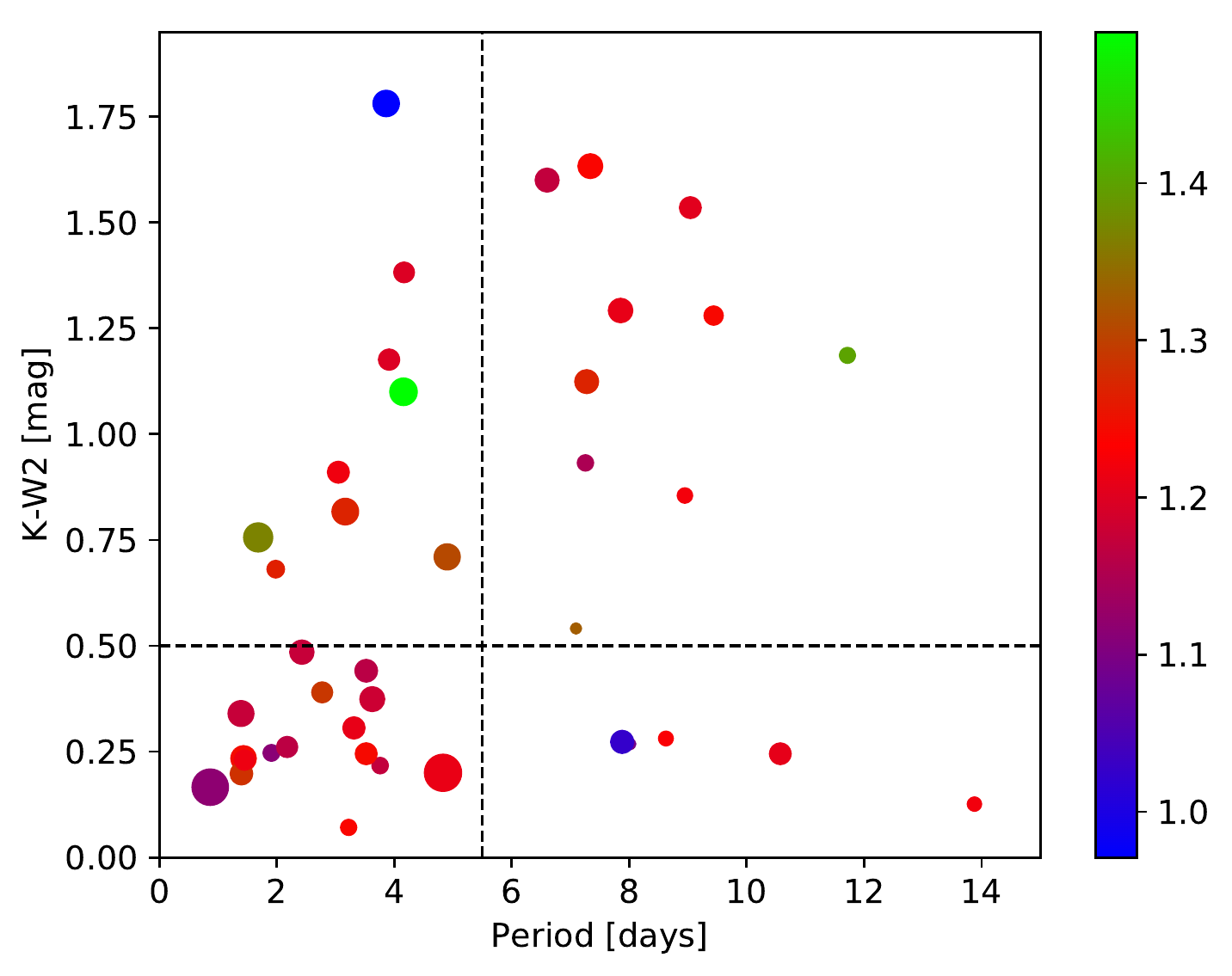}
\caption{{\bf Left:} Distribution of the periods in our sample, limited to 20 days. In blue we show all 59 sources, and in orange the IC\,5070 YSO sample is shown. The over plotted lines are CDFs of our YSO sample in IC\,5070 (solid), the $\rho$-Oph sample from \citet{2018AJ....155..196R} (dashed), and the Taurus sample from \citet{2020AJ....159..273R} (dot-dash). The vertical dashed line indicates adopted separation of fast and slow rotators at P\,=\,5.5\,d. {\bf Right:} Dependence of the period on the K-W2 colour of the YSO sample. The colour code represents the parallax in mas and the symbol size the median I-band brightness of the stars (small = faint). The two dashed lines indicate the adopted separation of objects with and without disk at K-W2\,=\,0.5\,mag (horizontal) and the separation of fast and slow rotators as in the left panel (vertical).  \label{per_hist}}
\end{figure*}

\subsection{Colour magnitude diagram}

In the right panel of Fig.\,\ref{i_ri_cmd} we show the R-I vs. absolute I-Band colour magnitude diagram of the HOYS field as black dots, with the periodic variables over plotted as coloured circles. The R and I magnitudes for all stars are determined as the median along each light curve. The absolute magnitudes are determined using the Gaia\,EDR3 parallaxes. No extinction correction has been applied. The symbol colour encodes the parallax in mas and the symbol size is proportional to the rotation period. One can clearly identify the main sequence population of stars in the field, as well as the YSO population to the top right of the main sequence. Indeed most of our sample of periodic variables belongs to the latter group. In the top right of the figure, we find reddened background giants, some of which have been cut off due to their extreme R-I colours. 

We matched all the objects in our master catalogue against the list of spectroscopic observations of YSOs in \citet{2020ApJ...904..146F} to obtain spectral types, effective temperatures, luminosities and extinction values. These are all listed in Table\,\ref{source-list}. Only one of the 19 sources removed from the sample as potential non-YSOs has a match in that catalogue. This is object ID 3197 (LkHA\,177), which has a very uncertain parallax (Renormalised Unit Weight Error: RUWE\,=\,12.8) and hence could still be a cluster member. Indeed, in the colour magnitude diagram in the right panel of Fig.\,\ref{i_ri_cmd}, it is situated just above the main cluster of YSOs. If one assumes a parallax of 1.2\,mas, as for the other sources, it would fall directly into the main group of the other IC\,5070 YSOs. Of our 40 YSOs, only 9 have no match in \citet{2020ApJ...904..146F}. Generally these are the fainter sources in our sample, with the exception of the two bright YSOs (IDs 3220, 9267) with an absolute I magnitude of about 2\,mag. It is not clear why they have not been included in the \citet{2020ApJ...904..146F} sample.

We list all our periodic YSOs with their properties in Table\,\ref{source-list}.We show the source ID number, the RA, DEC (J2000) of the GAIA cross-match, the period, the source properties (spectral type, effective temperature, optical extinction, and luminosity) from the cross-match to \citet{2020ApJ...904..146F} and additional notes. The properties of the 19 objects removed from our list are discussed in Sect.\,\ref{nonvariables} and their properties are listed in Table\,\ref{source-list-nonyso}. 

\subsection{Period Distribution}

We show the distribution of the detected periods in the left panel of Fig.\,\ref{per_hist}. In blue all 59 objects from the master list of periodic variables are plotted, while the orange overlay contains only the 40 YSOs identified in our sample. Two clear groups of objects are evident, one with short periods (1--5 days) and one with longer periods (6--10 days), with a clear gap without sources at P = 5--6 days. The figure also shows the cumulative distribution function (CDF) for the YSOs (solid line). The CDF illustrates that the gap in our YSO periods between five and six days is genuine and not a result of the histogram binning. On the other hand the apparent split of the short period objects into two groups in the histogram is not supported by the CDF and is likely a binning artefact. We use a KS-test to evaluate if the YSO period distribution could have been drawn from a homogeneous distribution. We find that the null hypothesis of uniformly distributed periods can be excluded with a 98.9\,\% probability.

We also show the CDF for YSO periods in the $\rho$-Oph and Taurus star forming region for comparison, as determined from Kepler/K2 data by \citet{2018AJ....155..196R, 2020AJ....159..273R}, in the left panel of Fig.\,\ref{per_hist}. Note that we only selected stars from these two regions which fell in the range of periods (0.5\,d\,$<$\,P\,$<$\,15\,d) and V-K colours (1.8\,mag\,$<$\,V-K\,$<$\,4.7\,mag) of our YSO sample. The K2 data does not feature the typical daytime and weather gaps of ground-based data and is therefore a good reference point. The bi-modality seen in our sample is also visible in the K2 periods and is thus considered a real feature of the period distribution of stars in IC\,5070. Moreover, bi-modality in YSO periods has been observed in numerous campaigns, going back to the 1990s \citep{1993AJ....106..372E, 2001ApJ...554L.197H, 2005A&A...430.1005L}. The exact position of the peaks will depend on mass and age of the population. In IC\,5070, YSO periods peak around 3\,d and 8\,d, which is comparable to period distribution of low-mass stars in other very young star forming regions, for example in the ONC \citep{2007prpl.conf..297H, 2010A&A...515A..13R}. The bi-modality is usually attributed to the fact that the presence of disks slows down the rotation. We further investigate the link between disks and rotation periods in the next subsection. 

\subsection{Infrared excess}

In the right panel of Fig.\,\ref{per_hist} we show the K-W2 colours of our objects against the period. This colour is an excellent indicator for the presence or absence of a disk -- stellar photospheres should have K-W2\,$<$\,0.5\,mag, i.e., higher values indicate infrared excess due to circumstellar dust (see \citet{2020A&A...642A..86T} for a discussion). According to this plot, the fast rotators with P\,$<$\,5.5\,d tend to have no disk, whereas the slow rotators with P\,$>$\,5.5\,d predominantly do have a disk. In particular there are 9 fast rotators with disks, 5 slow rotators without disk, 10 slow rotators with disk, and 16 fast rotators without disk. We investigated if this distribution can occur by chance. We assume a homogeneous distribution in period and K-W2 colour. A simple Monte Carlo simulation draws 40 objects randomly from the parameter space and we check how often the resulting distribution is as asymmetric, or more asymmetric, than the observed one. We find that the probability that our observed distribution is drawn by chance is of the order of 4.8\,$\times$\,10$^{-3}$.

Thus, our data are consistent with the idea that the presence of a disk slows down the rotation (e.g. \citet{2007prpl.conf..297H}). Fast rotators without disks have spun up due to their pre-main sequence contraction. The slow rotators without disks could be stars that have lost their disks recently and have not had time yet to spin up. The overall appearance of this plot of period vs. infrared colour is consistent with previous work in other regions (see in particular \citet{2006ApJ...646..297R}). If we use K-W2\,$=$\,0.5\,mag as the threshold for disks, we find that the disk fraction amongst our periodic YSO sample is 50\,\%. Using an approximate disk fraction age relation, such as in \citet{2009AIPC.1158....3M}, we find that this disk fraction is in good agreement with the age estimates for the IC\,5070 region of the order of 1\,Myr by \citet{2020ApJ...904..146F}.

\begin{figure}
\centering
\includegraphics[width=\columnwidth]{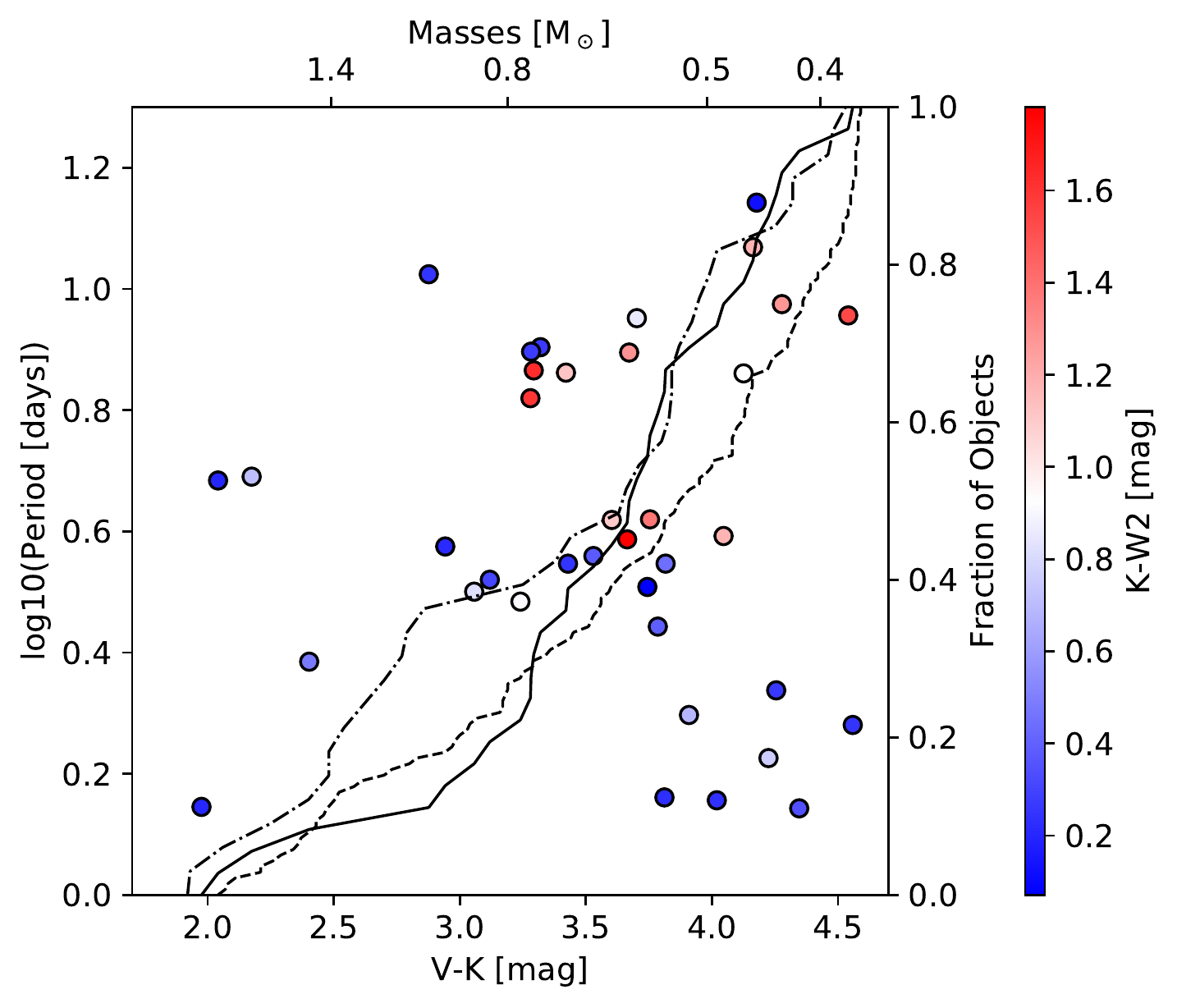}
\caption{V-K colour against period of the IC\,5070 YSO sample. The symbol colour represents the K-W2 colour. The mass estimates are based on 1\,Myr isochrones from \citet{2015A&A...577A..42B}. V-K colours have been de-reddened using the median Av values from the matches in \citet{2020ApJ...904..146F}. The colour code indicates the K-W2 colour. The over plotted lines are CDFs of our data in IC\,5070 (solid), the $\rho$-Oph sample from \citet{2018AJ....155..196R} (dashed), and the Taurus sample from \citet{2020AJ....159..273R} (dot-dash). \label{mass_period}}
\end{figure}

\subsection{Mass dependence}

To investigate the mass dependence of the YSO rotation periods we estimate de-reddened V-K colours as mass proxy for the stars. In principle we could use the individual A$_{\rm V}$ values from \citet{2020ApJ...904..146F} for this task. However, there seems to be a slight bias in those values that can be seen in Table\,\ref{source-list}. Our sample of matched objects mostly contains K-type stars with effective temperatures of about 4000\,K and typically about A$_{\rm V} = 2$\,mag. There are, on the other hand, three G-type stars in the sample with higher temperatures above 5000\,K. These three objects have A$_{\rm V}$ values of  5\,mag or slightly higher. This seems unusual and we hence decided to use the median A$_{\rm V}$ values of all matched sources to de-redden the V-K colours. We applied an extinction law of E(V-K)\,=\,0.89\,A$_{\rm V}$ \citep{1990ARA&A..28...37M}. The results are shown in Fig.\,\ref{mass_period}. We show the colour coded K-W2 values, i.e., the presence of a disk. Note three objects lack V or K, so are not shown. The mass estimates in Fig.\,\ref{mass_period} are based on 1\,Myr isochrones from \citet{2015A&A...577A..42B}. We also over plot the CDF of the V-K colour (mass proxy) as a solid line in Fig.\,\ref{mass_period}. Similarly, the CDFs for $\rho$-Oph and Taurus are also shown \citep{2018AJ....155..196R, 2020AJ....159..273R}. 

The plot demonstrates again that our period sample for YSOs is broadly consistent with reference samples for YSOs in other regions. Fast rotators are found at the high-mass end in our sample (above 1.4\,M$_\odot$) and at very low masses (below 0.7\,M$_\odot$). In between, the sample only contains stars with periods above $\sim 3$\,d. However, the statistics are too poor to draw definitive conclusions about mass-period trends from this plot, in particular given the incomplete knowledge of extinction values in this sample. 

\subsection{Non-YSO periodic variables}\label{nonvariables}

In our master list of periodic variables there are 19 objects with parallax and proper motion indicating they are likely not members of the cluster of young stars in IC\,5070. These sources stand out in position and colour code in the colour magnitude plot shown in Fig.\,\ref{i_ri_cmd}. Several sources have very small parallax values, are very red and are intrinsically bright. These are most likely background giants, potentially heavily reddened. The other obvious group are the stars on or slightly above the main sequence. As can be seen, this group splits into two populations: most are in the foreground, while the others are in the vicinity of the cluster. We summarise the properties of these 19 sources in Table\,\ref{source-list-nonyso}. We list our ID number, the RA, DEC (J2000) coordinates of the Gaia cross-match, the period, Gaia distance and additional notes. Below we briefly describe the three sub-groups of non-YSOs in our sample of periodic objects.

\subsubsection{Background Giants}

All seven stars in this group have distances above 2.7\,kpc and are thus background sources. The group contains the objects with the two longest periods (11246 - 41\,d, 7159 - 17.2\,d), both without IR excess. These are most likely pulsating giants. The star 9522 has a period of almost 5\,d. The shape of the light curve indicates the star might be a $\delta$\,Cep object. The amplitude of the variations is about 0.4\,mag. Three stars in this group (2287, 5715, 9324) have periods shorter than one day. These sources could be $\delta$\,Scuti stars.

Object 5686 (also know as V1706\,Cyg) has a light curve which clearly resembles an eclipsing binary and has a period of 1d 10hr 3m. This is the only object where our procedure has found half of the correct period. We have manually adjusted the period of the object by a factor of two. This is also the only obvious eclipsing binary light curve in our sample. The phase-folded light curve shows that the primary and secondary eclipses have an almost identical depth of about 0.75\,mag. Thus, the object seems to be an equal size binary, contrary to its classification as Orion type variable in SIMBAD.

\subsubsection{Foreground YSOs}

This group contains eight objects which are mostly near the bottom of the main sequence in our colour magnitude diagram. All are clearly in the foreground to the cluster. One object, 7896 (V1598\,Cyg), is situated clearly above the main sequence. This source has no infrared excess and \citet{2020MNRAS.497.4602F} concluded that it is most likely a close binary and/or a foreground YSO. Most of the sources have no or only marginal infrared excess. All but one of the others (6592), have longer periods between about five and eight days. These sources most likely represent a foreground population of young main sequence stars, binaries or older weak line T\,Tauri objects. 

\subsubsection{Potential Cluster Members}

The remaining group contains four sources. They are all significantly above the main sequence in the colour magnitude diagram. All but one (4421) have a clear infrared excess. In particular 3197 (LkHA\,177), is a known emission line star and treated as a cluster member in \citet{2020ApJ...904..146F}. All sources suffer from very uncertain parallax measurements, with large RUWE values. We also checked the distance estimates from \citet{2020BAAA..61C...38A} using Gaia DR2 parallax plus additional colours from Pan-STARRS-1, 2MASS, and AllWISE. But only objects 4286 and 4421 are matched. They have similar distance uncertainties as in Gaia EDR3. The periods of the objects in this group range from three to six days. It is conceivable that all of them are potential YSOs and cluster members. However, given the uncertain parallax values, we have not included them in the YSO sample. 

\begin{table*}
\caption{\label{source-list}List of all 40 periodic YSO variables identified in our work. We list the following properties: Our object ID number; RA, DEC the equatorial (J2000) coordinates coordinates of the cross matched GAIA source; Period in days; Spectral type, effective temperature, optical extinctions and luminosities from cross matches to \citet{2020ApJ...904..146F}; Additional notes such as the SIMBAD identifier (if it exists) and periods found in other surveys. For objects with uncertain parallax we also note the RUWE value from GAIA\,EDR3.}
\centering
\begin{tabular}{rrrrcccr|p{7.5cm}}
ID & RA & DEC & Period & SpT & Teff & Av & L & Notes \\ \hline
 & [deg] & [deg] & [days] & & [K] & [mag] & [L$_\odot$] & \\ \hline
 3220 & 313.37768 & 44.69840 &  0.8661 & -    & -    & -   & -      & RUWE\,=\,3.544 \\
 3314 & 313.38780 & 44.70225 & 13.8783 & -    & -    & -   & -      & -  \\
 3988 & 312.72581 & 44.63562 &  9.4377 & K8.2 & 3928 & 2.4 &  1.025 & 2MASS\,J20505418+4438083 \\
 4097 & 313.25278 & 44.61654 &  1.6827 & G6.4 & 5350 & 5.2 & 23.630 & RUWE\,=\,5.459 \\
 4101 & 312.99370 & 44.62339 &  1.9081 & K7.8 & 3946 & 3.9 &  1.541 & - \\
 4198 & 312.68440 & 44.61357 &  1.9818 & -    & -    & -   & -      & - \\
 4446 & 313.10924 & 44.57396 &  1.4334 & G5.0 & 5500 & 5.0 & 11.088 & - \\
 4476 & 312.34682 & 44.57701 & 11.7162 & -    & -    & -   & -      & 2MASS J20492323+4434373 \\
 4766 & 312.75374 & 44.53048 &  6.6024 & K5.4 & 4091 & 2.0 &  0.856 & V* V1701\,Cyg \\
 5119 & 313.14145 & 44.48802 &  8.6252 & K5.4 & 4091 & 3.6 &  1.256 & - \\
 5535 & 312.83745 & 44.43877 &  3.8622 & K8.3 & 3921 & 2.2 &  2.948 & V* V1703\,Cyg, EM* LkHA\,153; V\,121 in {\citet{2019A&A...627A.135B}}, P\,=\,3.840\,d; RUWE\,=\,8.379 \\
 5548 & 312.99661 & 44.42881 &  4.1573 & K7.9 & 3943 & 2.4 &  3.000 & EM* LkHA\,164; RUWE\,=\,3.215 \\
 5559 & 312.93711 & 44.43862 &  3.7590 & K8.0 & 3940 & 1.6 &  0.480 & - \\
 5575 & 313.41700 & 44.43060 &  1.3901 & G8.7 & 5150 & 5.0 & 11.347 & 2XMM\,J205340.1+442550; RUWE\,=\,2.659 \\
 5886 & 312.12005 & 44.40321 &  9.0413 & -    & -    & -   & -      & 2MASS\,J20482880+4424115 \\
 6060 & 312.81885 & 44.38279 &  2.4266 & K4.2 & 4291 & 1.4 &  1.130 & V* V1702\,Cyg, EM* LkHA\,152; V\,128 in {\citet{2019A&A...627A.135B}}, P\,=\,2.420\,d \\
 6149 & 312.94395 & 44.37257 &  2.1763 & K8.2 & 3928 & 2.9 &  1.935 & - \\
 6259 & 312.77765 & 44.36132 &  1.3979 & K1.9 & 4775 & 2.3 &  1.128 & not in {\citet{2019A&A...627A.135B}} but in survey area \\
 6315 & 313.07439 & 44.35443 &  3.2233 & K7.6 & 3952 & 2.6 &  0.873 & - \\
 6337 & 312.84446 & 44.35212 &  3.9113 & K7.2 & 3964 & 2.1 &  1.258 & EM* LkHA\,154; V\,178 in {\citet{2019A&A...627A.135B}}, but no period found \\
 6393 & 313.35269 & 44.34279 &  2.7728 & K6.6 & 3990 & 2.6 &  1.539 & - \\
 6620 & 313.17748 & 44.32255 &  8.9487 & M0.8 & 3657 & 2.0 &  0.601 & - \\
 6813 & 312.81307 & 44.30490 &  4.1672 & K7.8 & 3946 & 2.1 &  0.913 & EM* LkHA\,150; V\,182 in {\citet{2019A&A...627A.135B}}, but no period found \\
 6856 & 312.80221 & 44.30257 &  8.0136 & K8.0 & 3940 & 1.8 &  0.347 & 2MASS\,J20511252+4418093; V\,105 in {\citet{2019A&A...627A.135B}}, P\,=\,7.954\,d \\
 6861 & 313.04822 & 44.29854 &  3.5217 & K4.2 & 4292 & 2.9 &  1.993 & - \\
 6929 & 312.74460 & 44.29190 &  7.2758 & K8.4 & 3916 & 1.7 &  1.256 & EM* LkHA\,145; V\,107 in {\citet{2019A&A...627A.135B}}, P\,=\,7.223\,d; RUWE\,=\,3.565 \\
 7181 & 312.75654 & 44.26168 &  7.3382 & K6.8 & 3979 & 1.5 &  1.562 & EM* LkHA\,146; P\,=\,7.365\,d in {\citet{2018RAA....18..137I}}; V\,106 in {\citet{2019A&A...627A.135B}}, P\,=\,7.359\,d \\
 7422 & 312.74312 & 44.24232 &  4.9010 & K3.8 & 4373 & 1.5 &  1.868 & 2MASS\,J20505834+4414323; RUWE\,=\,2.102 \\
 7465 & 313.14529 & 44.23348 & 10.5727 & K4.6 & 4216 & 2.1 &  1.172 & - \\
 7472 & 313.09386 & 44.23339 &  3.0487 & K4.1 & 4311 & 2.2 &  1.285 & 2MASS\,J20522252+4414002 \\
 7566 & 312.75600 & 44.22497 &  7.0943 & K9.6 & 3814 & 2.4 &  0.413 & - \\
 7609 & 313.30836 & 44.21606 &  7.2549 & K6.2 & 4010 & 2.9 &  0.851 & 2MASS\,J20531400+4412577 \\
 7632 & 312.82600 & 44.21895 &  7.8531 & K7.1 & 3966 & 1.9 &  1.968 & 2MASS\,J20511824+4413082 \\
 7954 & 313.35736 & 44.17926 &  1.4492 & K6.2 & 4010 & 2.7 &  1.433 & - \\
 8025 & 312.45491 & 44.17952 &  3.3130 & -    & -    & -   & -      & - \\
 8038 & 312.78141 & 44.17628 &  3.5221 & -    & -    & -   & -      & - \\
 8249 & 312.76358 & 44.15360 &  7.8800 & K8.2 & 3928 & 1.9 &  1.543 & - \\
 9267 & 312.87064 & 44.07309 &  4.8298 & -    & -    & -   & -      & P\,=\,4.825535\,d in {\citet{2018MNRAS.477.3145J}} \\
 9321 & 312.87737 & 44.06251 &  3.1660 & K4.5 & 4235 & 2.1 &  3.131 & 2MASS\,J20513057+4403449 \\
 9961 & 313.09561 & 44.01582 &  3.6251 & -    & -    & -   & -      & - \\
\hline
\end{tabular}
\end{table*}

\begin{table*}
\caption{\label{source-list-nonyso} List of all 19 periodic variables that have not been selected as clear YSO members of the IC\,5070 cluster in our work. The table is split into the same three sub-categories of sources as discussed in Sect.\,\ref{nonvariables}. We list the following properties: Our object ID number; RA, DEC the equatorial (J2000) coordinates of the cross matched GAIA source; Period in days; Approximate distances based on the GAIA\,EDR3 parallax; Additional notes such as the SIMBAD identifier (if it exists), periods found in other surveys, and information relating to the placement of the object in the colour magnitude diagram as well as IR excess in the spectral energy distribution obtained via the \href{http://vizier.unistra.fr/vizier/sed/}{VizieR Photometry viewer}. For objects with uncertain parallax we also note the RUWE value from GAIA\,EDR3. }
\centering
\begin{tabular}{rrrrr|p{10.7cm}}
ID & RA & DEC & Period & d & Notes \\ \hline
 & [deg] & [deg] & [days] & [pc] & \\
\hline
\multicolumn{5}{l}{Background  Giants} & \\
\hline
 2287 & 313.39885 & 44.78253 &  0.5433 & 3200 & excess in W3/W4; RUWE\,=\,1.873 \\
 5686 & 312.94814 & 44.41962 &  1.1768 & 3200 & V* V1706\,Cyg; ECB light curve; period manually multiplied by two; P\,=\,0.5883893\,d in {\citet{2018MNRAS.477.3145J}}; non member {\citet{2020ApJ...904..146F}} (A8.6); W3/W4 excess  \\
 5715 & 312.10525 & 44.42062 &  0.9843 & 3600 & slight W4 excess \\
 7159 & 313.24747 & 44.26019 & 17.1910 & 3400 & no IR excess \\
 9324 & 312.15591 & 44.06474 &  0.9980 & 2700 & IRAS 20469+4352; IRAS\,60/100 excess \\
 9522 & 312.15007 & 44.04641 &  4.9559 & 3000 & $\delta$\,Cep like light curve; very weak W4 excess \\
11246 & 312.04652 & 43.93232 & 41.0289 & 2700 & no IR excess \\
\hline
\multicolumn{5}{l}{Foreground YSOs} & \\
\hline
 3791 & 312.38457 & 44.65784 &  5.8050 &  614 & no IR excess; bottom of the MS in R-I vs I CMD \\
 4656 & 312.40003 & 44.54503 &  8.6737 &  316 & slight W3/W4 excess; bottom of the MS in R-I vs I CMD \\
 6592 & 312.73459 & 44.32398 &  1.8893 &  476 & 2MASS\,J20505630+4419262; P\,=\,1.8997111\,d in {\citet{2018MNRAS.477.3145J}}; V\,145 in {\citet{2019A&A...627A.135B}} P\,=\,0.678\,d; slight W4 excess; $\sim$\,1.0\,mag above MS in R-I vs I CMD \\
 7639 & 312.76398 & 44.21933 &  5.1603 &  585 & no IR excess; near bottom of the MS in R-I vs I CMD \\
 7896 & 312.76634 & 44.19463 &  0.8254 &  394 & V* V1598\,Cyg; P\,=\,0.8246 in {\citet{2020MNRAS.497.4602F}}; no IR excess; $\sim$\,1.1\,mag above MS in R-I vs I CMD; RUWE\,=\,1.870 \\
 8151 & 312.04659 & 44.16378 &  4.7818 &  345 & slight W3/W4 excess; $\sim$\,0.5\,mag above MS in R-I vs I CMD \\
 9155 & 313.14977 & 44.06884 &  3.0597 &  552 & W3/W4 excess; $\sim$\,0.6\,mag above MS in R-I vs I CMD \\
10116 & 312.95423 & 44.00872 &  5.7876 &  348 & W3/W4 excess; bottom of the MS in R-I vs I CMD \\
\hline
\multicolumn{5}{l}{Potential Cluster Members} & \\
\hline
 3197 & 313.27392 & 44.71030 &  4.1188 & 1700 & EM* LkHA\,177; Cluster member in {\citet{2020ApJ...904..146F}} (SpT\,=\,K4.5, T$_{\rm eff}$\,=\,4235\,K, A$_V$\,=\,4.2\,mag, L\,=\,5.59\,L$_\odot$); RUWE\,=\,12.811 \\
 4286 & 312.64742 & 44.60108 &  5.1693 &  654 & W3/W4 excess; $\sim$\,0.9\,mag above MS in R-I vs I CMD; RUWE\,=\,2.988 \\
 4421 & 312.44426 & 44.57928 &  2.8862 &  565 & slight W4 excess; $\sim$\,1.7\,mag above MS in R-I vs I CMD; RUWE\,=\,30.629 \\
 5419 & 312.89642 & 44.45004 &  5.6908 &  992 & slight W3/W4 excess; $\sim$\,1.8\,mag above MS in R-I vs I CMD; RUWE\,=\,23.371 \\
\hline
\end{tabular}
\end{table*}

\section{Conclusions}

We have utilised U, B, V, R, I data from the HOYS \citep{2018MNRAS.478.5091F} project to identify a sample of periodic variables in a 1$^\circ \times 1^\circ$ degree field centred around the Pelican Nebula IC\,5070. High cadence data spanning a duration of 80 days in the summer of 2018 have been used. From an initial list of just over 6000 light curves in the field, we identified 59 periodic objects. Using Gaia\,EDR3 parallax and proper motion, 40 sources have been identified as YSO members of the IC\,5070 region. The remaining sources are either background giants, foreground YSOs or potential cluster members with uncertain parallax measurements.

To identify periodic signals in the light curves, nine different periodogram methods have been tested. They rely on fitting sine functions or splines either to the light curves directly or in phase space. Establishing the sample of periodic variables was done in a double-blind manner. The identification of potential periods using the different periodogram methods has been done by two members of the team, without knowledge of the scientific aims of the project, or the expected shape of the light curves. The candidate periodic objects from this first step have been verified independently by two other team members, without the possibility of identifying the nature of the objects or the used period finding method. During this eyeballing process, periodic objects were selected as real if they showed a clear, consistent, periodic behaviour in the phase folded data in at least two filters. Only objects for which the two team members agreed in their selection have been added to our master list of periodic variables.

Based on this master list of periodic variables, we have determined the completeness and contamination of each of the periodogram methods. We find that none of the individual methods clearly outperforms all other methods. The best completeness achieved by any method is 64\,\%, with three others reaching above 50\,\%. The lowest contamination of any method is 10\,\% with some others achieving between 25\,\% and 40\,\%. The common feature of the best performing methods is that they all rely on sine function fitting. They differ however, in the way the sine function parameters are determined. We conclude that for heterogeneous data sets such as from our HOYS project, one should combine period searches using at least the GLS and one other sine fitting periodogram method to obtain as complete a list as possible of periodic variables. Manual quality checks still need to be employed to remove false positives.

We have investigated the properties of our unbiased sample of periodic variable YSOs. They form a clearly identifiable group of stars located above the main sequence in the R-I vs. I colour magnitude diagram. With a probability of 98.9\,\% we can exclude a homogeneous period distribution. Instead a clear split into fast and slow rotators with typical periods of three and eight days, respectively, can be seen. Utilising the K-W2 colour as an indicator for the presence of a disk shows that the fast rotators are predominantly disk-less, while the disk-harbouring objects are mostly part of the group of slow rotators. The probability that the observed distribution in period vs. K-W2 space occurs by chance is determined to be 4.8\,$\times$\,10$^{-3}$. We find a disk fraction of 50\,\% in our YSO sample. De-reddened V-K colours as mass proxy show that fast rotators (P\,$<$\,3\,d) are found at the high mass and low mass ends of our sample, while for roughly solar mass stars only periods above three days are found. All properties of our sample  are in good agreement with studies of samples of periodic YSOs from star forming regions of comparable age, such as $\rho$-Oph or Taurus.

\section*{Acknowledgements}

We would like to thank all contributors of observational data for their efforts towards the success of the \hc\ project.
A.\,Scholz acknowledges support through STFC grant ST/R000824/1. J.\,Campbell-White acknowledges support through STFC grant ST/S000399/1. 
S.\,Vanaverbeke would like to thank Prof. Dr.  Dominik Schleicher (University of Concepcion, Chile) for providing access to the cluster Leftraru on which the GLS computations were done. This research was partially supported by the supercomputing infrastructure of the NLHPC (ECM-02). 


\section*{Data Availability Statement}

The data underlying this article are available in the HOYS database at http://astro.kent.ac.uk/HOYS-CAPS/.


\bibliographystyle{mnras}
\bibliography{bibliography} 


\clearpage\newpage

\appendix
\clearpage\newpage

\section{Periodogram Methods}\label{per_meth_details}

In this Appendix the details of the period search methods used in the analysis will be presented. For each method we will show how the periodogram was obtained. Details specific to each method will also be provided for the initial selection of potential periods. We will use the following notation to describe a light curve for the methods mentioned below:
\begin{equation}
m_{i}=g(t_{i};p)+\epsilon_{i}
\end{equation} 
where $m_{i}$ denotes the magnitude for the $i^{\rm th}$ data point (observed data) at time $t_i$, $g$ is the signal to be estimated given a period $p$ and finally with $\epsilon_{i}$ we denote the error component of our observations that can be assumed to follow a Gaussian white noise distribution, and we can write $\epsilon_{i} \sim N(0,\tau^{2})$. With $N(\cdot)$ we denote the Gaussian distribution, and with $\tau ^{2}$ we denote the variance of the observations.
For some methods the phased data have been used, calculated according to Eq.\,\ref{phase} for each trial period: 

\begin{equation} \label{phase}
f_{ij}=\frac{t_{i}}{p_{j}}-\lfloor \frac{t_{i}}{p_{j}} \rfloor \in [0,1), \quad i=1,...,n \quad j=1,\dots,J 
\end{equation}
where $\lfloor \kappa \rfloor$ denotes the largest integer smaller or equal to $\kappa$ (floor), e.g. $\lfloor 3.7 \rfloor=3$. Furthermore, the measurement errors ($\sigma$) have been taken into account for some methods in the form of weighted regression as seen in Eq.\,\ref{wlc}. We denote the weights as $w_{i}=1/\sigma_{i}$ such that:

\begin{equation} \label{wlc}
m_{i}w_{i}=g(t_{i};p)w_{i}+\epsilon_{i}
\end{equation} 

\subsection{Determination of valid periods}

Any periodogram will always have a value that has a higher power than the rest, and that holds true even if there is no periodic signal in the data. We need methods that can distinguish between periodogram entries corresponding to real periods, and periodogram entries that are the outcome of noisy data. This problem is mostly addressed from the hypothesis test point of view. The general form of the test is the following: $[H_{0}$: The Periodogram entry for a given period comes from data with no signal$]$ vs. $[H_{1}$: otherwise$]$.

As we mentioned previously, many periodograms are based on least squares regression. In this case, it is very common to calculate the periodogram entries using the coefficient of determination ($R^{2}$) statistic. For this particular scenario, the periodogram can be assumed to follow a Beta distribution under the null hypothesis. This result can be found in \cite{1998MNRAS.301..831S} for example, and can be used to determine the probability of observing a periodogram entry as extreme or more extreme from the one observed, under the assumption of only noisy data ($H_{0}$). This probability is usually referred to as the p-value, and small p-values could be an indication of valid periods.

There are cases, however, where the distribution of the periodogram under the null hypothesis is not known or cannot be easily defined. For example, when robust regression is used it is not easy to define the distribution of the statistic used. This problem can be addressed by calculating the desired p-value using Monte Carlo approaches which are based on simulating many noise-only time series (see, for example, \cite{2009ApJ...691.1021D}). A different approach can be found in \cite{thieler2013periodicity} where the authors robustly fit a Beta distribution to the periodogram.

At this point we should note that these tests are conducted multiple times depending on the number of trial periods considered. This should be taken into account when defining the significance level, $\alpha$, or calculating the p-values. There are many methods to correct for the multiple tests; for example, when the tests are independent, one approach can be found in \cite{vsidak1967rectangular}. In our case the tests cannot be considered independent, and for that reason the local maxima of the periodogram entries, corresponding to periods the tests did not reject, are treated as valid. When the tests cannot be viewed as independent, some more sophisticated methods can be employed. An interesting discussion regarding this subject can be found in \cite{algeri2016methods}. 

\subsection{Methods \textbf{L2Beta}, \textbf{L2Boot}}

These methods are based on fitting a sinusoidal wave to the period folded data and least squares regression. The model is shown in Eq.\,\ref{sinmod} given a trial period $j$:

\begin{equation} \label{sinmod}
m_{i}=\beta_{0}+\beta_{1} \sin(2\pi f_{ij}) +\beta_{2}\cos(2 \pi f_{ij})+\epsilon_{i}
\end{equation}

The fitted values are calculated according to $\hat{m}=\hat{\beta_{0}}+\hat{\beta_{1}}\sin(2\pi f_{ij})+\hat{\beta_{2}}\cos(2\pi f_{ij})$, where:

\begin{equation} \label{L2}
\hat{\beta}=argmin_{\beta}\left( \sum_{i=1}^{n}w_{i}(m_{i}-\hat{m}_{i})^{2} \right) 
\end{equation}

The periodogram entries are calculated according to the coefficient of determination statistic $R^{2}$ as seen below, with $\bar{m}$ denoting the arithmetic mean of our observations:

\begin{equation} \label{R2}
Per(p)=\dfrac{\sum\limits_{i=1}^{n}(m_{i}-\bar{m})^{2}-\sum\limits_{i=1}^{n}(m_{i}-\hat{m})^{2}}{\sum\limits_{i=1}^{n}(m_{i}-\bar{m})^{2}}
\end{equation}

For method \textbf{L2Beta} the periodogram and critical value for the hypothesis test are obtained from the R package {\tt Robper} \citep{thieler2016robper}. The periodogram for \textbf{L2Boot} is also obtained from the same package, but the critical value is obtained by bootstrapping the periodogram. See more details in (Derezea et al. in prep). Specifically, the periodogram entries $Per(p)$ for both methods are exactly the same. The only difference between them is the way the hypothesis testing for valid periods was performed. In the first case the test is based on the Beta distribution, whereas for method \textbf{L2Boot} the test is based on re-sampling the periodogram entries. 

\subsection{Methods \textbf{L1Beta}, \textbf{L1Boot}}

For these methods the sinusoidal model is used again as seen in Eq.\,\ref{sinmod} on the phased data. The measurement errors have been taken into account in the form of a weighted regression. The difference is that absolute deviations regression is used to obtain the estimates of the coefficients, $\beta$, according to:
 
\begin{equation} \label{L1}
\hat{\beta}=argmin_{\beta}\left( \sum\limits_{i=1}^{n}w_{i}|m_{i}-\hat{m}_{i}| \right) 
\end{equation}
and the periodogram entries are obtained according to Eq.\,\ref{robR} with $\bar{m}$ now denoting the median:

\begin{equation} \label{robR}
Per(p)=\dfrac{\sum\limits_{i=1}^{n}|m_{i}-\bar{m}|-\sum\limits_{i=1}^{n}|m_{i}-\hat{m}|}{\sum\limits_{i=1}^{n}|m_{i}-\bar{m}|}
\end{equation}

These approaches fall under the umbrella of Robust regression and an analytical discussion of periodograms based on these methods can be found in \cite{thieler2013periodicity}. Using Robust regression methods can prove advantageous, especially under the presence of outliers in our data. For the \textbf{L1Beta} method, the periodogram and critical value are obtained from {\tt Robper}. For the \textbf{L1Boot} method the periodogram is obtained from {\tt Robper}, and the critical value by bootstrapping the periodogram.

\subsection{Methods \textbf{L2spB}, \textbf{L1spB}}

Methods here are based on fitting splines to the phased data instead of a sinusoidal wave and the coefficients are obtained with the use of least squares and absolute deviation regression respectively. The measurement errors have been taken into account in the form of a weighted regression here as well. More details on periodograms based on splines can be found in \cite{oh2004period} and \cite{akerlof1994application}. A simple way to view splines is as fitting piece-wise polynomials to our data, taking into account constraints for the continuity etc.. It could be seen as a combination of fitting polynomials and a step function, thus allowing the capture of more complex shapes. The points at which the piecewise polynomials meet are called knots ($K$). This is a more general model with fewer assumptions on the shape of the data. 

A simple cubic spline model with two knots at $\xi_{k}$ has the form:
 
\begin{equation}
m=\sum_{q=1}^{K+4}\beta_{q}h_{q}(t) +\epsilon
\end{equation}
 
Where, $h_{k}(t)=t^{k-1},\quad k=1,\dots,4$
$$h_{4+k}(t)=(t-\xi_{k})_{+}^{3}, \quad k=1,\dots,K $$
Or simply:
$$m=\beta_{1}+\beta_{2}t+\beta_{3}t^{2}+\beta_{4}t^{3}+\beta_{5}(t-\xi_{1})_{+}^{3}+\beta_{6}(t-\xi_{2})_{+}^{3}+\epsilon_{i}$$
 Where $(t-\xi_{k})_{+}=t-\xi_{k}$ if $t>\xi_{k}$ and zero otherwise. Here B-cubic splines have been used with 4 knots.The $t$ in our case would be the phased data points $f_{ij}$. The periodogram for these methods are obtained from {\tt Robper}. The initially valid periods were determined by bootstrapping the periodograms.

\subsection{Method L2persp}

For this approach cubic splines with 4 knots plus a sinusoidal component (as seen in methods \textbf{L2beta}) were used to fit the data. In this case the model was fitted in time and not phase space. The measurement errors were not taken into account. The periodogram entries were calculated according to Eq.\,\ref{R2} and the critical value to determine potential periods was obtained by bootstrapping the periodogram.

\subsection{GLS}

The method \textbf{L2Beta} described above is conceptually very similar to the generalized Lomb-Scargle periodogram introduced by \citet{2009A&A...496..577Z}. This method also fits a model (like in Eq.\,\ref{sinmod}) to the data using least squares regression, but does not phase fold the data before doing the fit. The generalized Lomb-Scargle periodogram (referred to as \textbf{GLS}) has been developed as an extension of the Lomb-Scargle periodogram \citep{1982ApJ...263..835S}. The Lomb-Scargle periodogram does not include the constant term in the model (Eq.\,\ref{sinmod}). For that reason the term floating-mean periodogram is also often used in the literature to refer to the \textbf{GLS} method. 
 
Our data are taken by many observatories who repeatedly observe almost every clear night. The effect of the observing cadence on the periodogram analysis therefore becomes important with many time series having their strongest peaks at periods of 1 day and corresponding aliases. To mitigate the effect of the observing cadence, we compute the spectral window function as follows:

\begin{equation} \label{spectralwindow}
P_{W}(p)=  \frac{1}{n^{2}} \left( \left[ \sum_{i=1}^{n} \cos\left(\frac{2\pi t_{i}}{p}\right) \right]^2 + \left[ \sum_{i=1}^{n} \sin\left(\frac{2\pi t_{i}}{p}\right) \right]^2 \right).
\end{equation} 

We then select all peaks in the spectral window function which are at least 3 standard deviations above the median and discard all frequencies in the \textbf{GLS} periodogram which are within 0.05 Hz of the frequencies corresponding to the significant peaks in the spectral window function.  

Finally, we assess the statistical significance of the remaining strongest peak in the \textbf{GLS} periodogram by using a bootstrapping method. We re-sample each light curve by applying 1000 random permutations of the data points keeping the time stamps fixed and compute the periodograms for each re-sampled time series. The false alarm probability (FAP) is then computed as:

\begin{equation} \label{FAP}
FAP=\dfrac{n_{e}}{n_{p}},
\end{equation}
where $n_{p}$ denotes the total number of permutations and $n_{e}$ is the number of times the strongest peak in a permuted time series exceeds the maximum periodogram power in the original time series.  

\subsection{LS}

A standard Lomb-Scargle periodogram \citep{1982ApJ...263..835S} was also performed. This version of LS is functionally the same as the GLS method except the GLS also fits a floating mean to the data. The standard LS method assumes that the periodic signal is centered around the mean value of the light curve. Therefore, we have even sampled the light curve for all values in phase space. For light curves where this is true, fitting a floating mean is unnecessary and the standard Lomb-Scargle periodogram can be used. For this method the Python package {\tt astropy.stats} Lomb-Scargle periodogram was used. 

\clearpage\newpage

\begin{figure*}
\section{Phase folded light curves}\label{appb}
\centering
\includegraphics[width=\columnwidth]{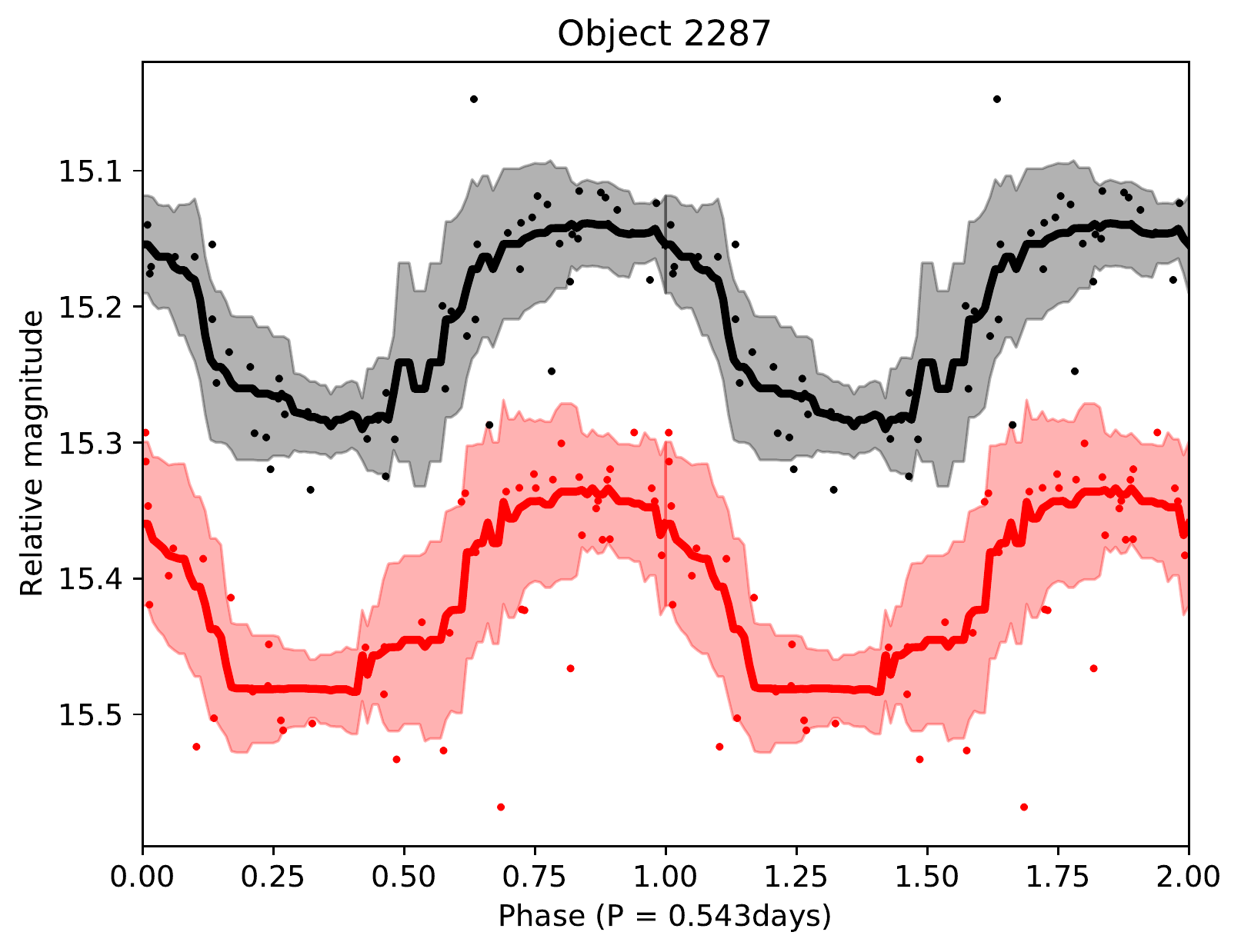} \hfill
\includegraphics[width=\columnwidth]{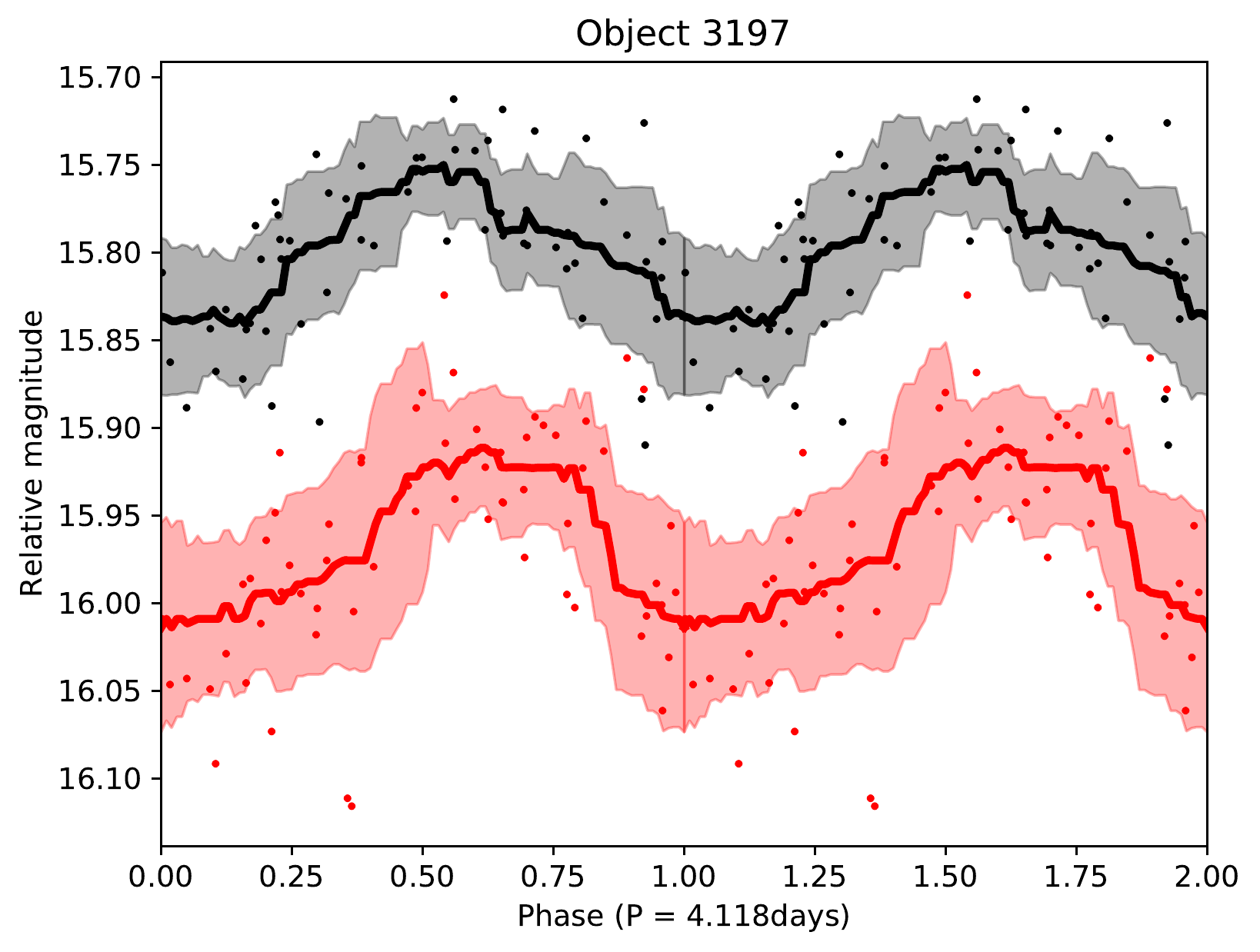} \\
\includegraphics[width=\columnwidth]{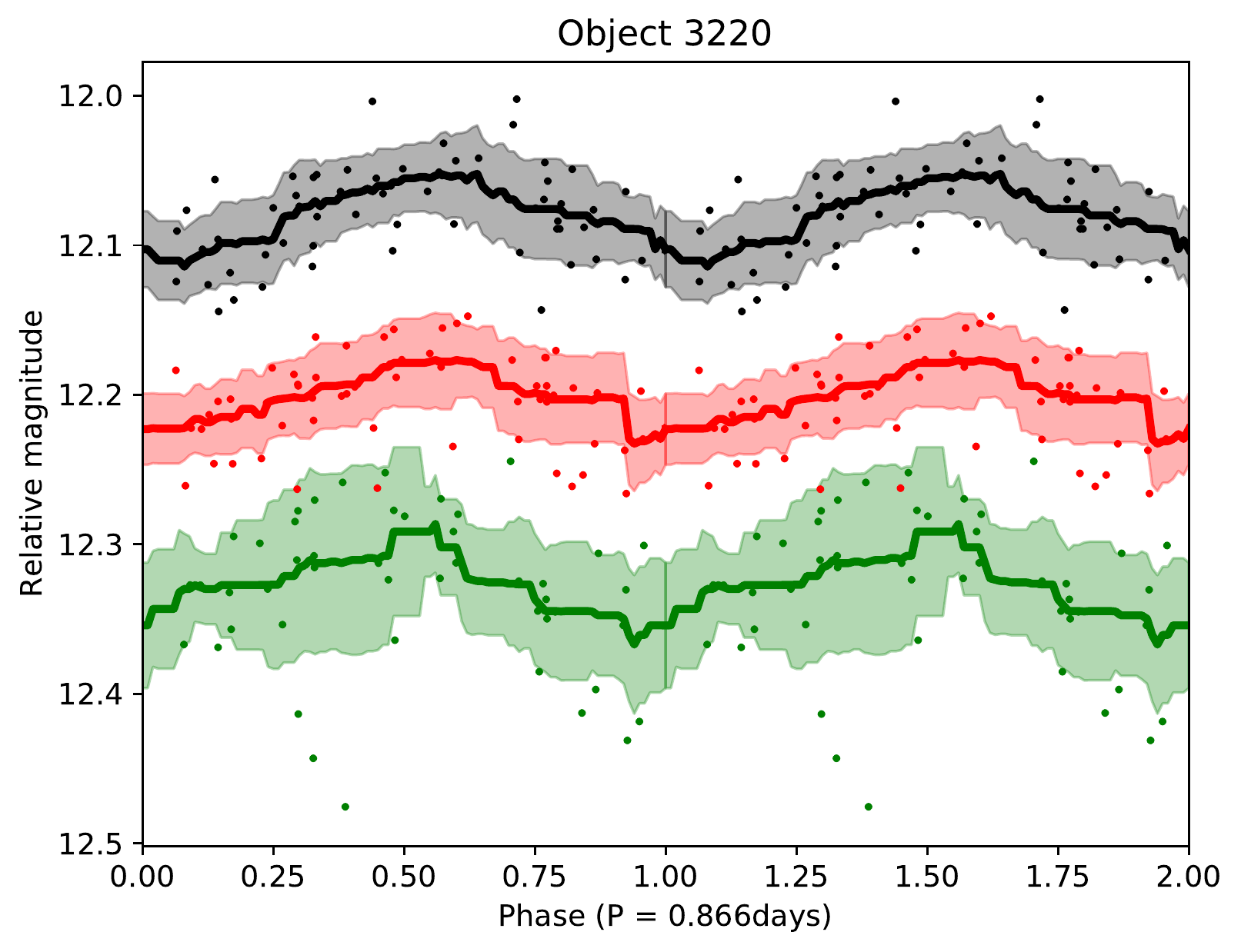} \hfill
\includegraphics[width=\columnwidth]{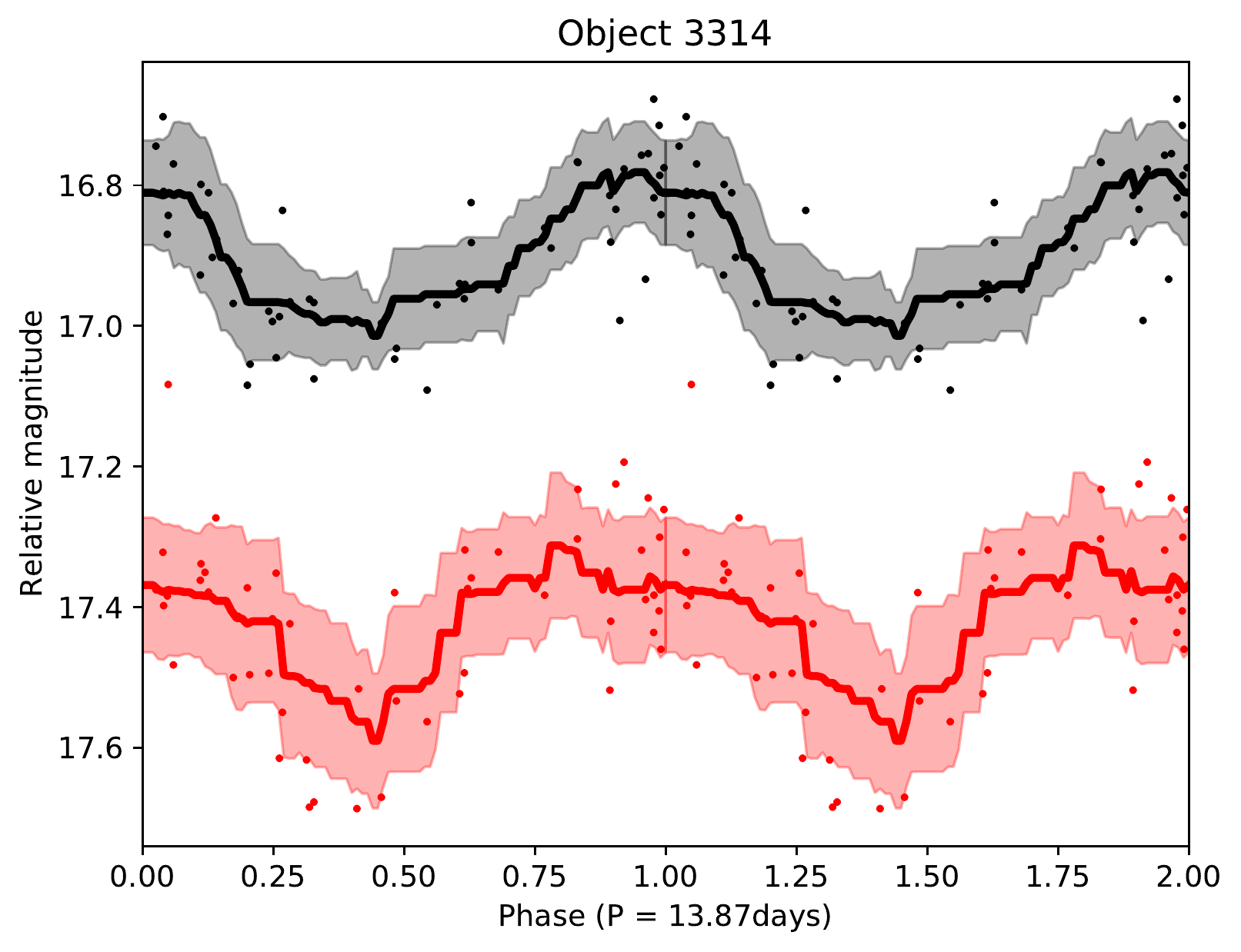} \\
\includegraphics[width=\columnwidth]{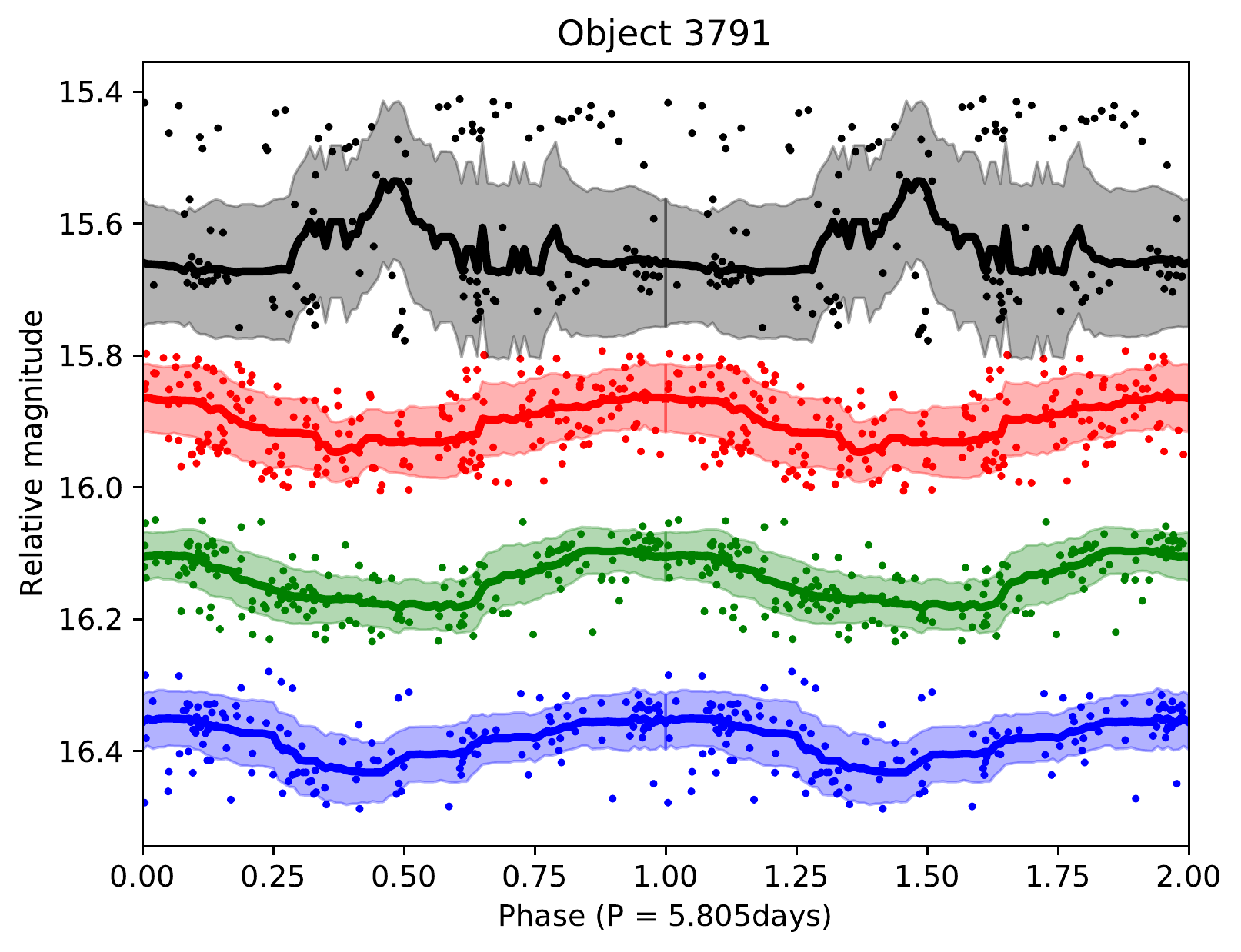} \hfill
\includegraphics[width=\columnwidth]{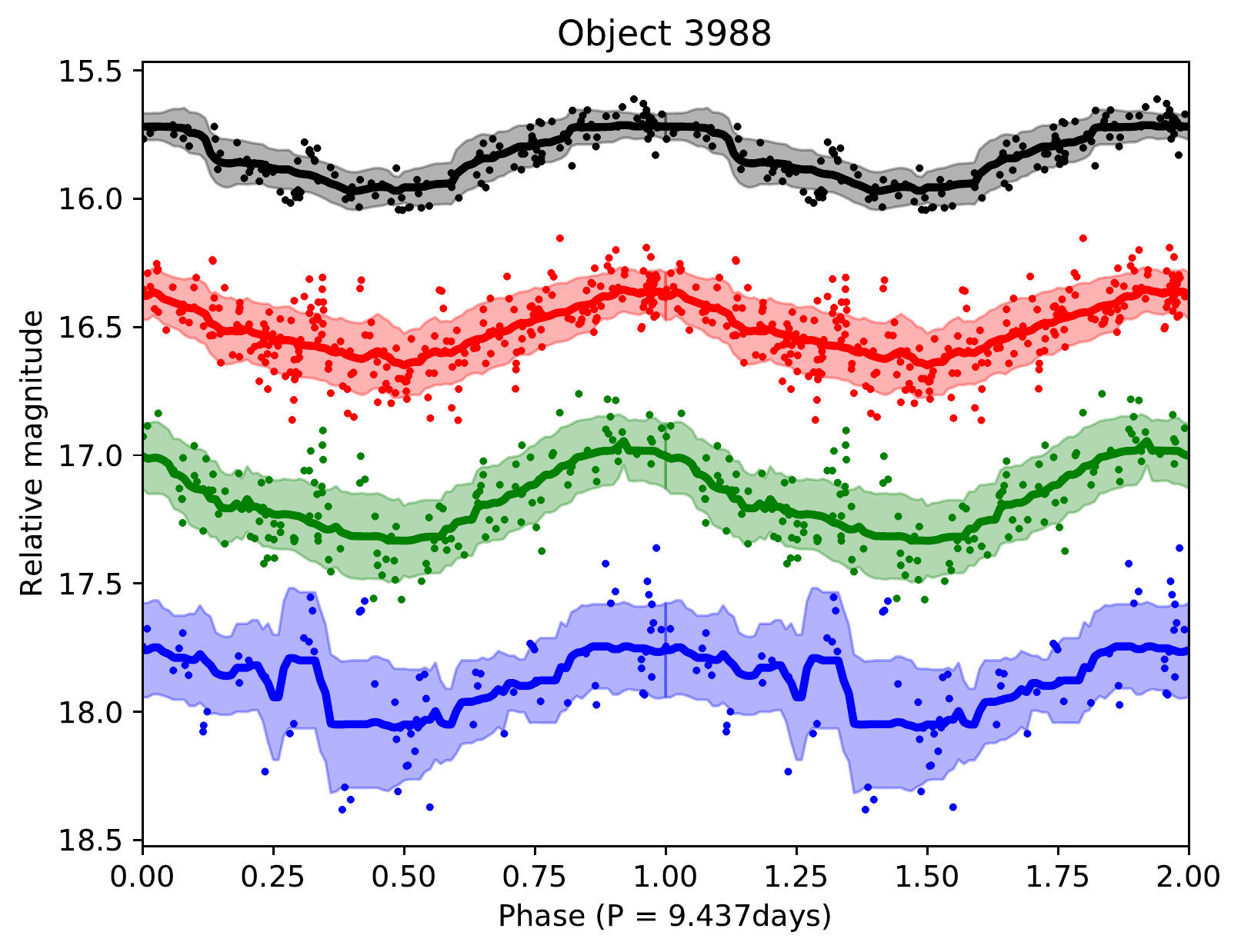} \\
\caption{Phase-folded light curves for objects 2287, 3197, 3220, 3314, 3791, and 3988. We show the individual photometric data points as small dots with the U data in purple, B data in blue, V data in green, R data in red, and I data in black. The R data is shown at the correct magnitudes, and the data for all other filters are arbitrarily shifted for better visibility. The solid overlaid lines represent a running median and the shaded regions are the one sigma deviations of the data from the running median. We do not have data in all filters for each object. The period is indicated below each plot. There are four objects where the I-band data seems to be split into a bright and a faint component. These are stars where our colour term correction of the photometry has not worked correctly. The data is however shown for completeness. One such source is object 3791 (bottom left) in this plot. \label{phaseplots}}
\end{figure*}

\clearpage\newpage

\begin{figure*}
\centering
\includegraphics[width=\columnwidth]{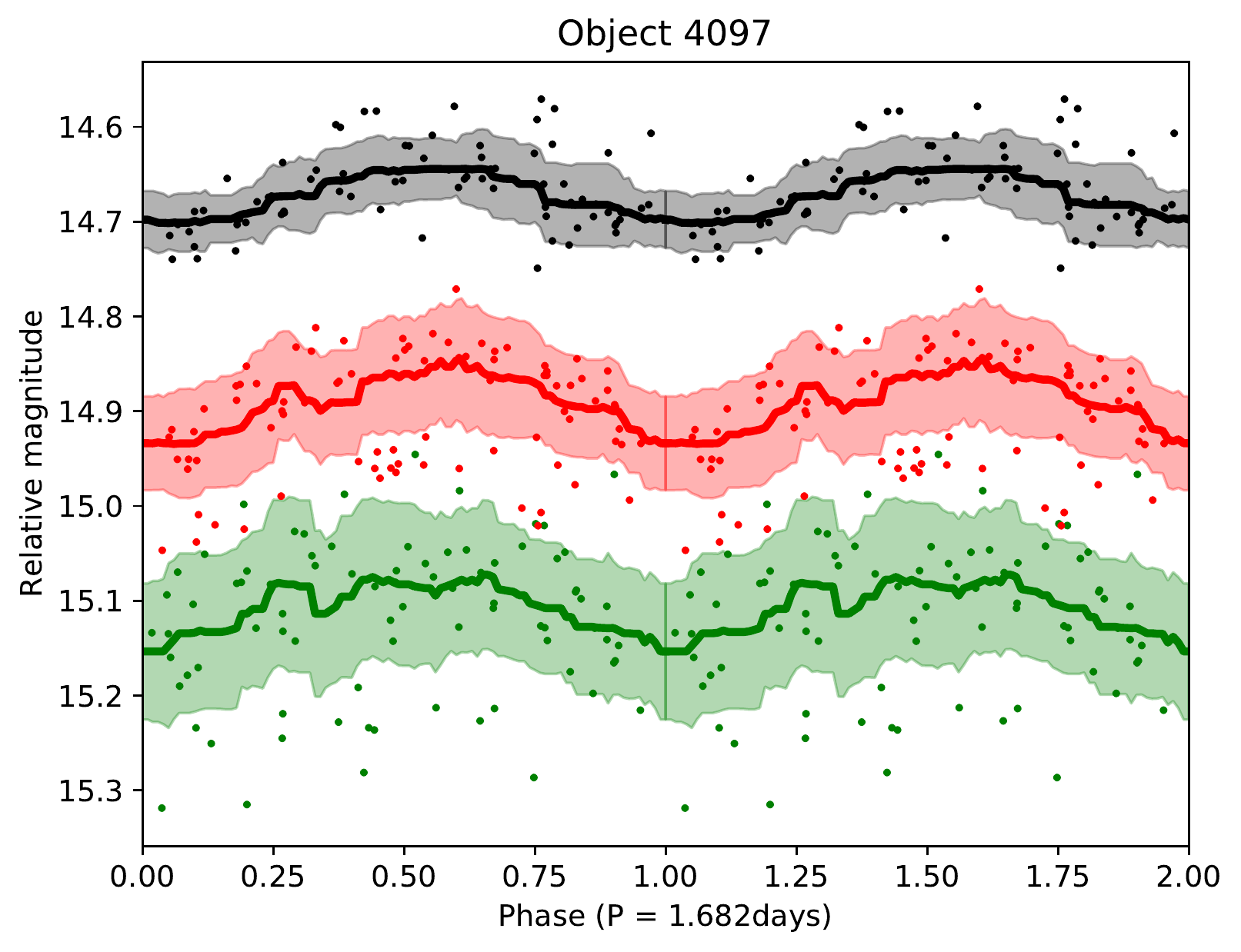} \hfill
\includegraphics[width=\columnwidth]{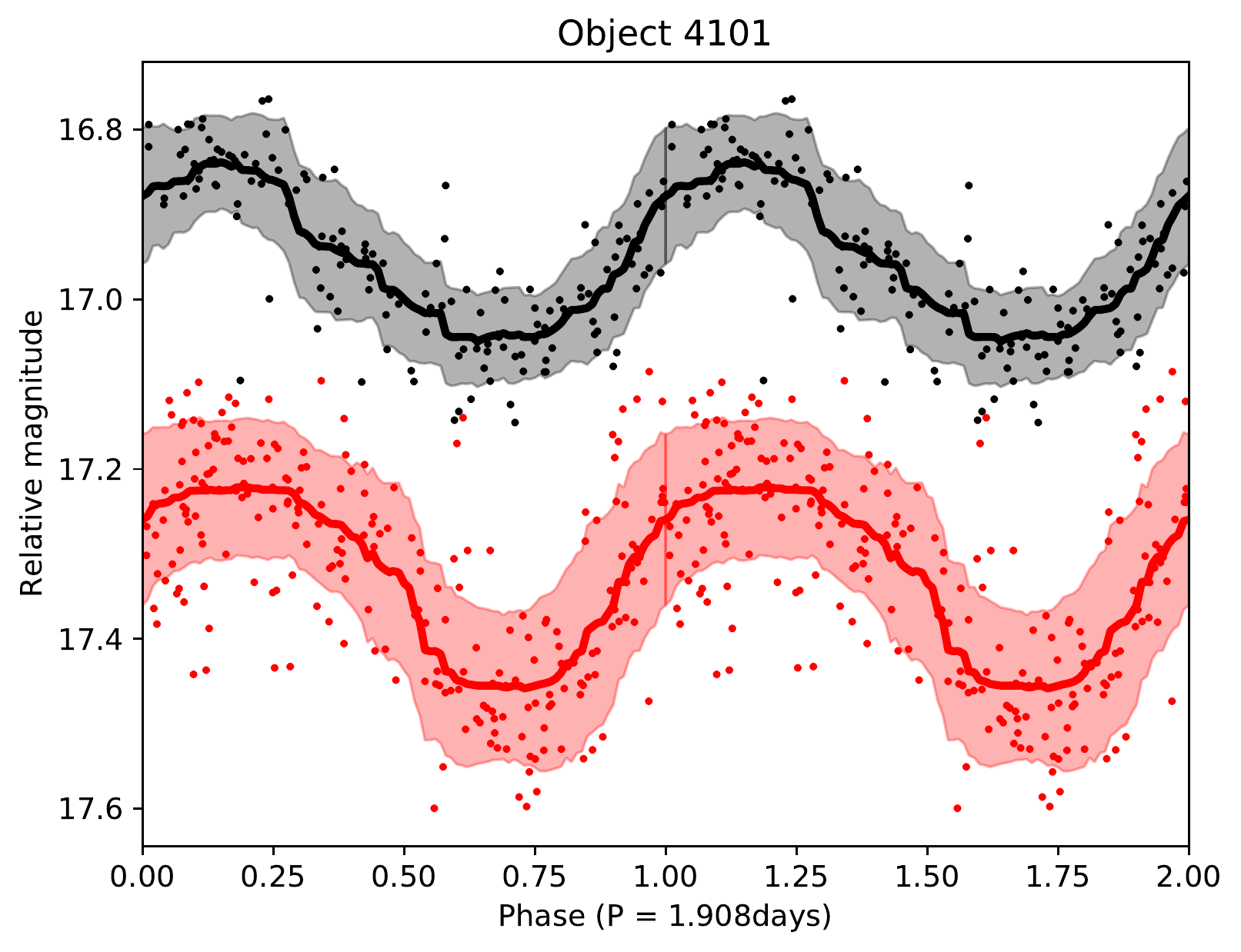} \\
\includegraphics[width=\columnwidth]{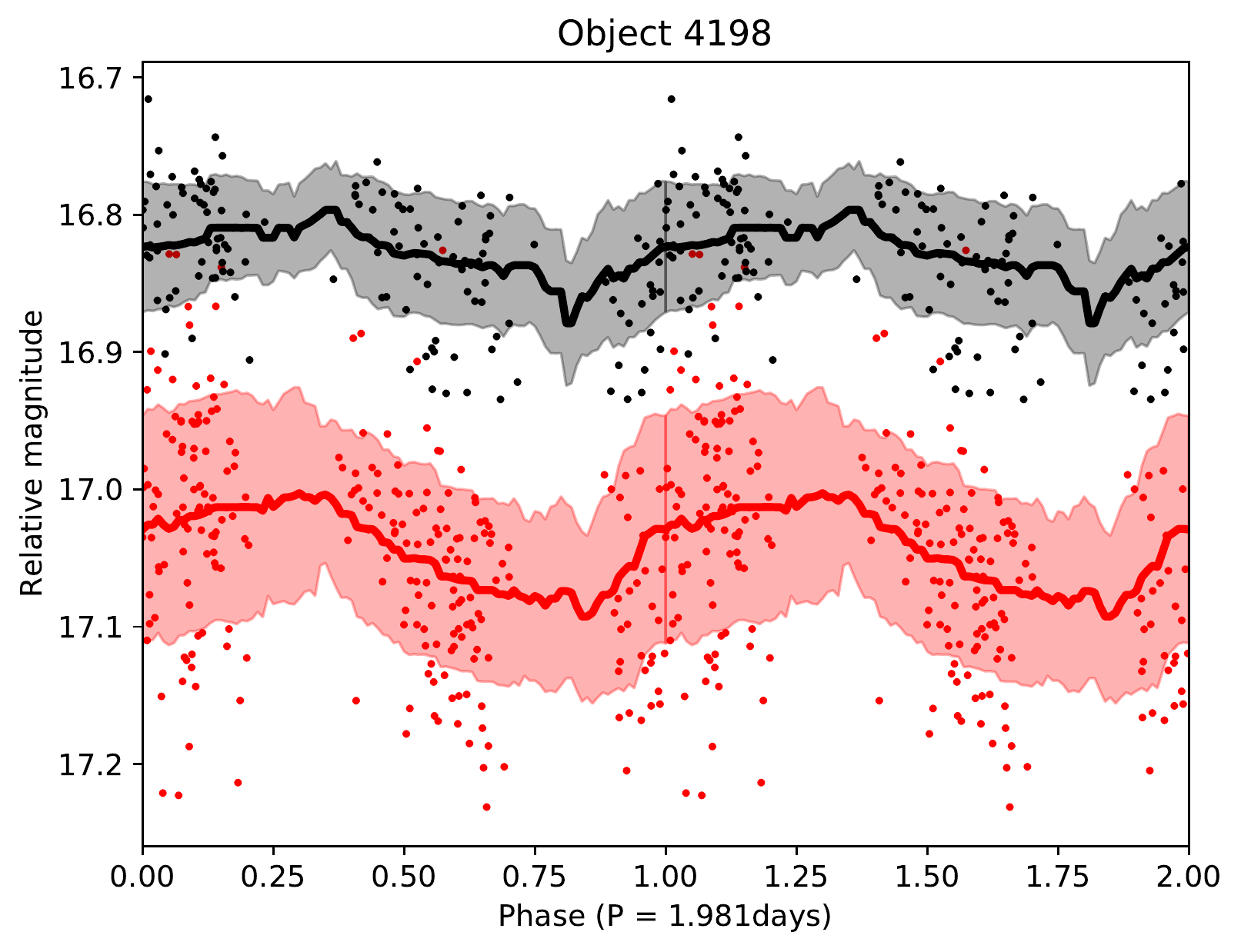} \hfill
\includegraphics[width=\columnwidth]{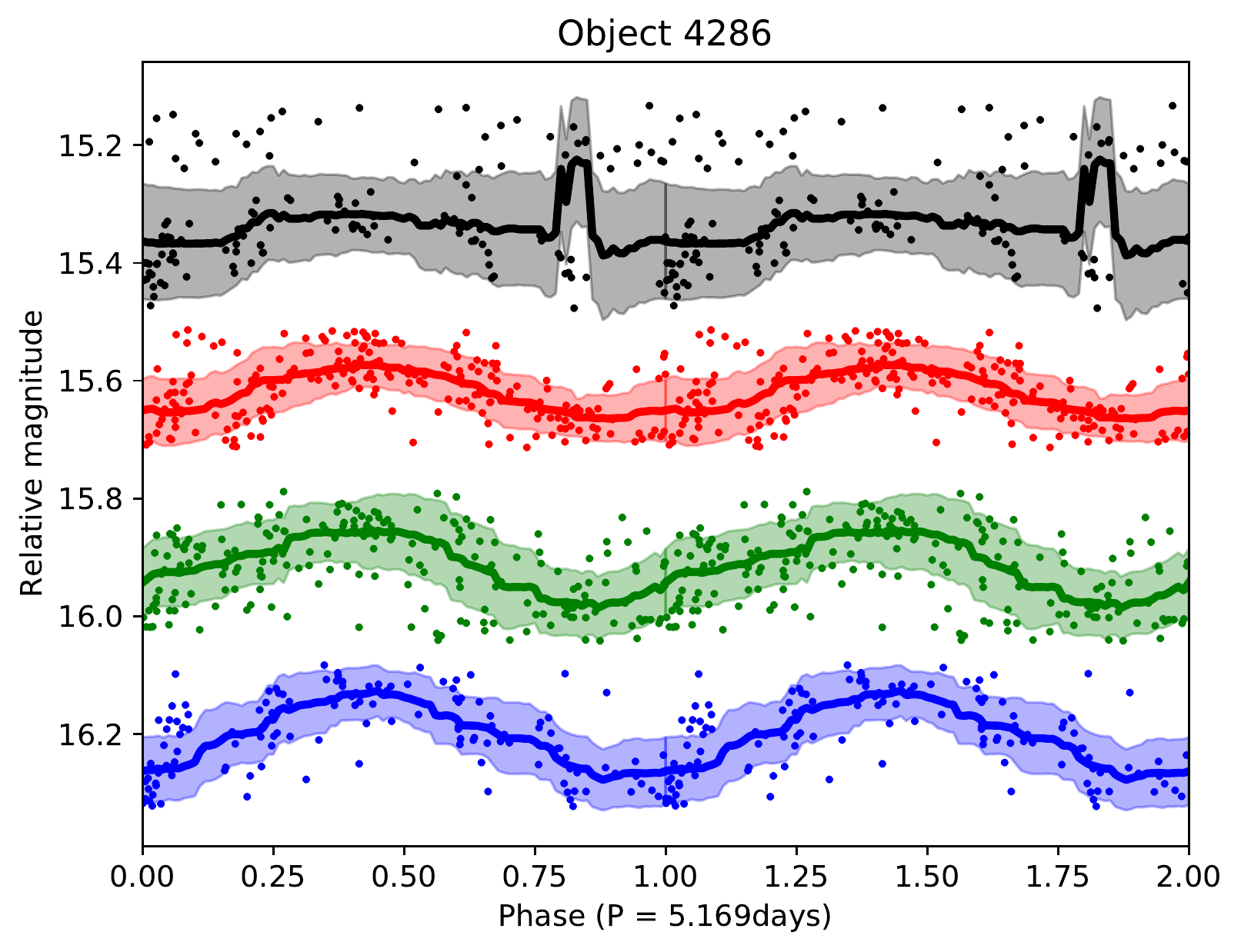} \\
\includegraphics[width=\columnwidth]{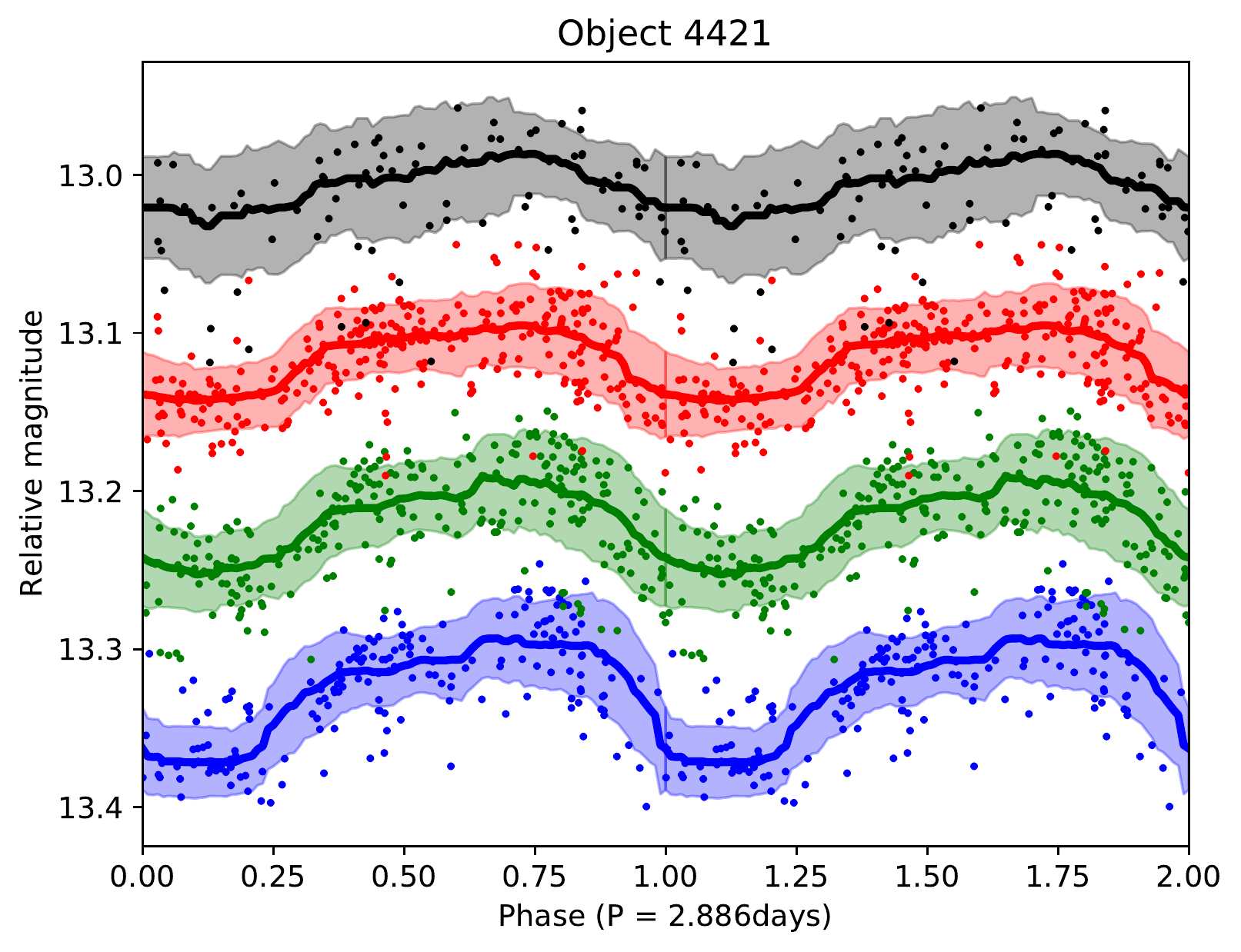} \hfill
\includegraphics[width=\columnwidth]{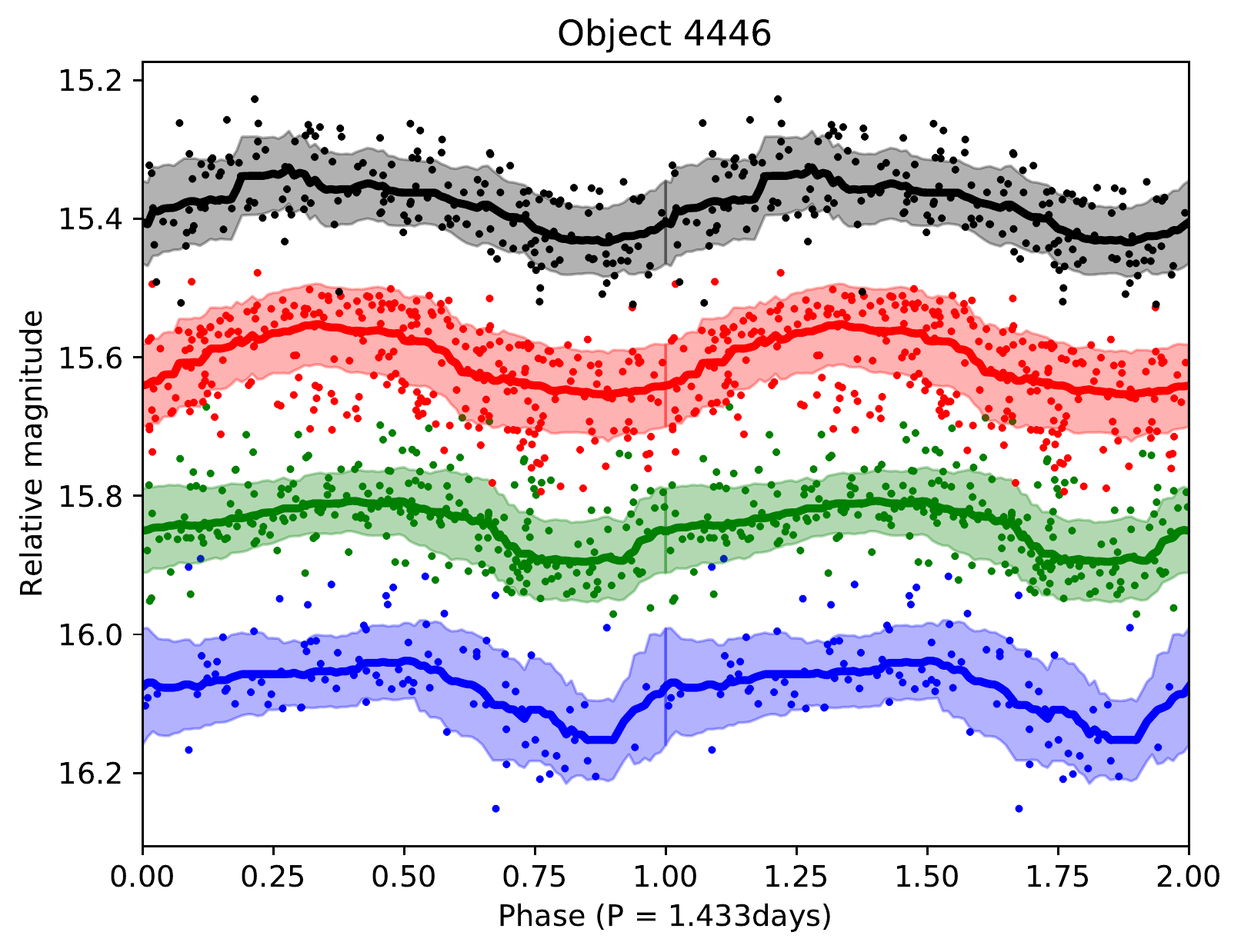} \\
\caption{As Fig.\,\ref{phaseplots} but for objects 4097, 4101, 4198, 4286, 4421, and 4446.}
\end{figure*}

\clearpage\newpage

\begin{figure*}
\centering
\includegraphics[width=\columnwidth]{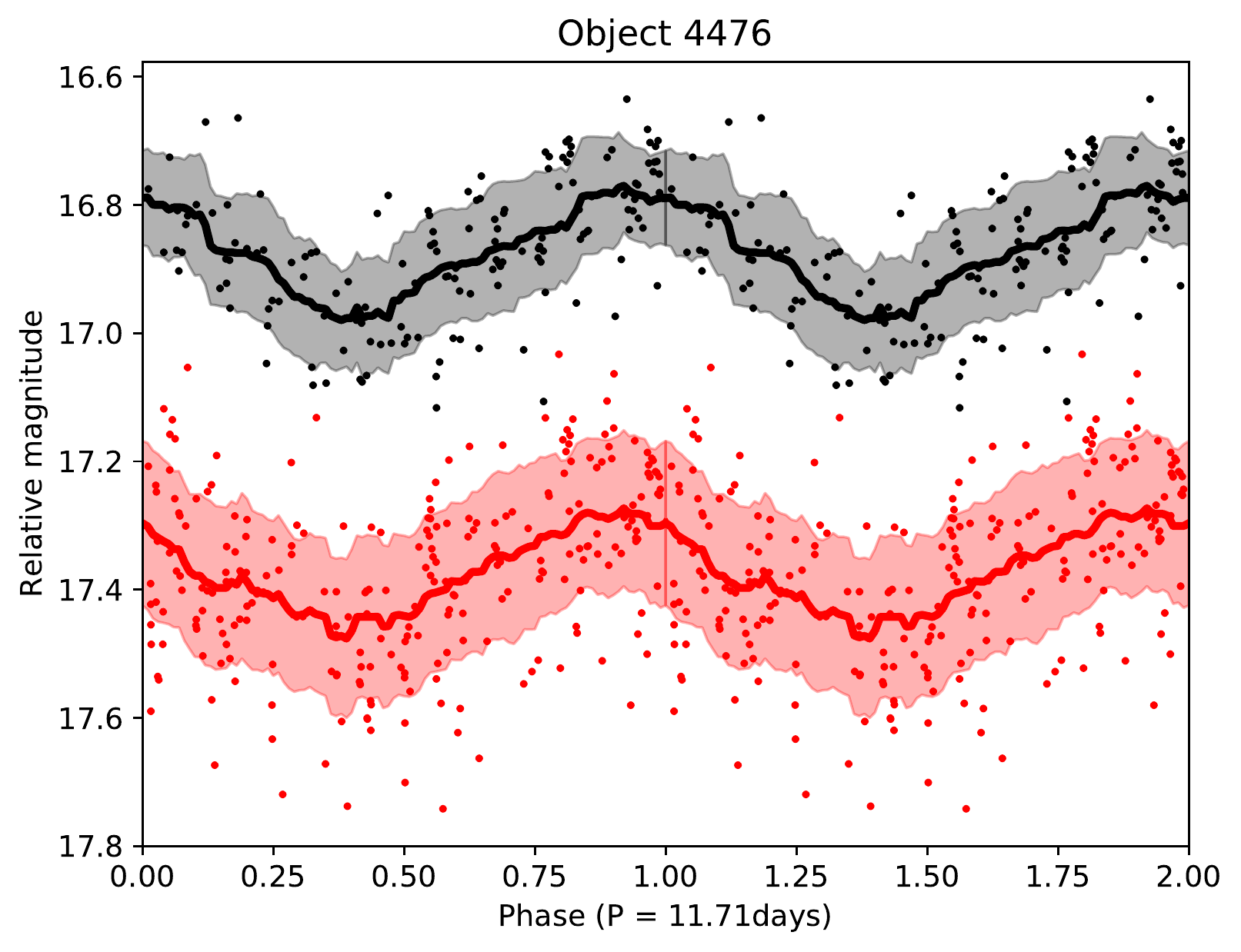} \hfill
\includegraphics[width=\columnwidth]{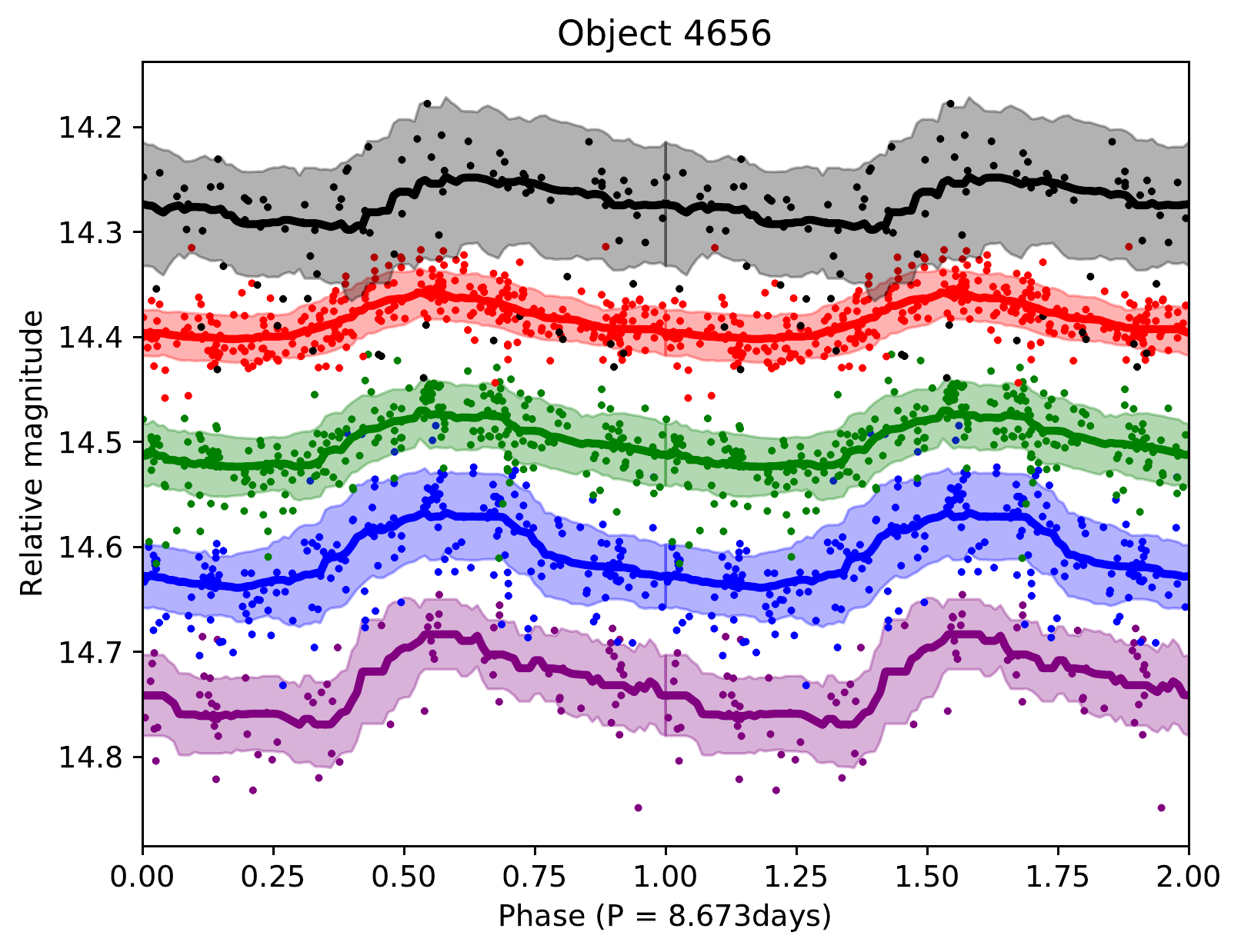} \\
\includegraphics[width=\columnwidth]{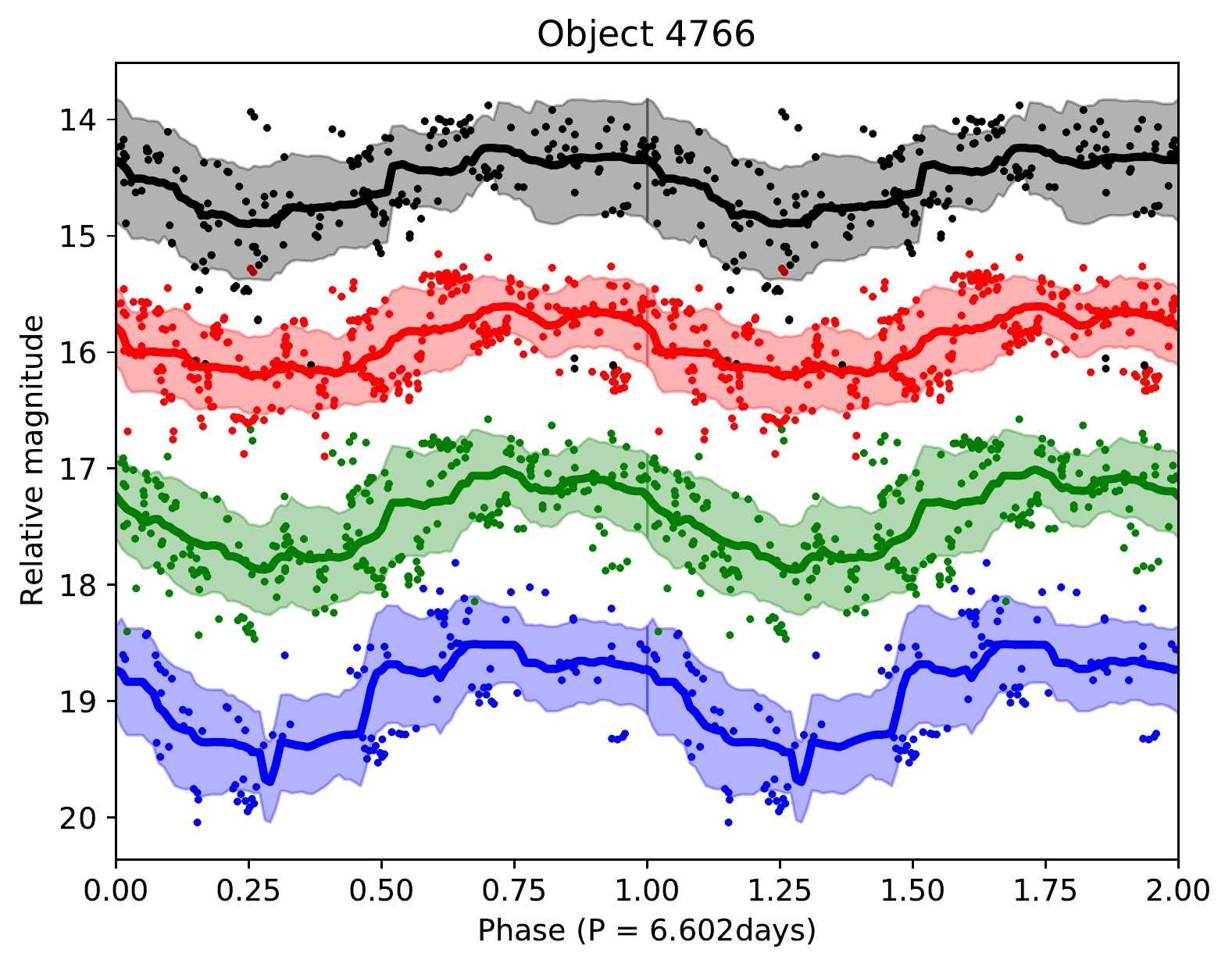} \hfill
\includegraphics[width=\columnwidth]{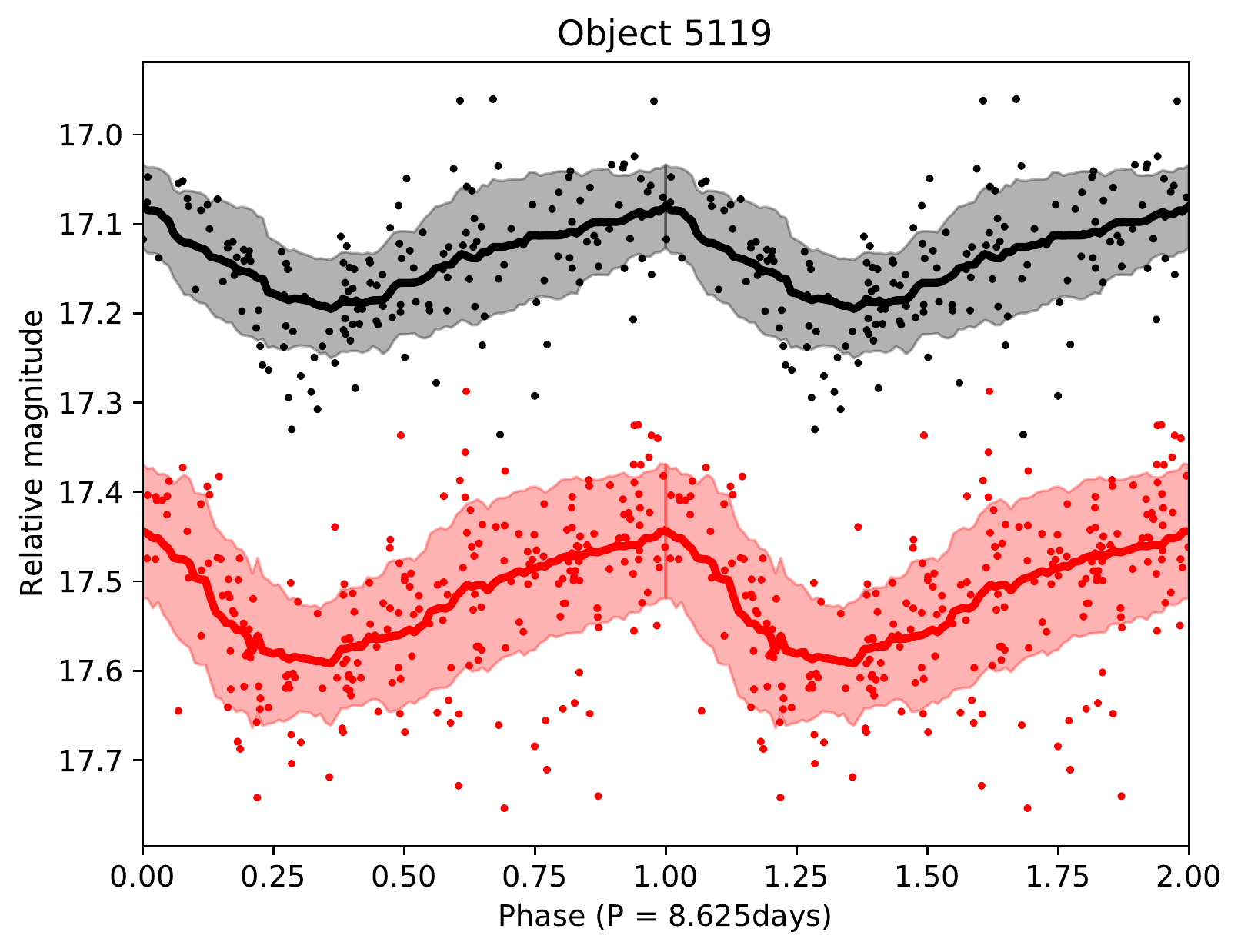} \\
\includegraphics[width=\columnwidth]{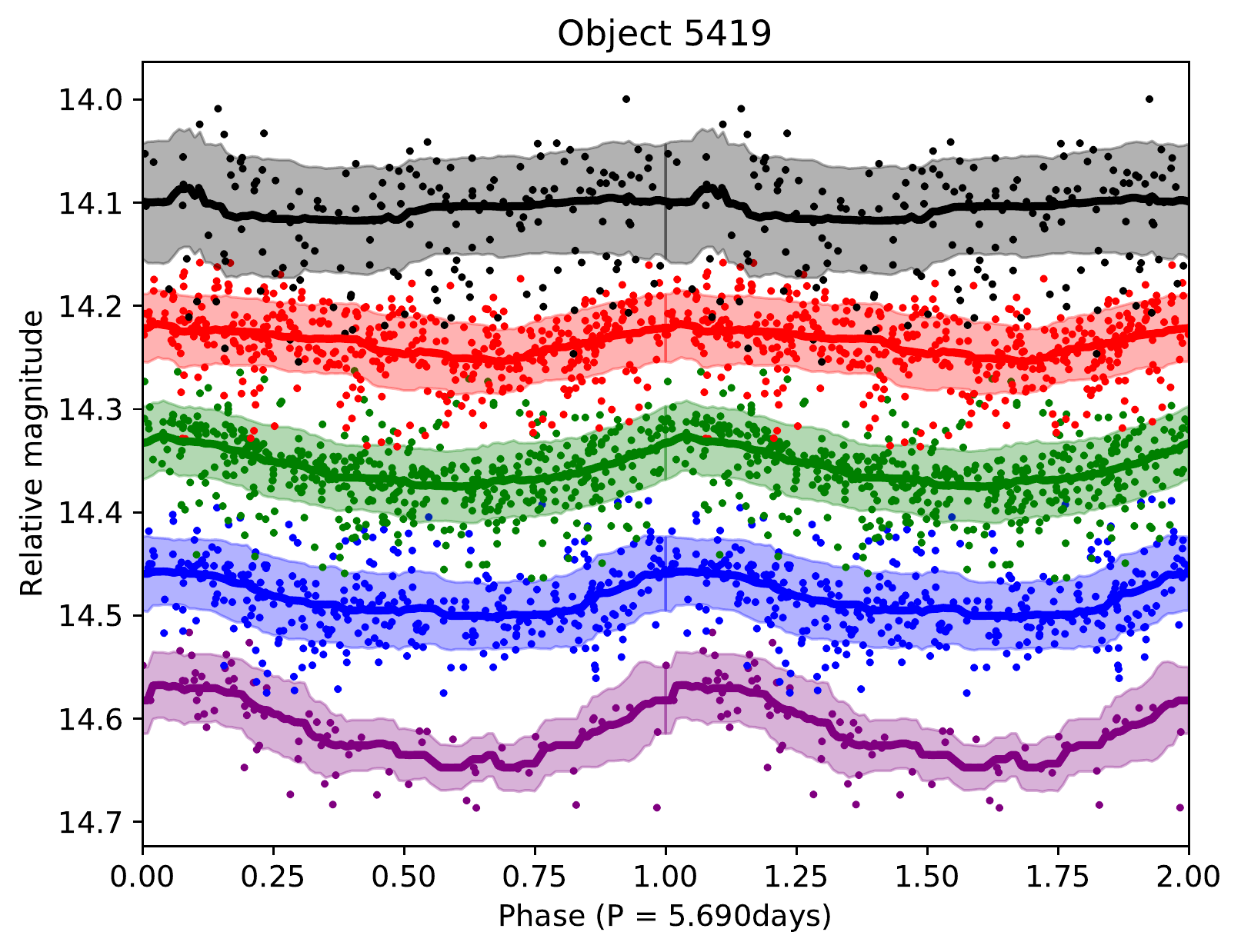} \hfill
\includegraphics[width=\columnwidth]{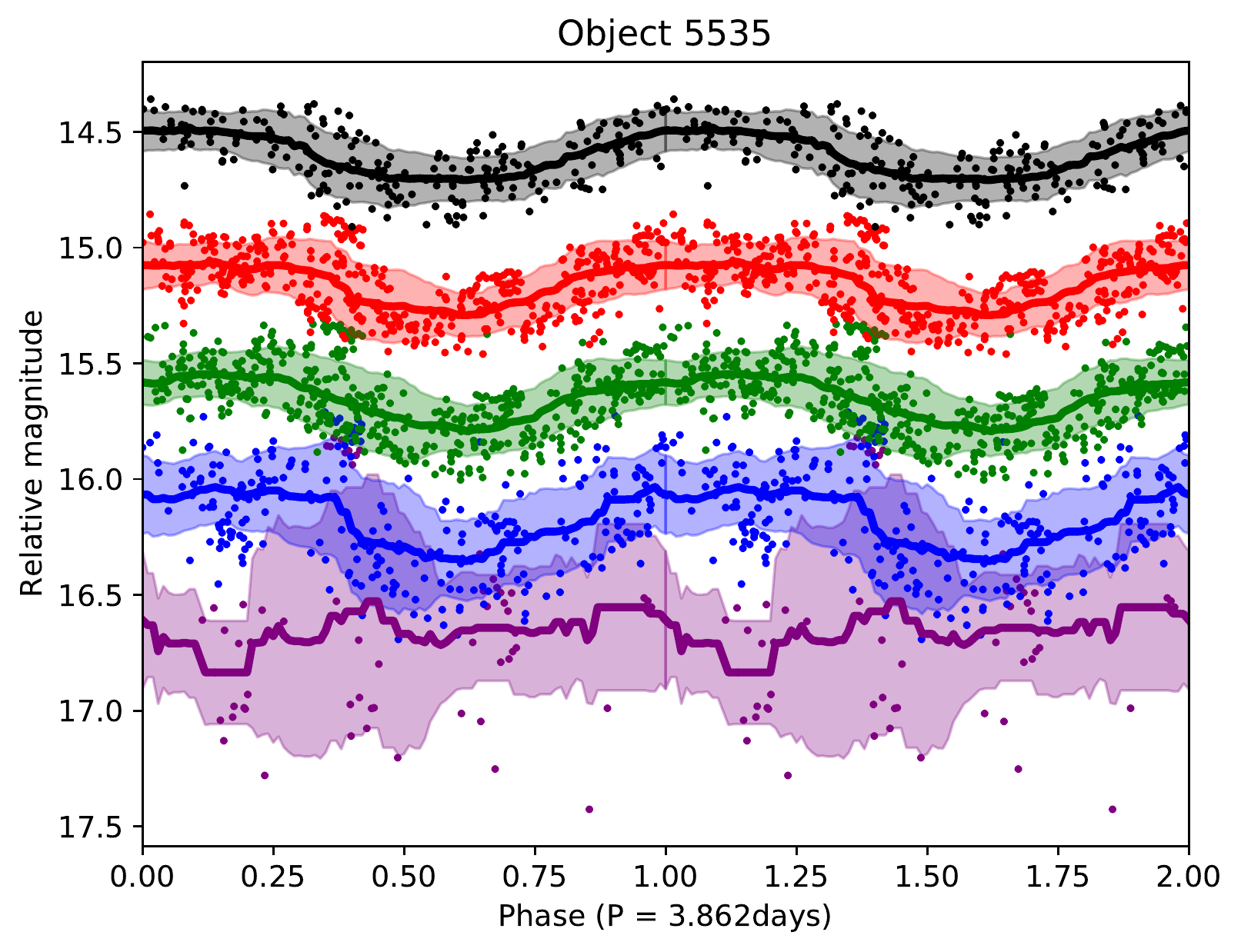} \\
\caption{As Fig.\,\ref{phaseplots} but for objects 4476, 4656, 4766, 5119, 5419, and 5535.}
\end{figure*}

\clearpage\newpage

\begin{figure*}
\centering
\includegraphics[width=\columnwidth]{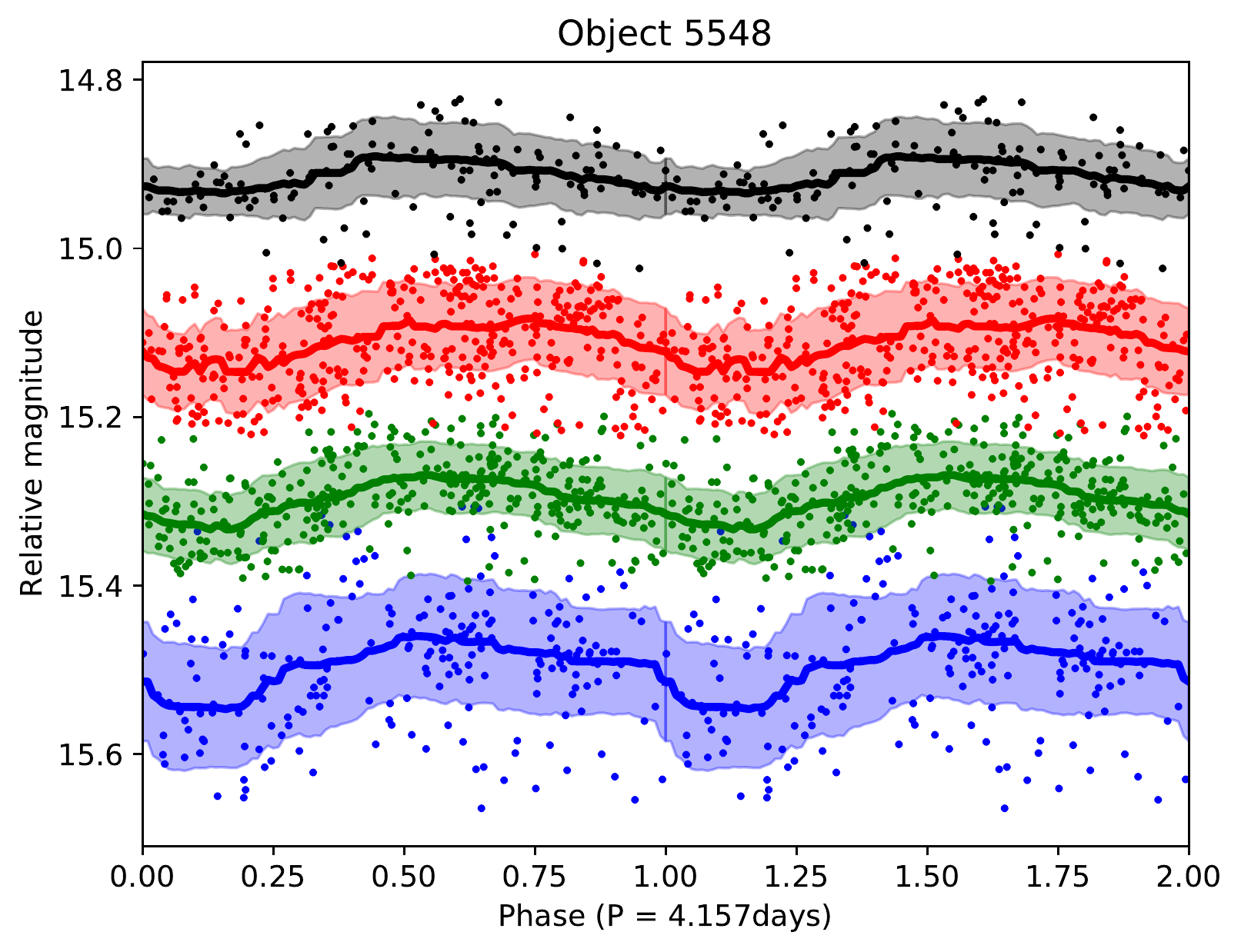} \hfill
\includegraphics[width=\columnwidth]{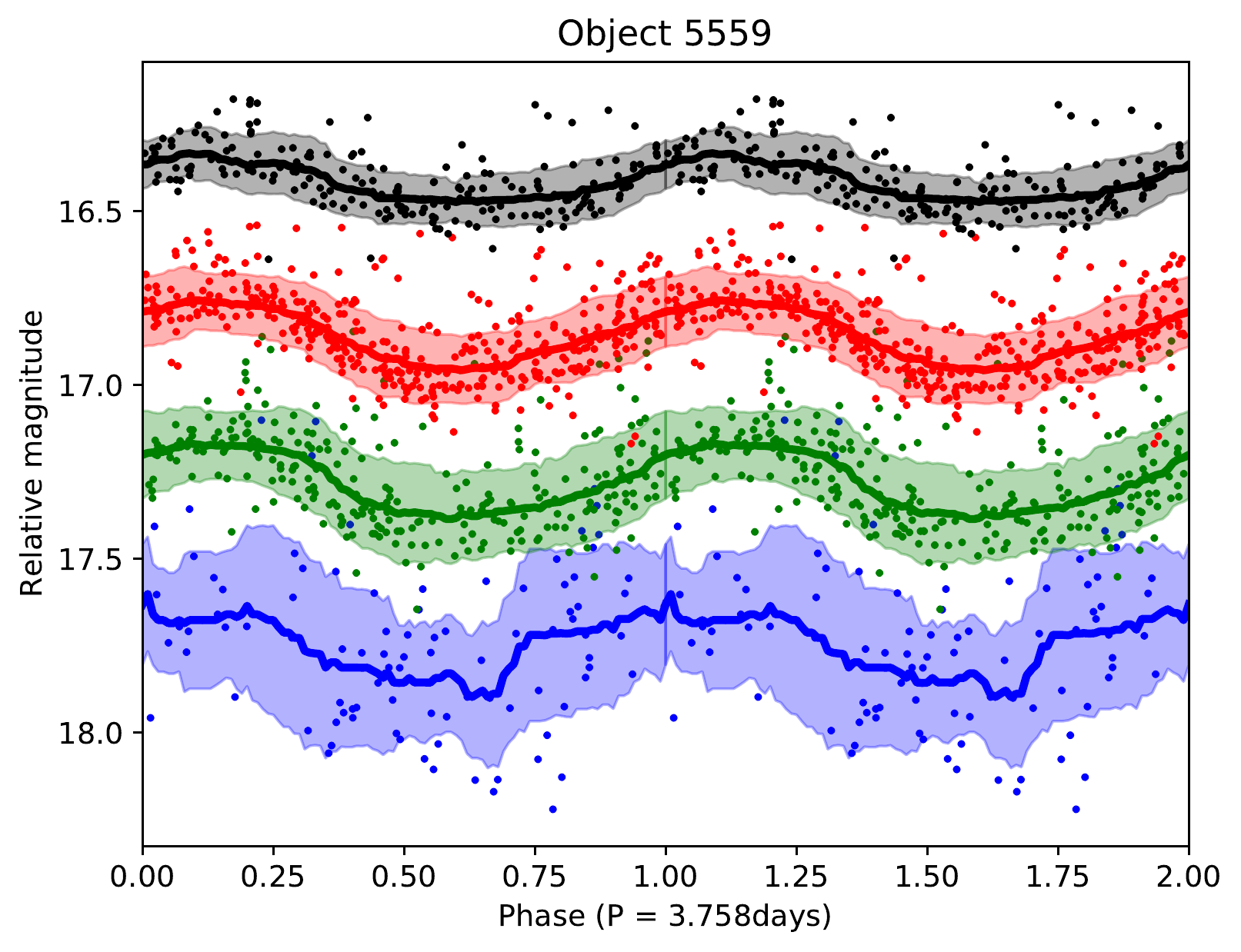} \\
\includegraphics[width=\columnwidth]{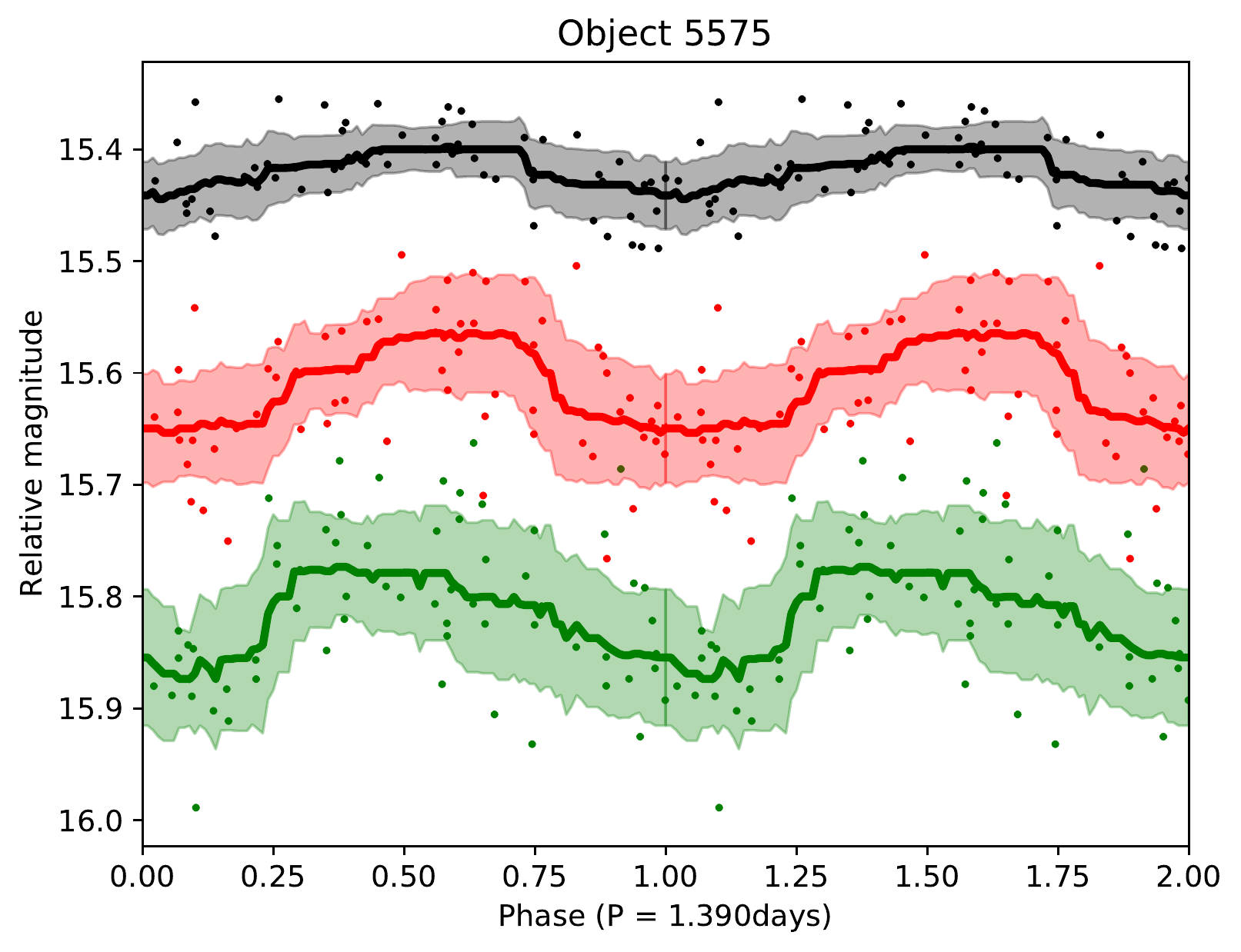} \hfill
\includegraphics[width=\columnwidth]{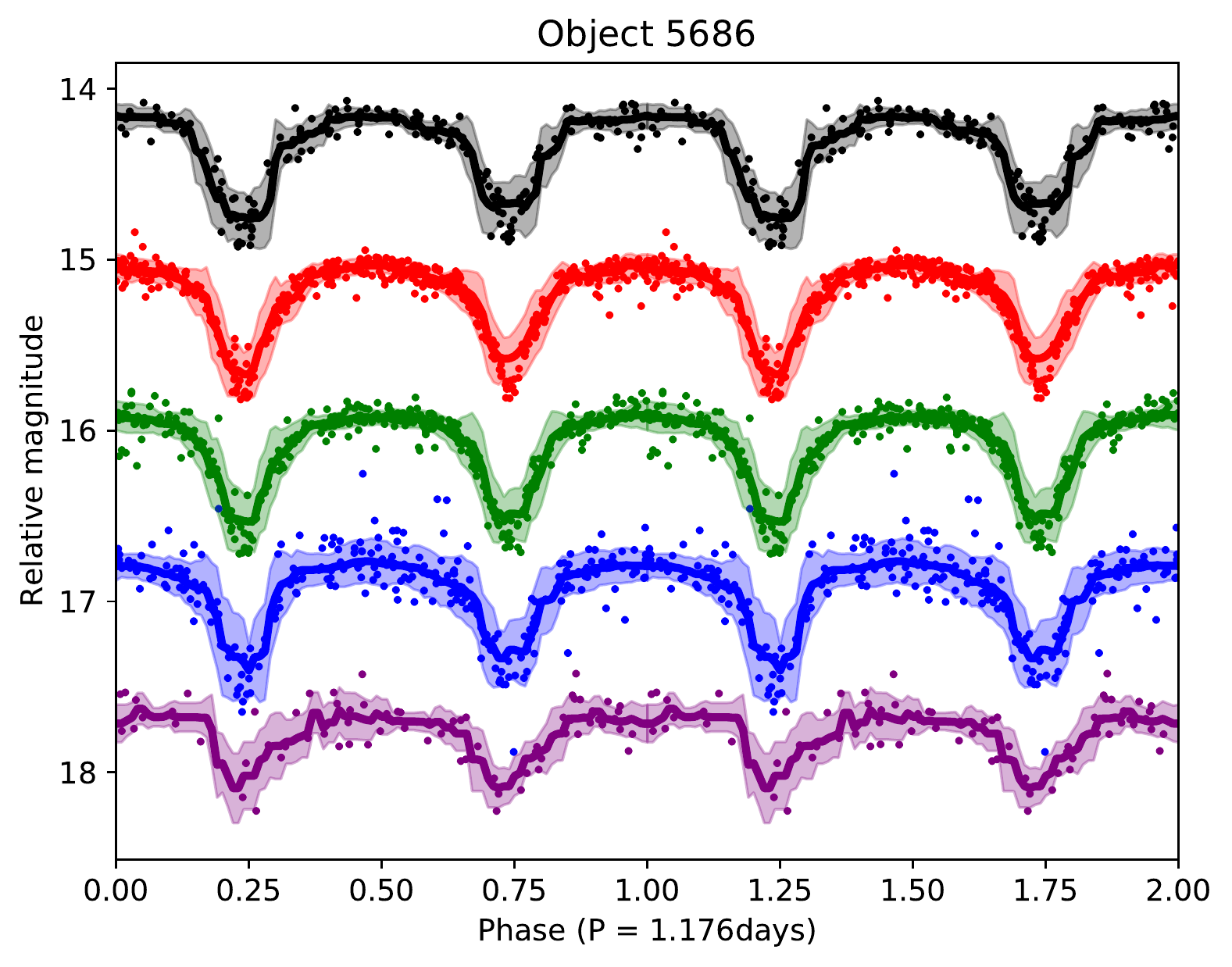} \\
\includegraphics[width=\columnwidth]{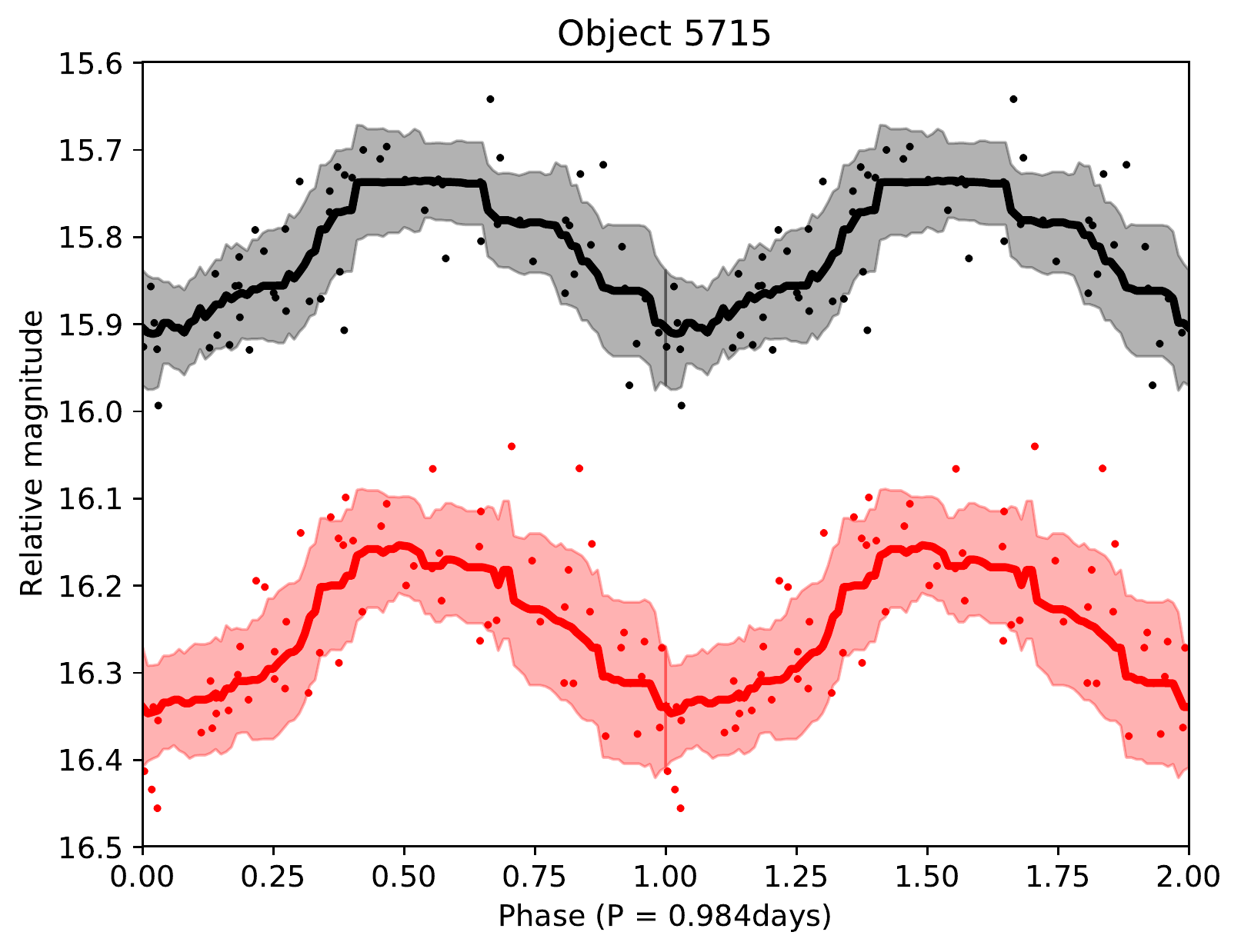} \hfill
\includegraphics[width=\columnwidth]{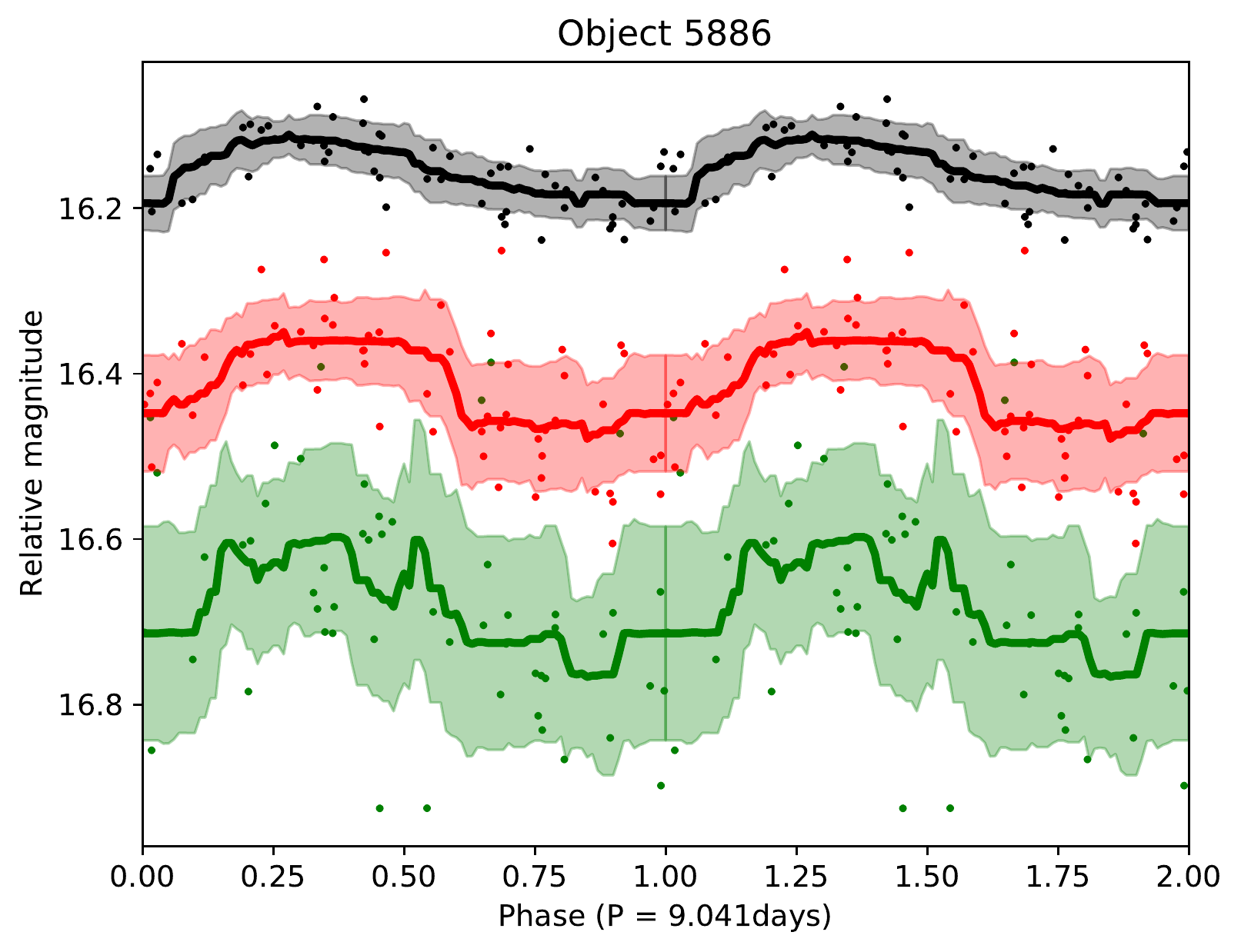} \\
\caption{As Fig.\,\ref{phaseplots} but for objects 5548, 5559, 5575, 5685, 5715, and 5886.}
\end{figure*}

\clearpage\newpage

\begin{figure*}
\centering
\includegraphics[width=\columnwidth]{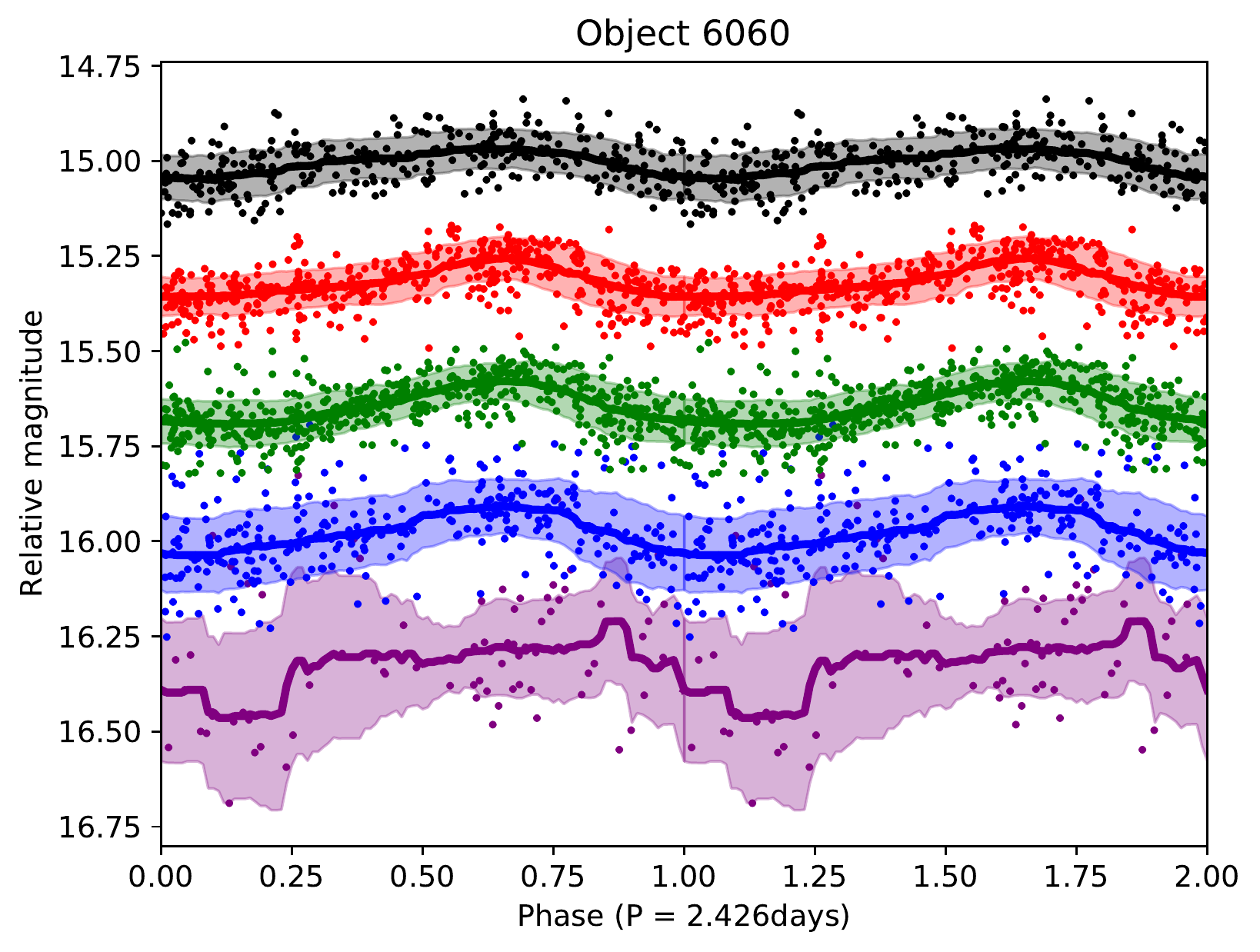} \hfill
\includegraphics[width=\columnwidth]{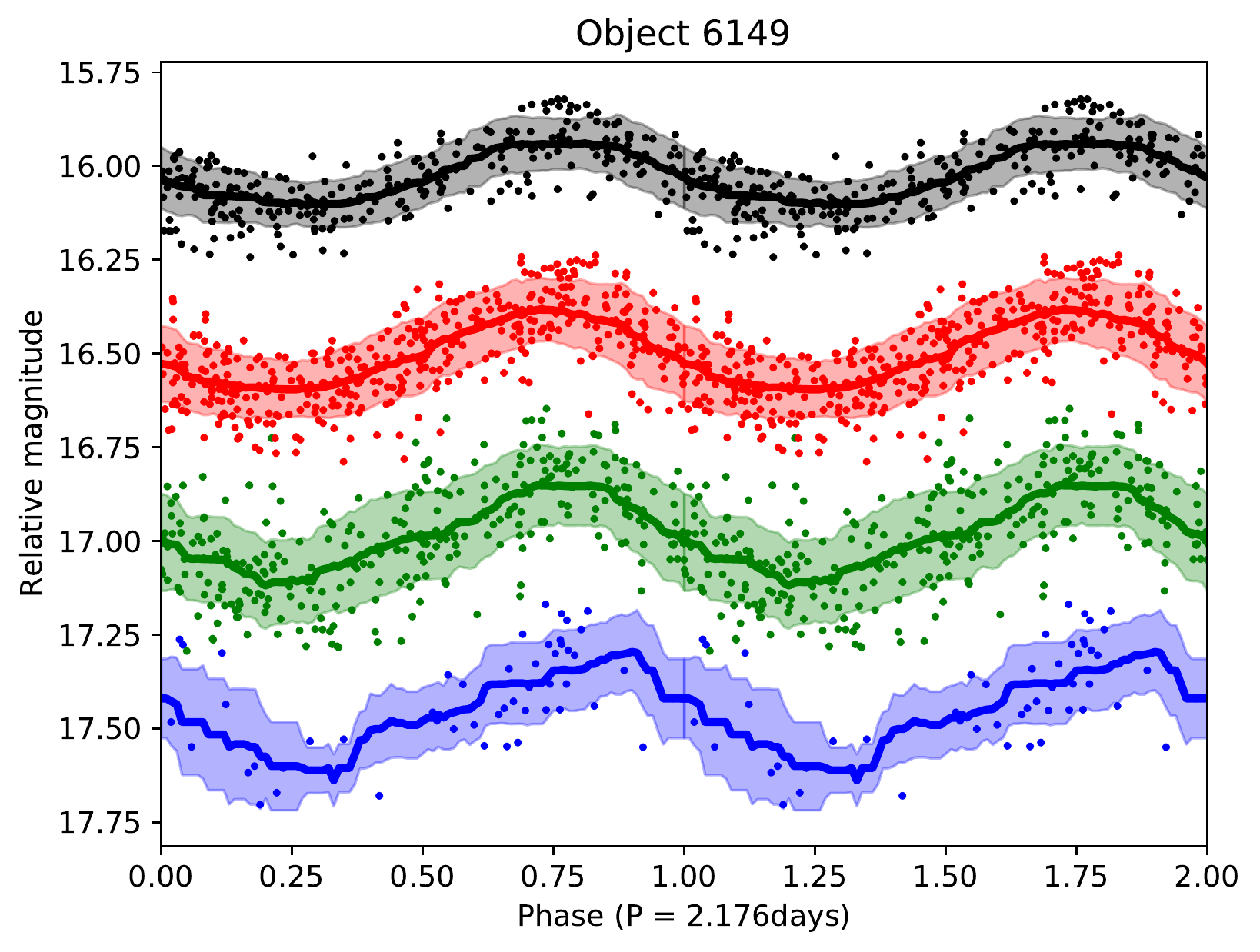} \\
\includegraphics[width=\columnwidth]{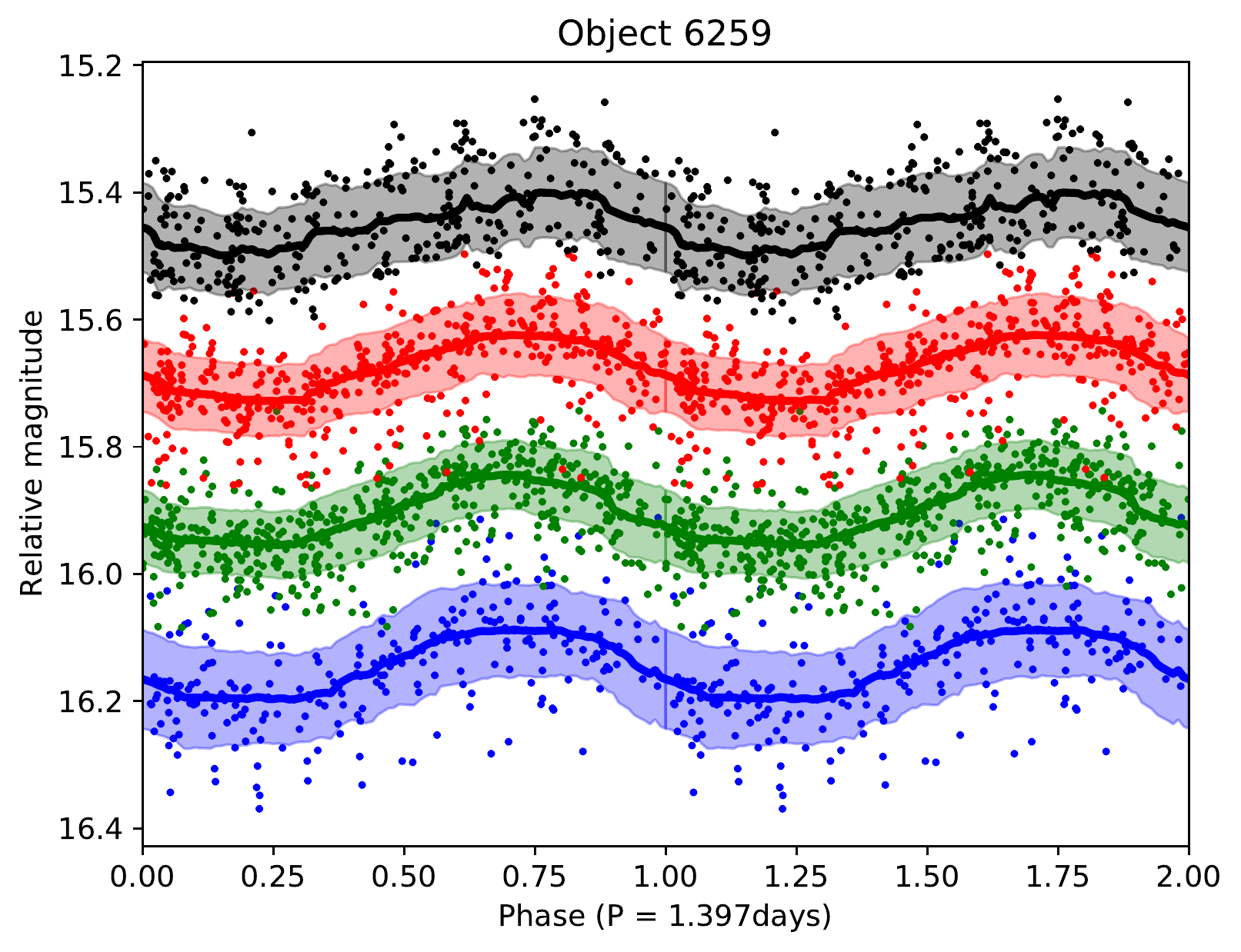} \hfill
\includegraphics[width=\columnwidth]{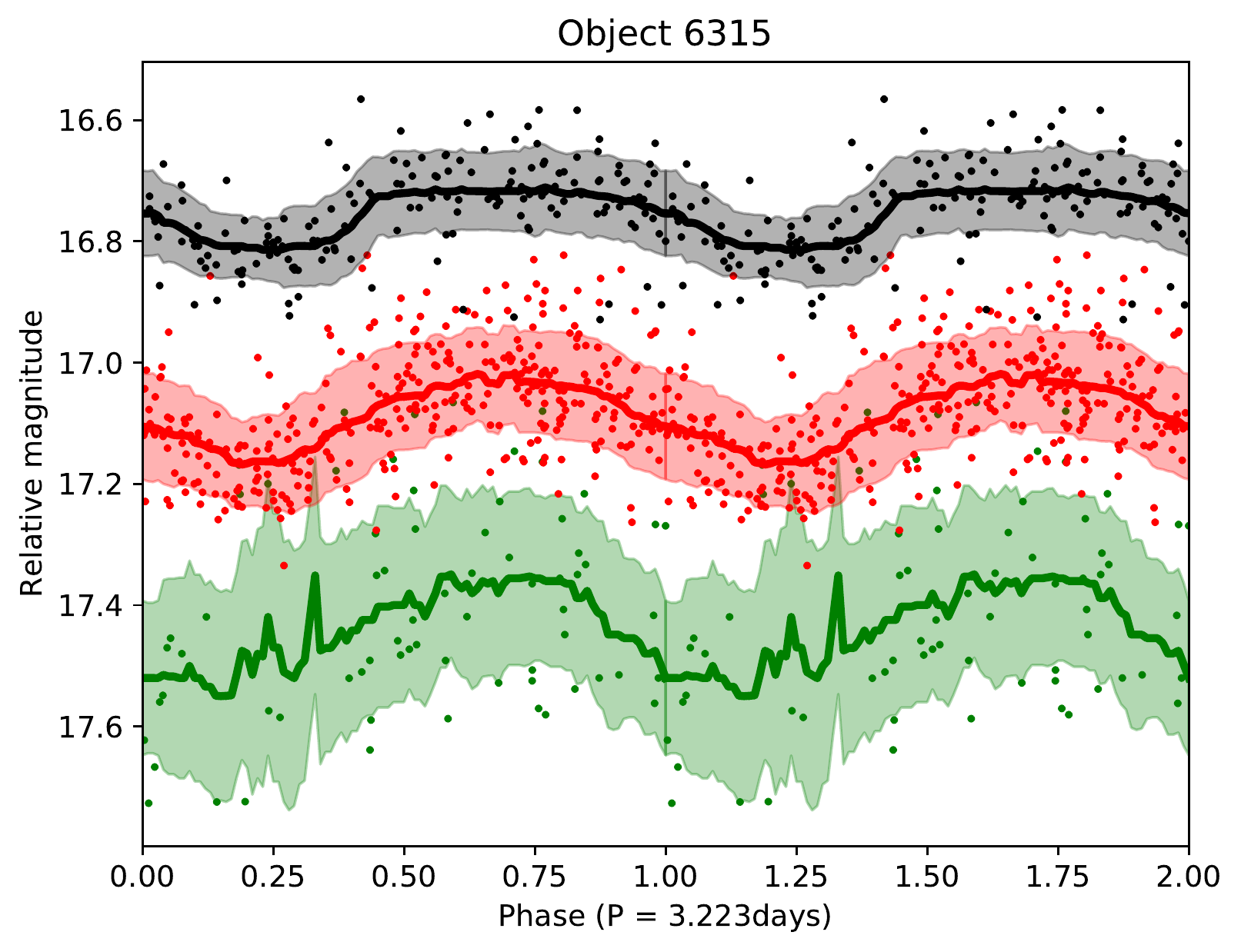} \\
\includegraphics[width=\columnwidth]{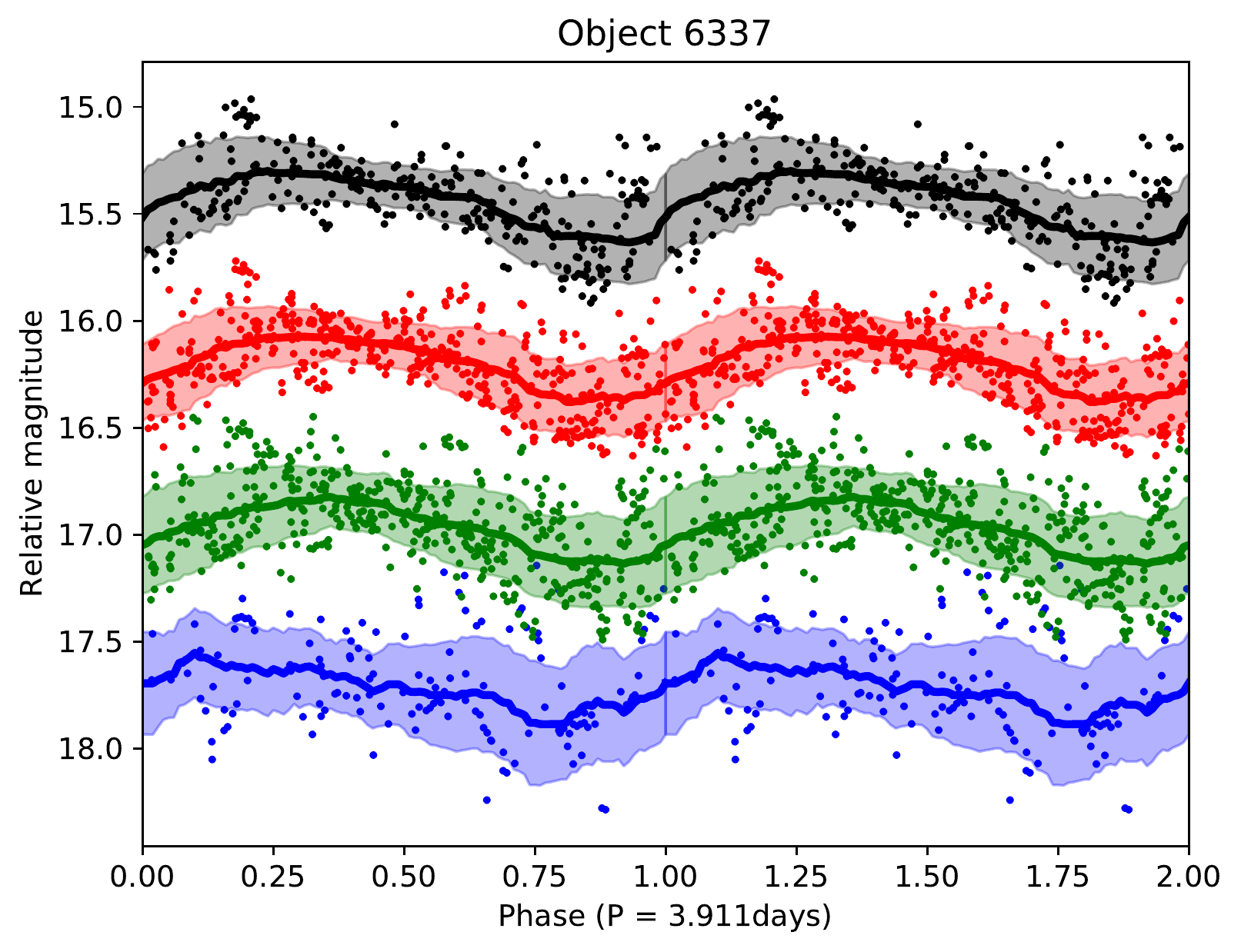} \hfill
\includegraphics[width=\columnwidth]{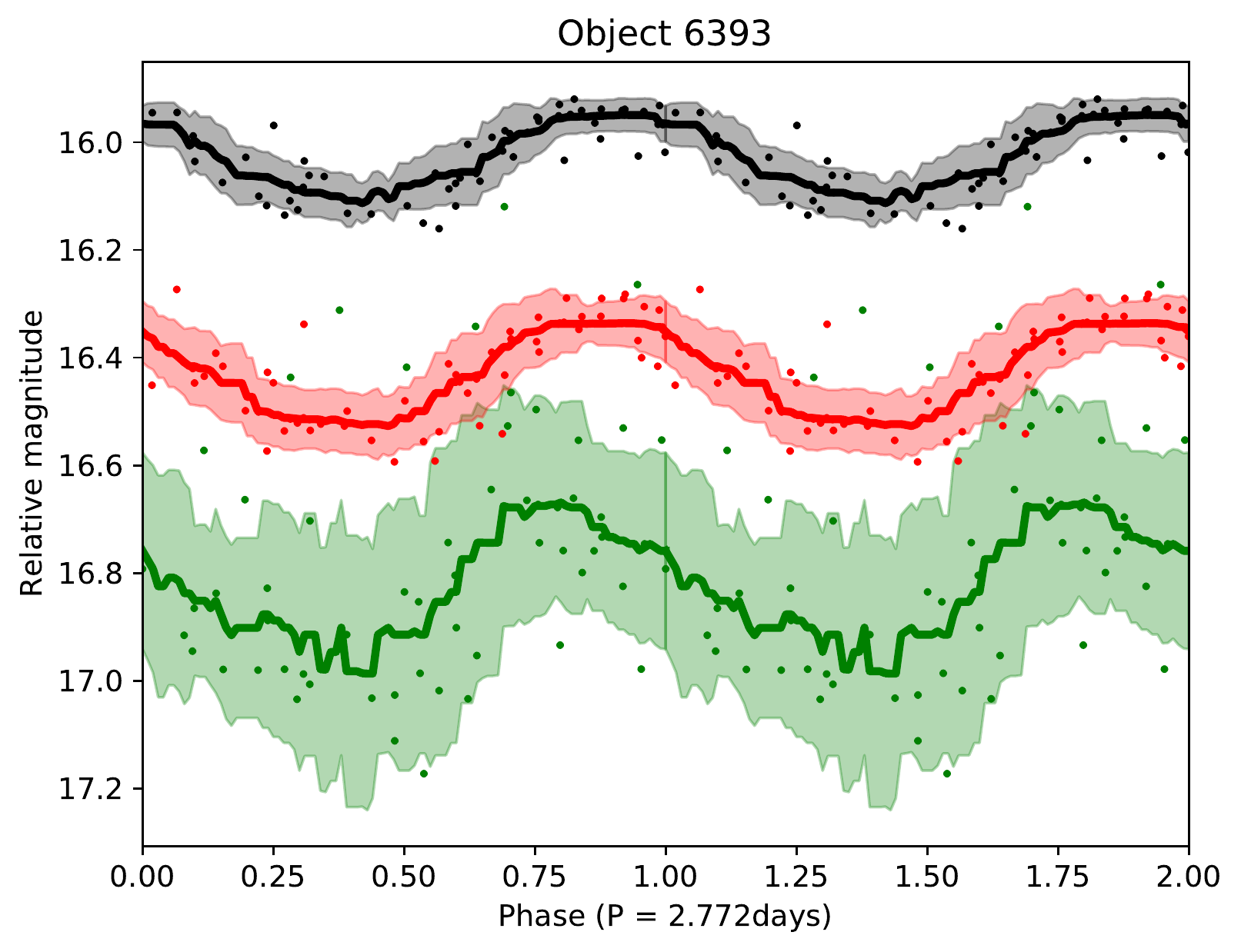} \\
\caption{As Fig.\,\ref{phaseplots} but for objects 6060, 6149, 6259, 6315, 6337, and 6393.}
\end{figure*}

\clearpage\newpage

\begin{figure*}
\centering
\includegraphics[width=\columnwidth]{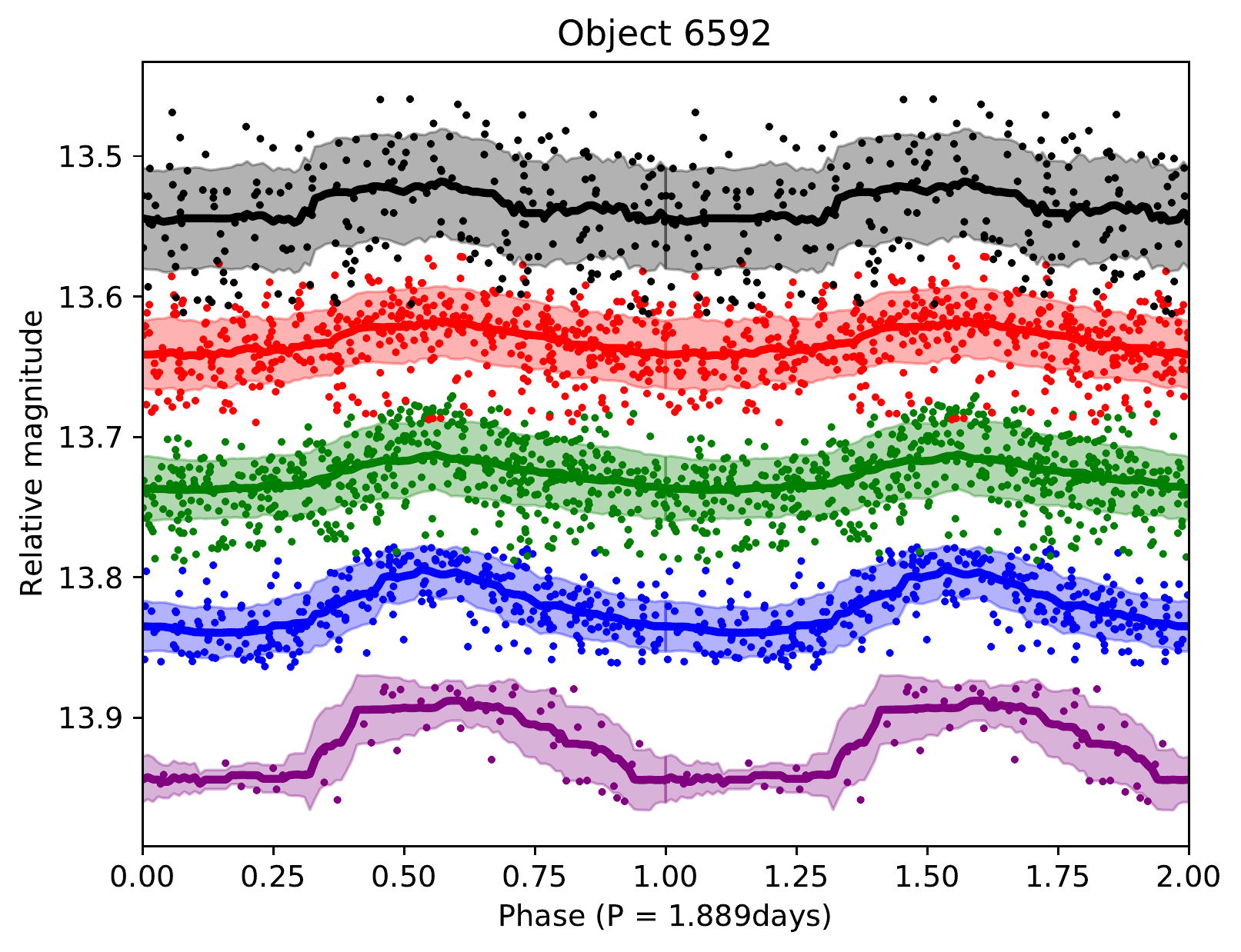} \hfill
\includegraphics[width=\columnwidth]{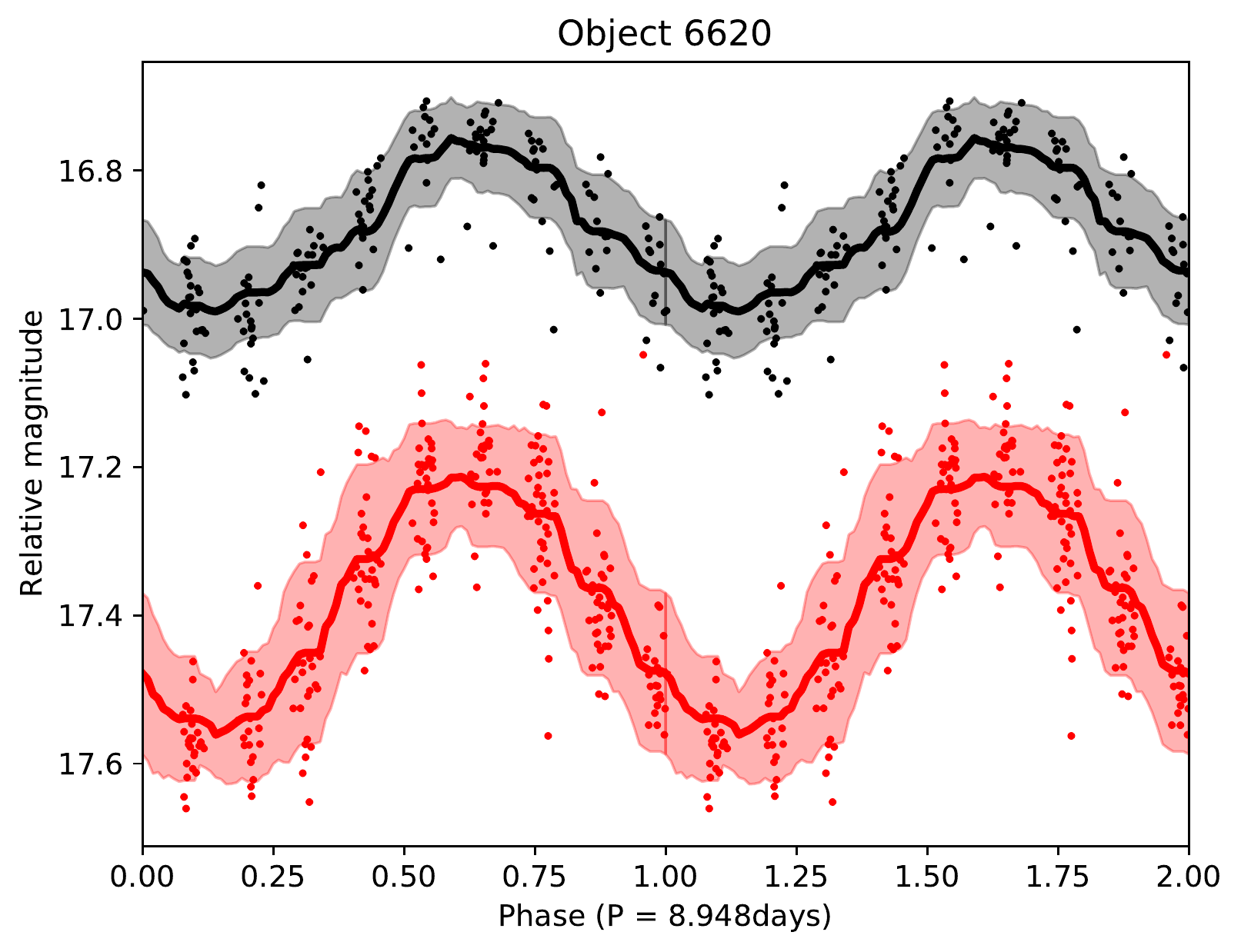} \\
\includegraphics[width=\columnwidth]{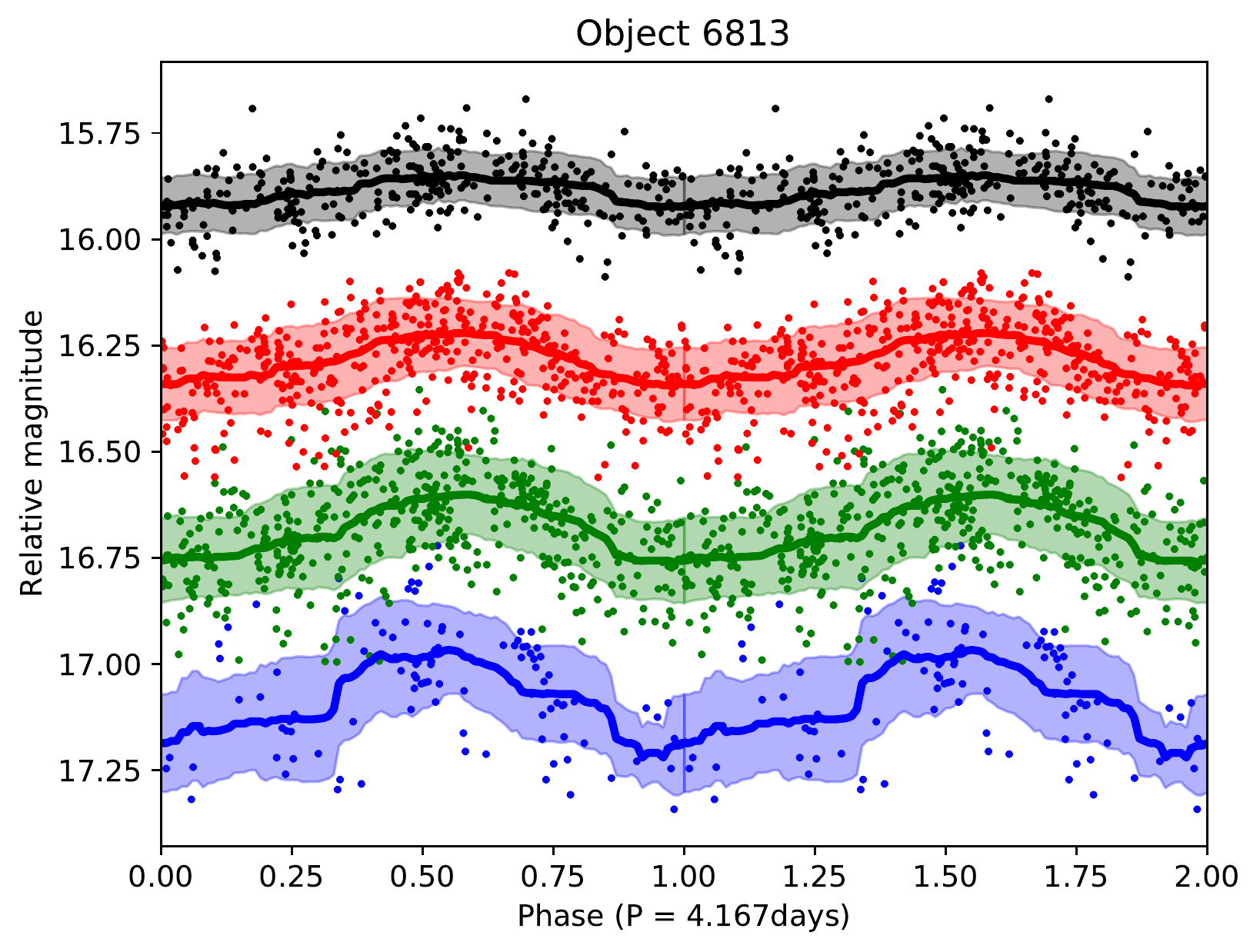} \hfill
\includegraphics[width=\columnwidth]{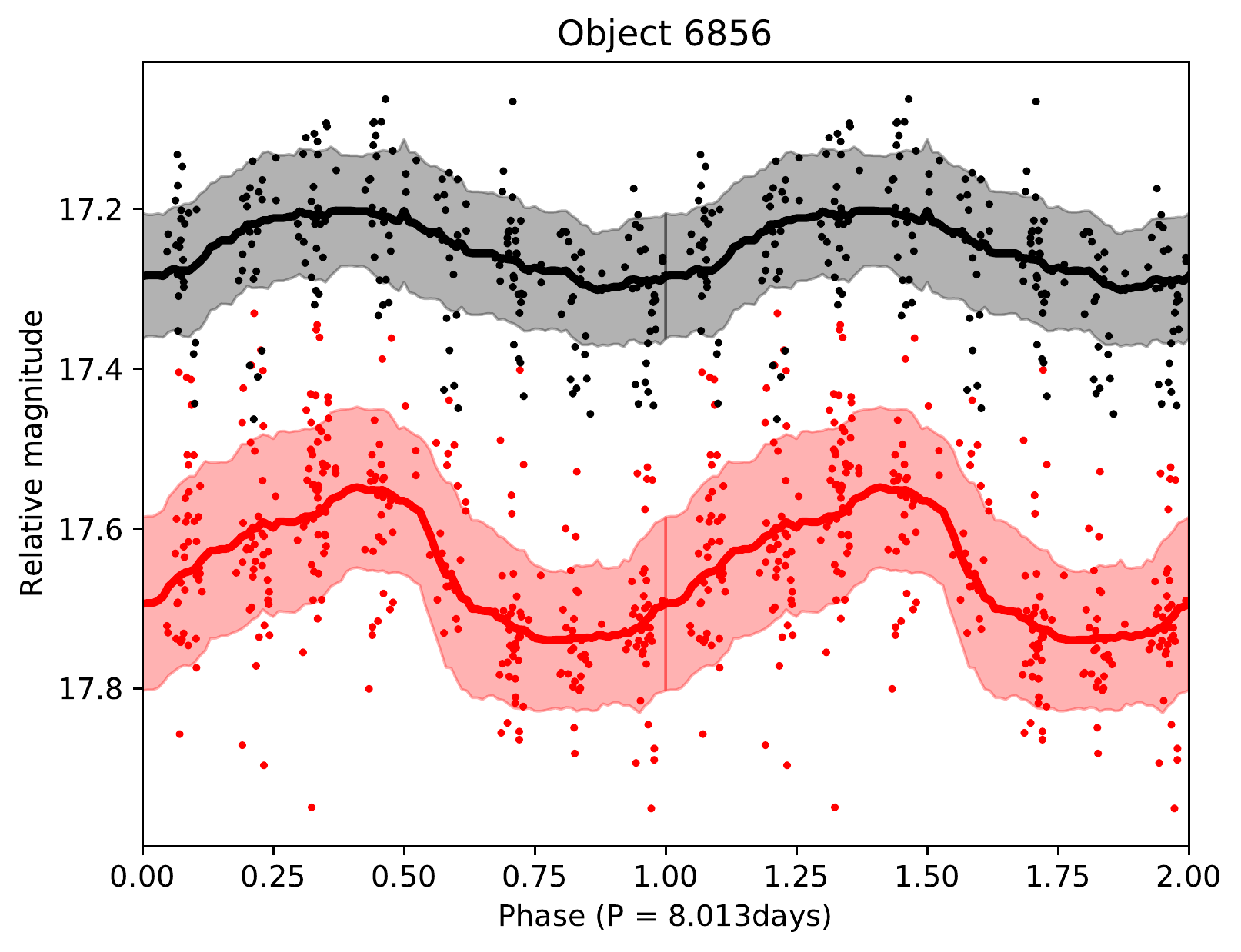} \\
\includegraphics[width=\columnwidth]{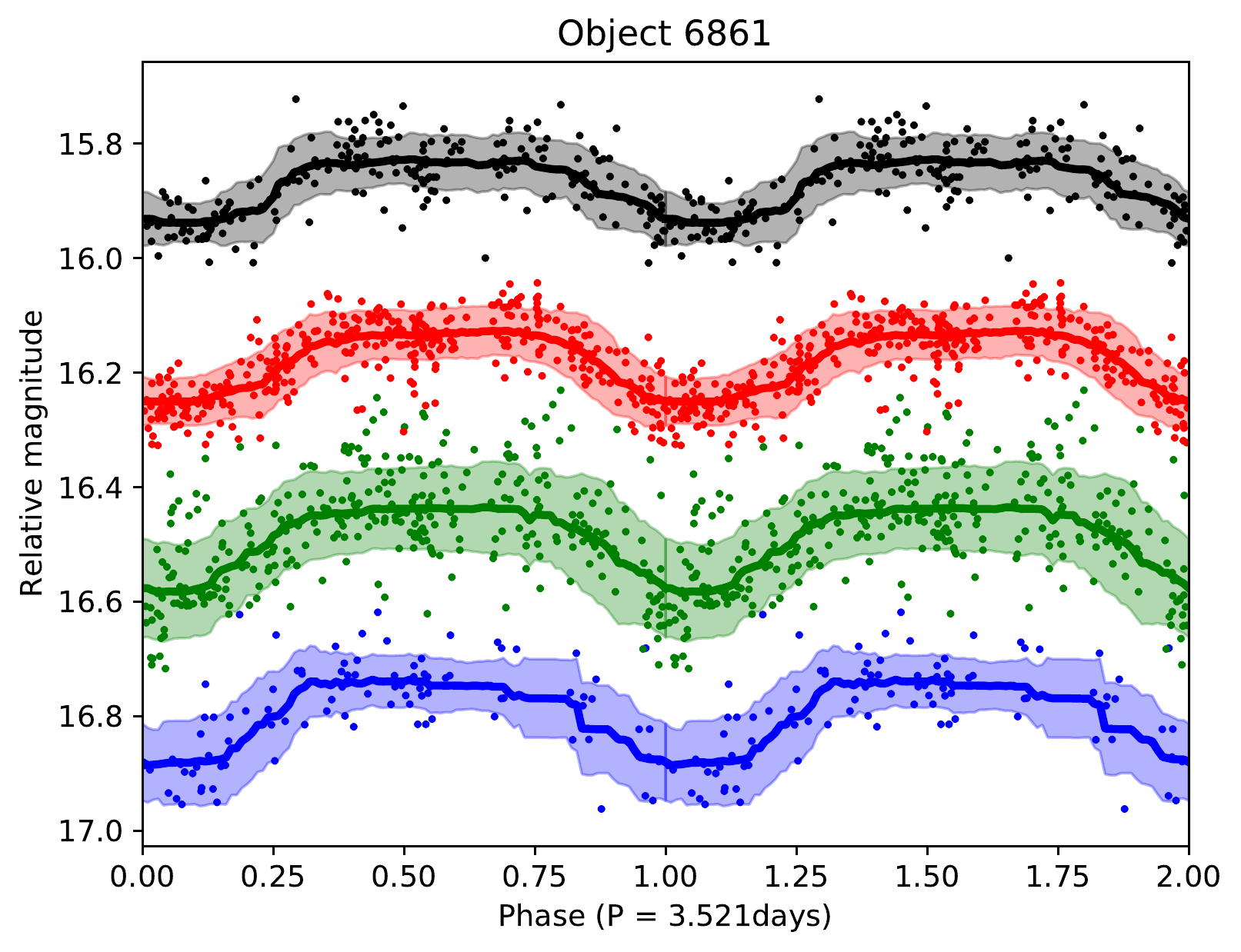} \hfill
\includegraphics[width=\columnwidth]{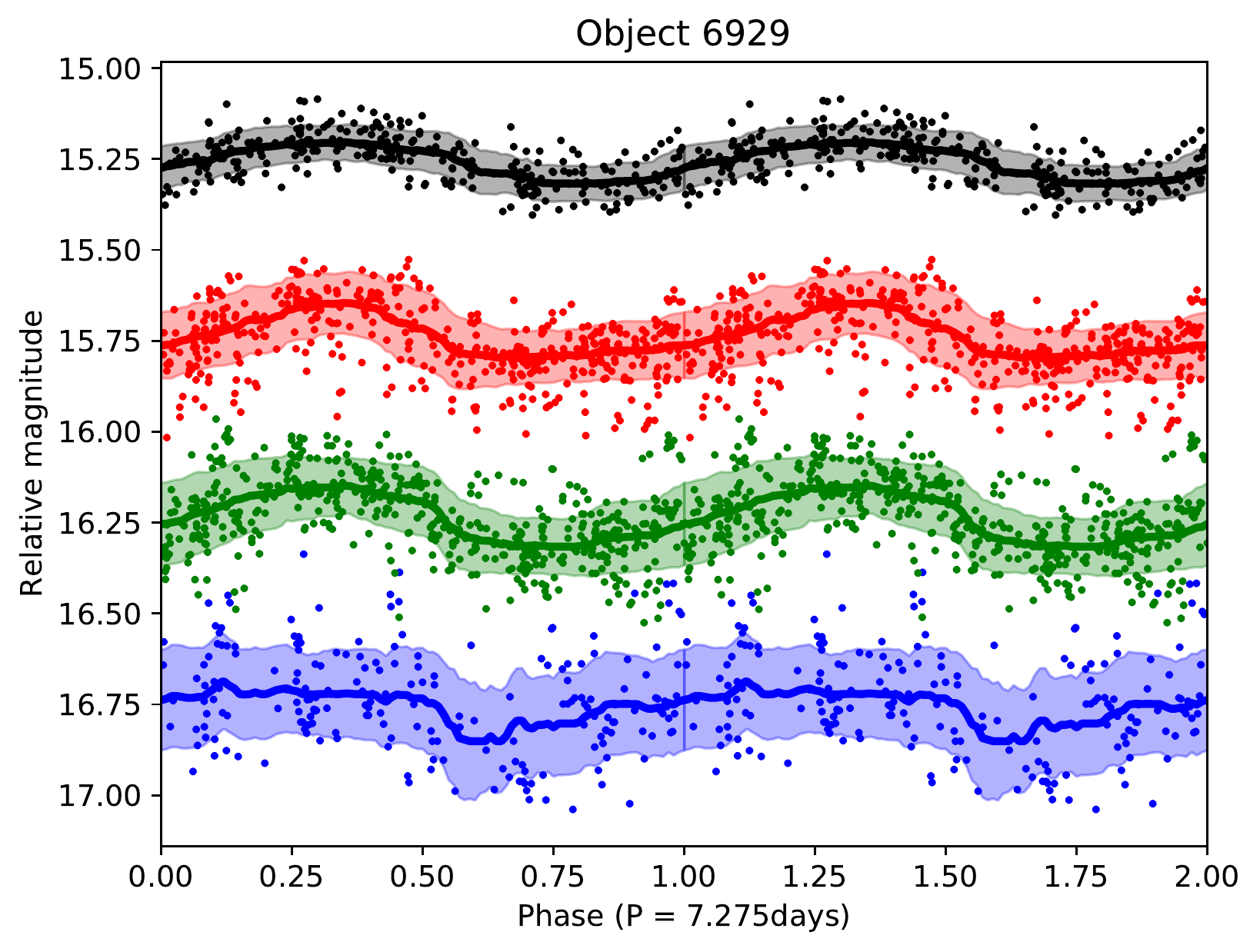} \\
\caption{As Fig.\,\ref{phaseplots} but for objects 6592, 6620, 6813, 6856, 6861, and 6929.}
\end{figure*}

\clearpage\newpage

\begin{figure*}
\centering
\includegraphics[width=\columnwidth]{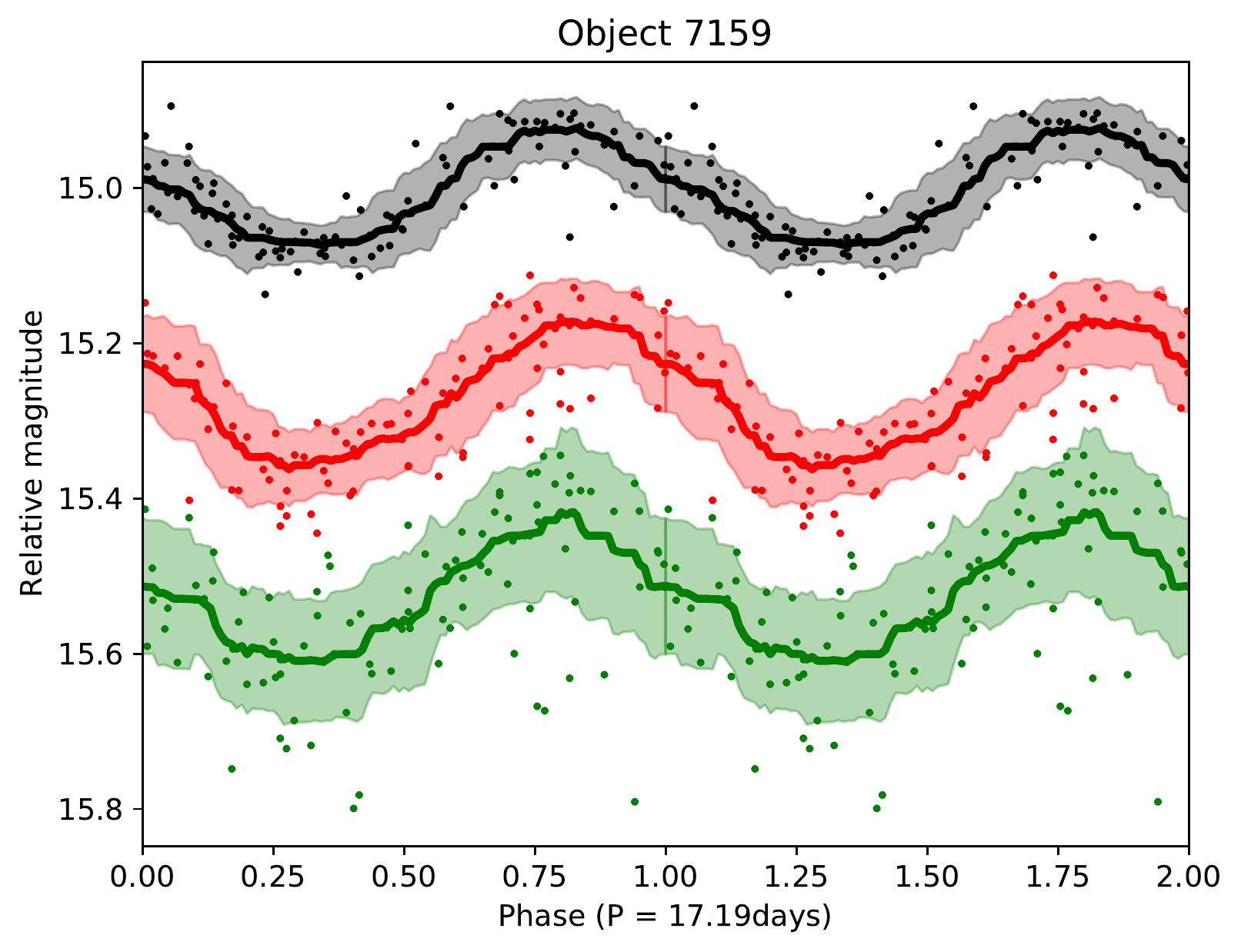} \hfill
\includegraphics[width=\columnwidth]{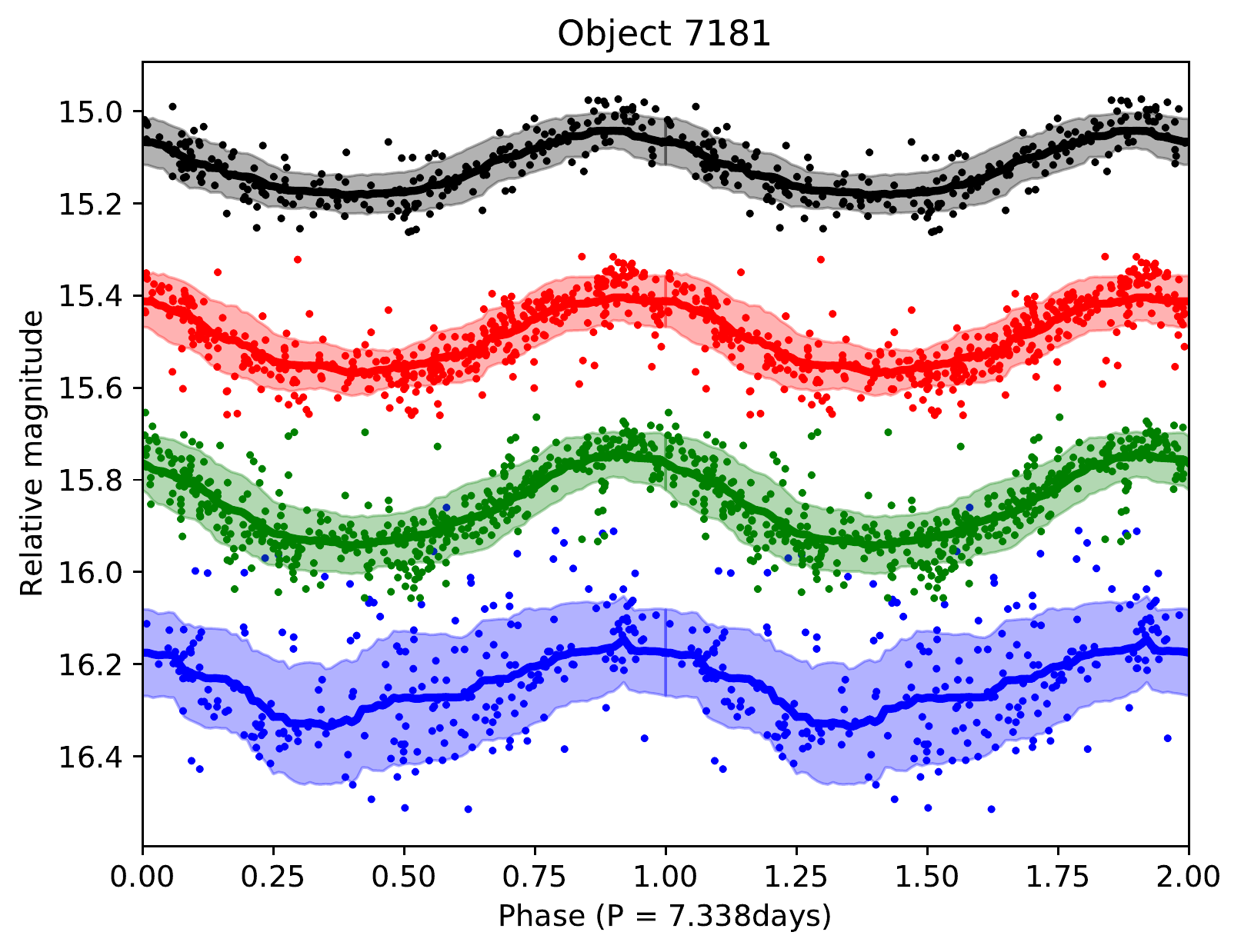} \\
\includegraphics[width=\columnwidth]{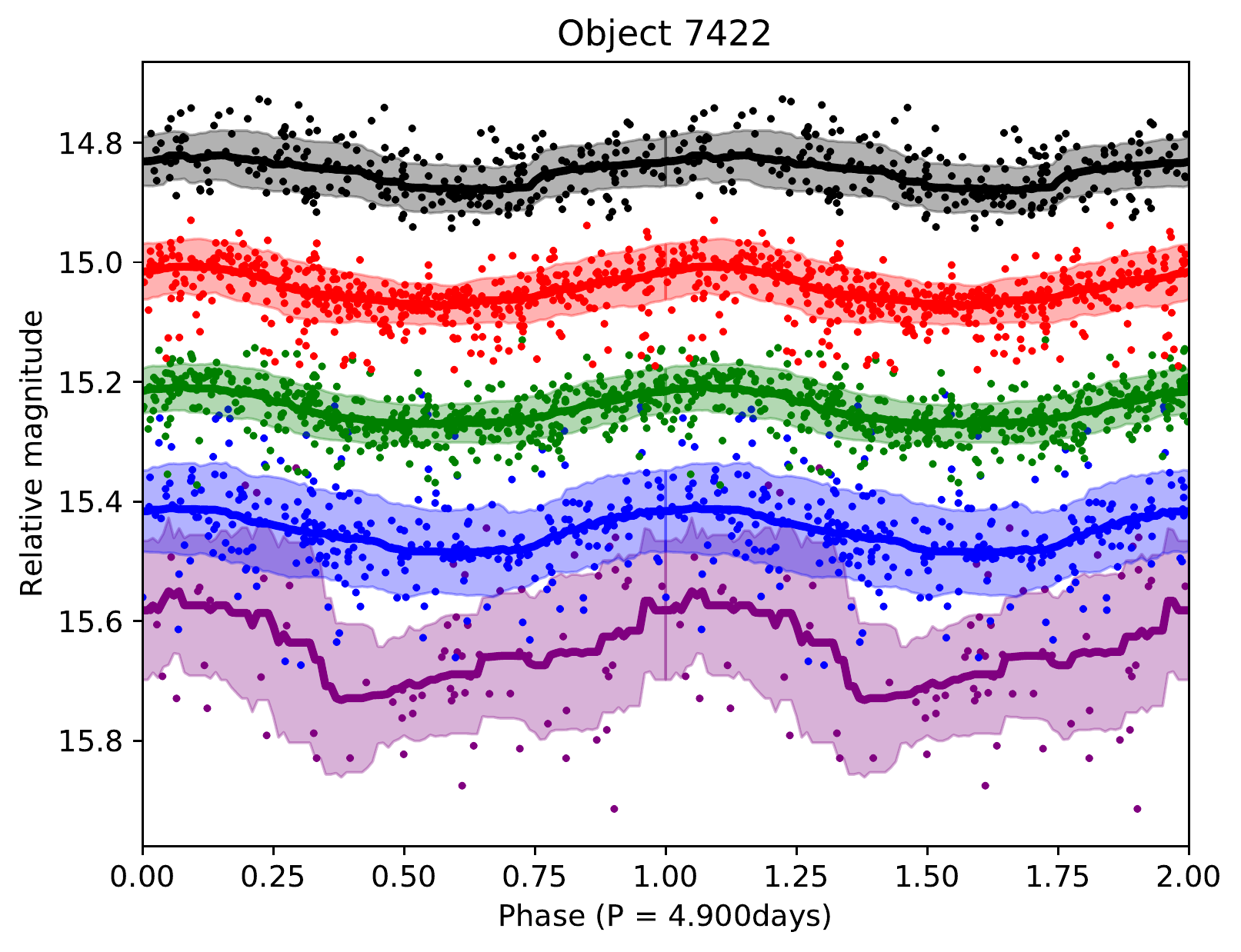} \hfill
\includegraphics[width=\columnwidth]{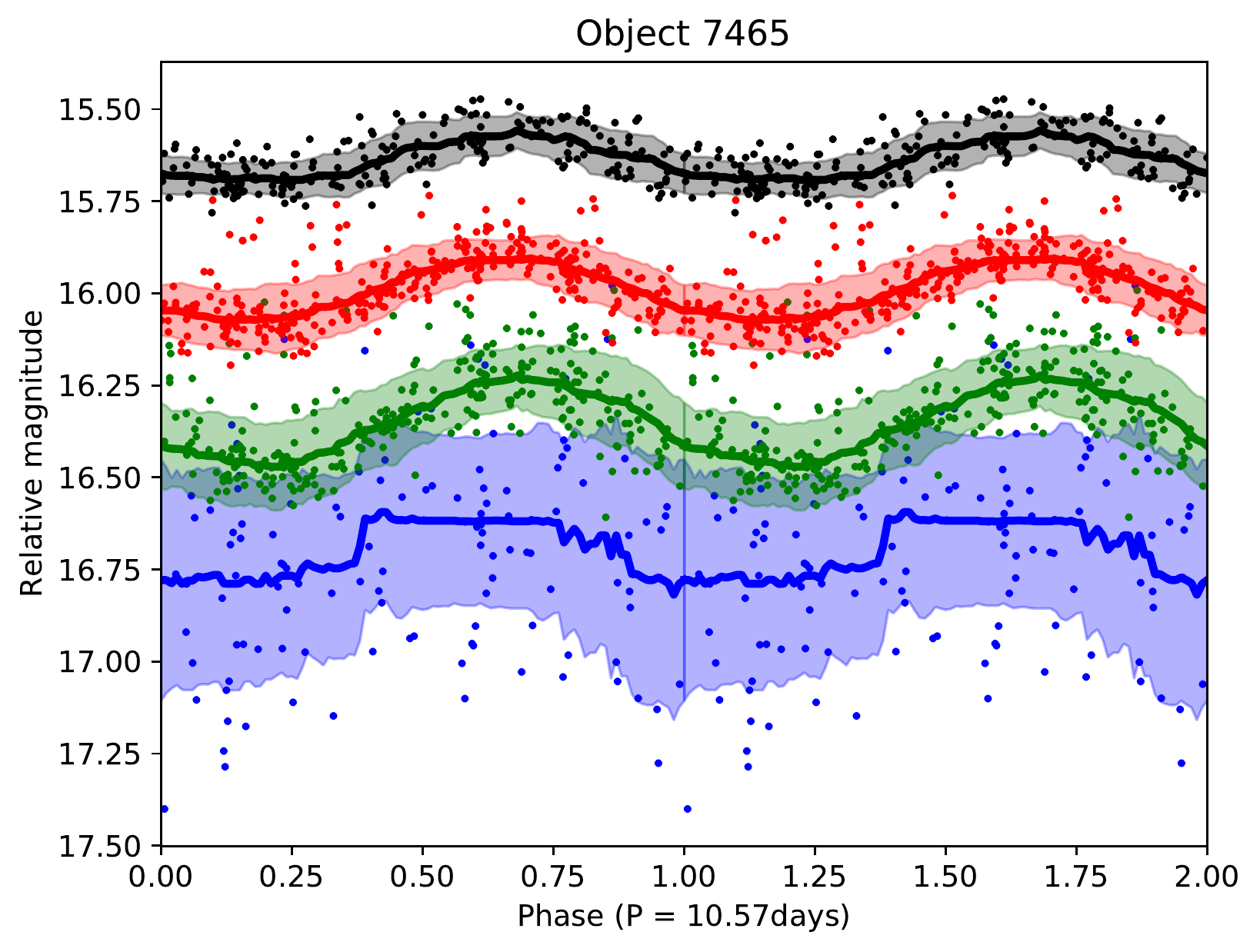} \\
\includegraphics[width=\columnwidth]{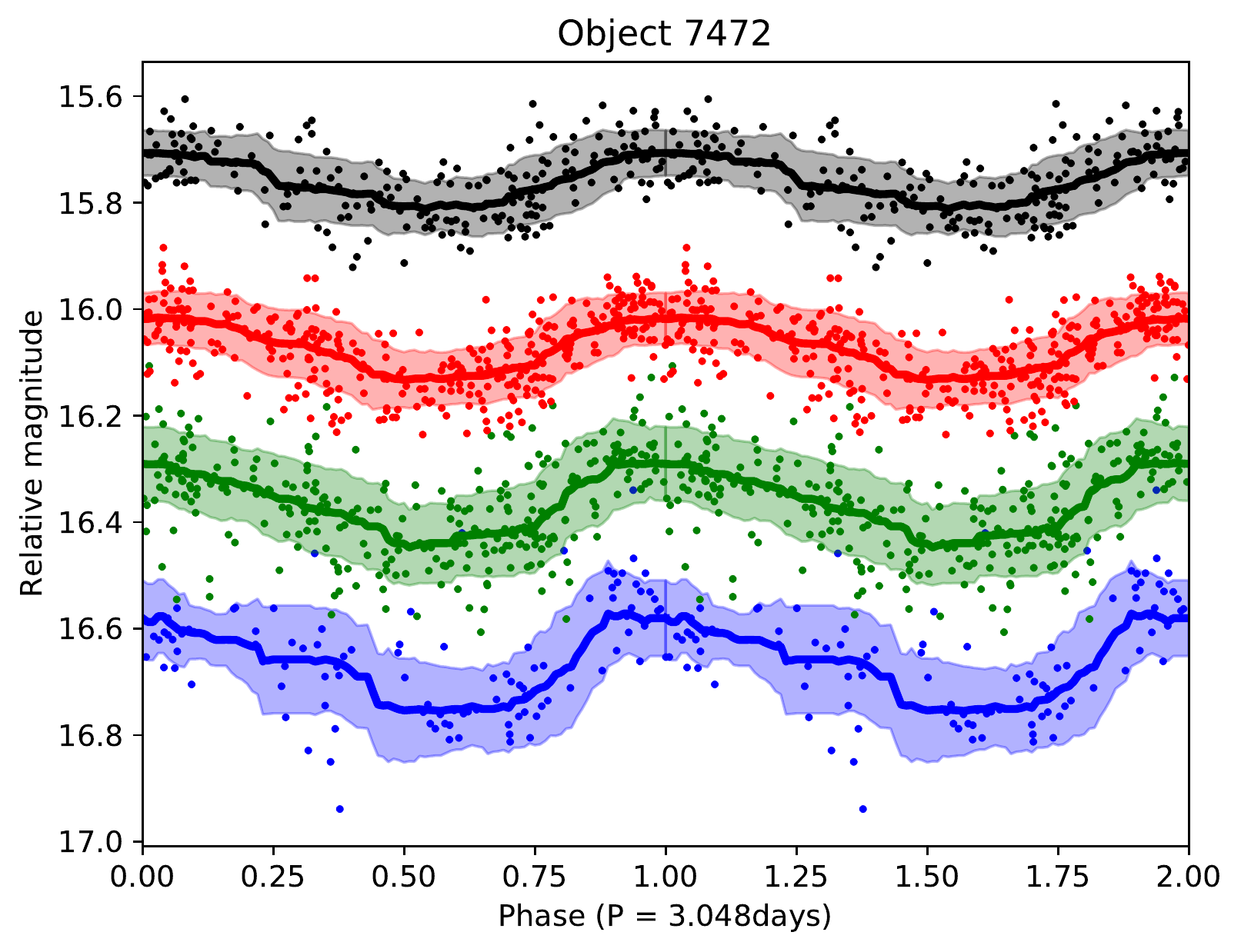} \hfill
\includegraphics[width=\columnwidth]{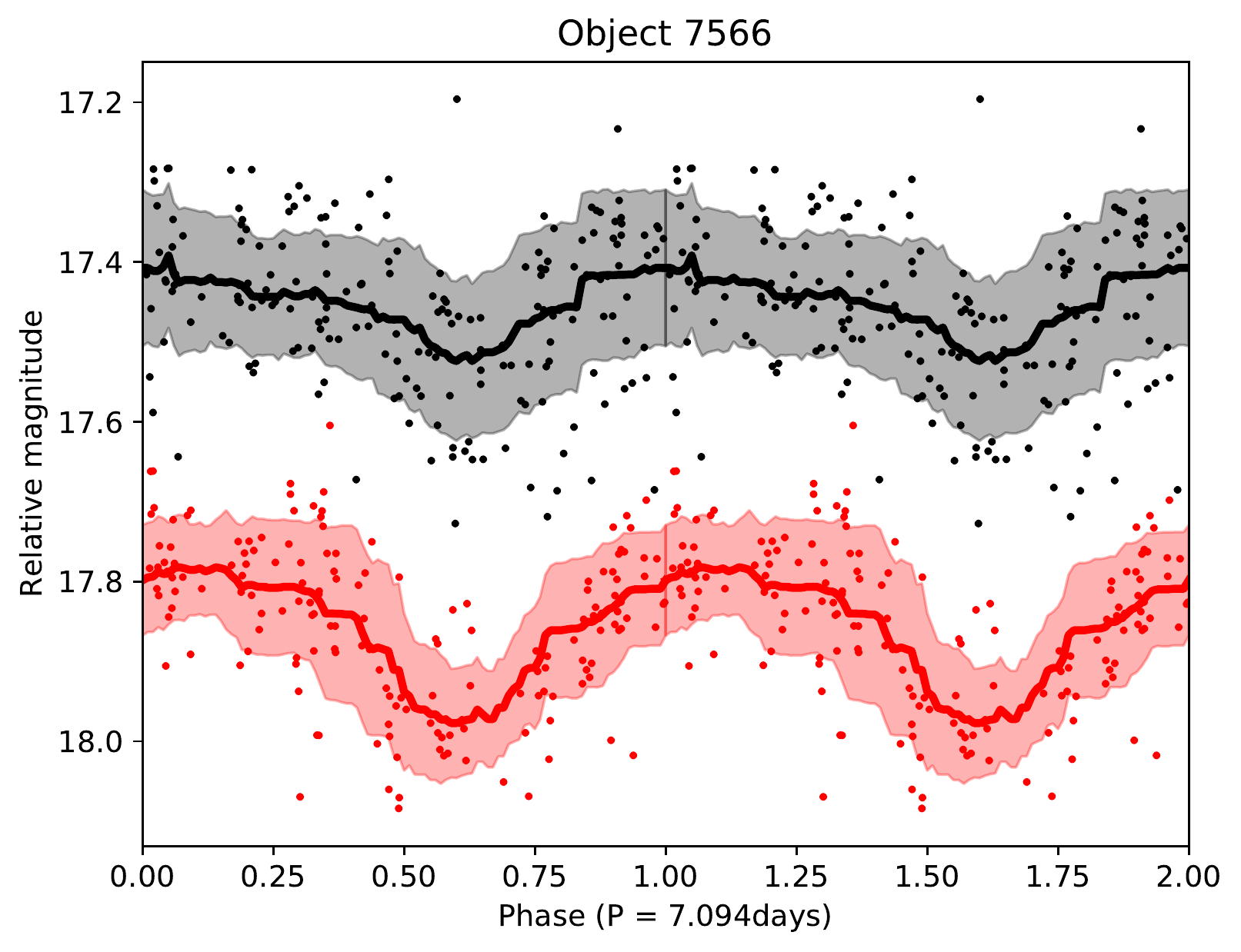} \\
\caption{As Fig.\,\ref{phaseplots} but for objects 7159, 7181, 7422, 7465, 7472, and 7566.}
\end{figure*}

\clearpage\newpage

\begin{figure*}
\centering
\includegraphics[width=\columnwidth]{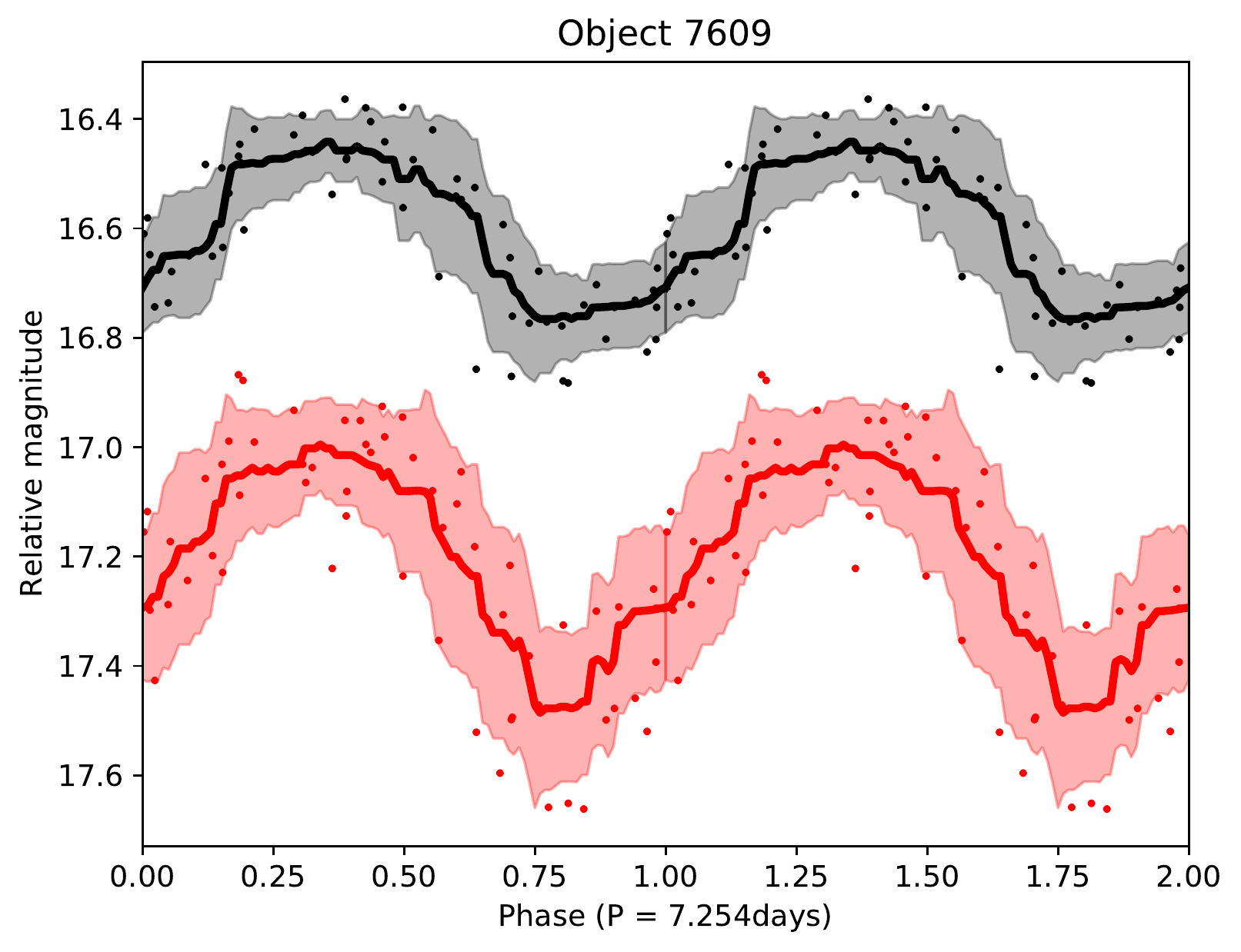} \hfill
\includegraphics[width=\columnwidth]{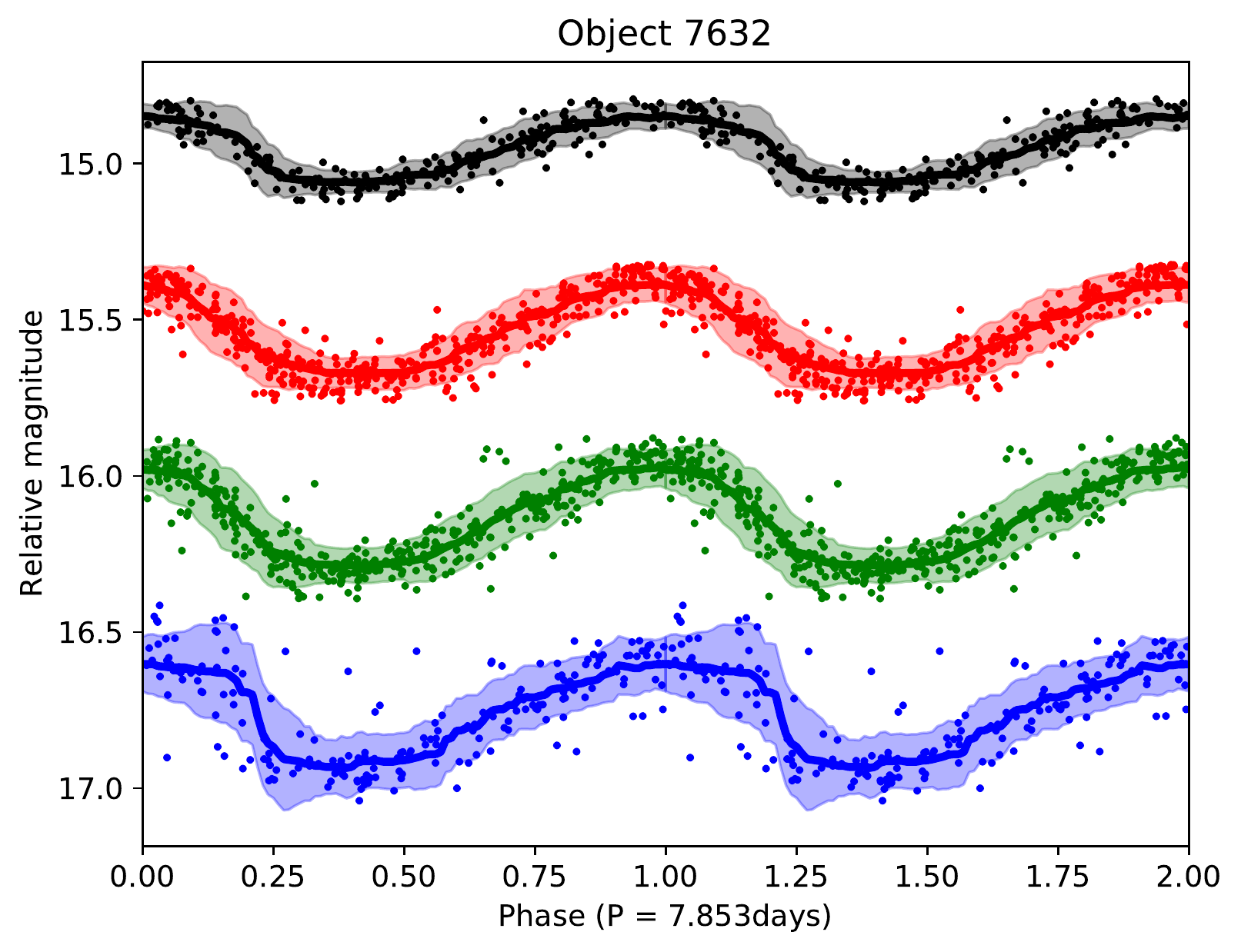} \\
\includegraphics[width=\columnwidth]{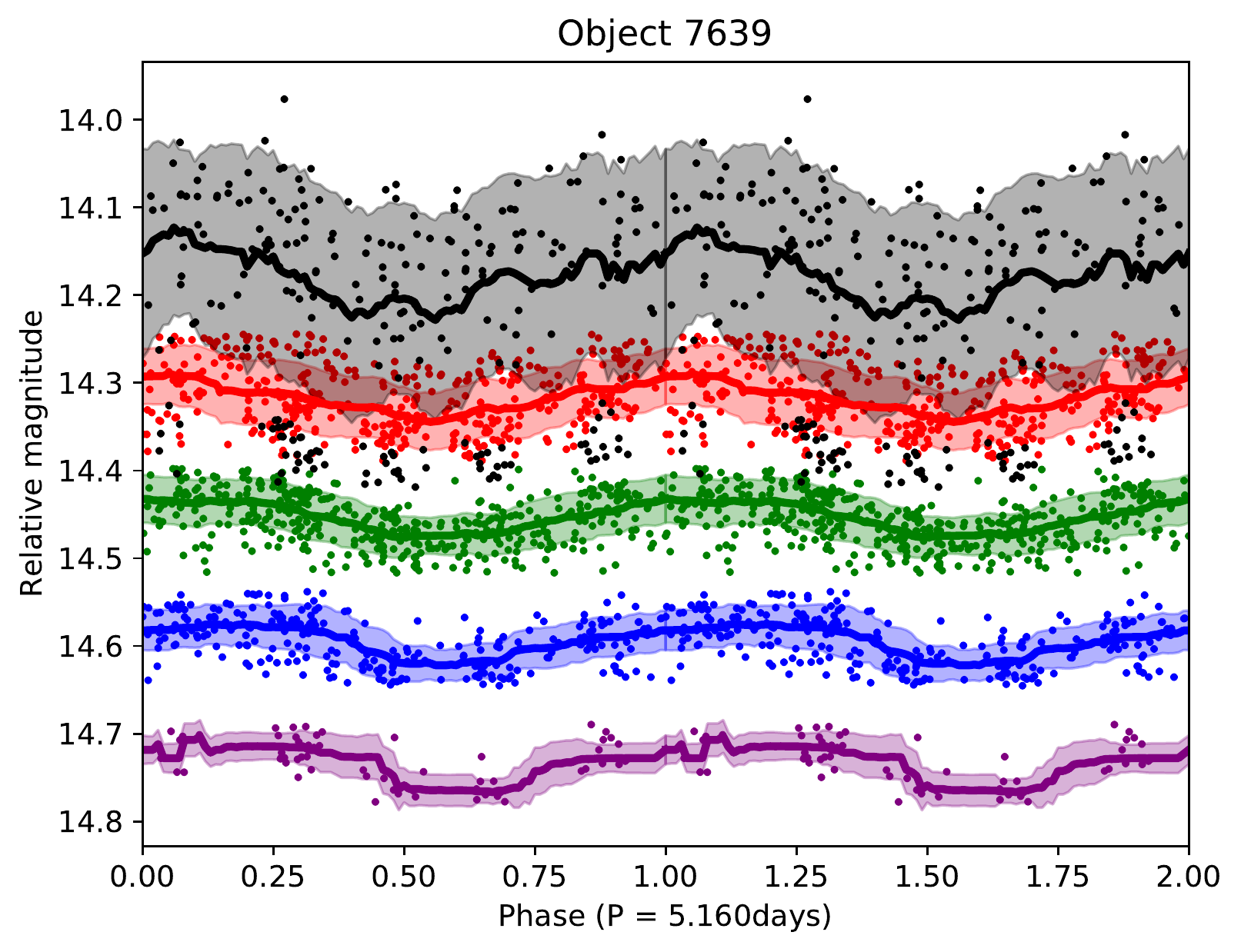} \hfill
\includegraphics[width=\columnwidth]{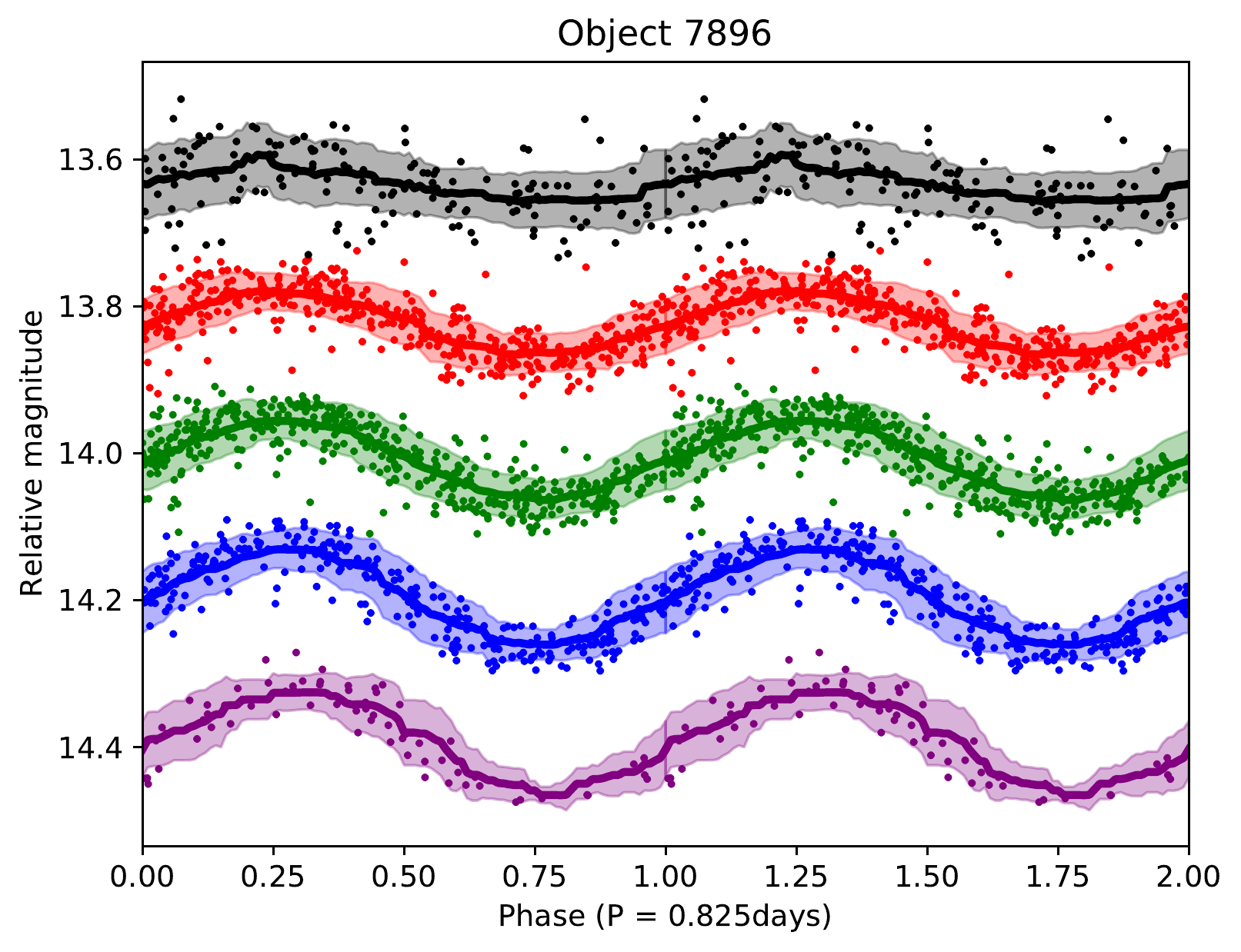} \\
\includegraphics[width=\columnwidth]{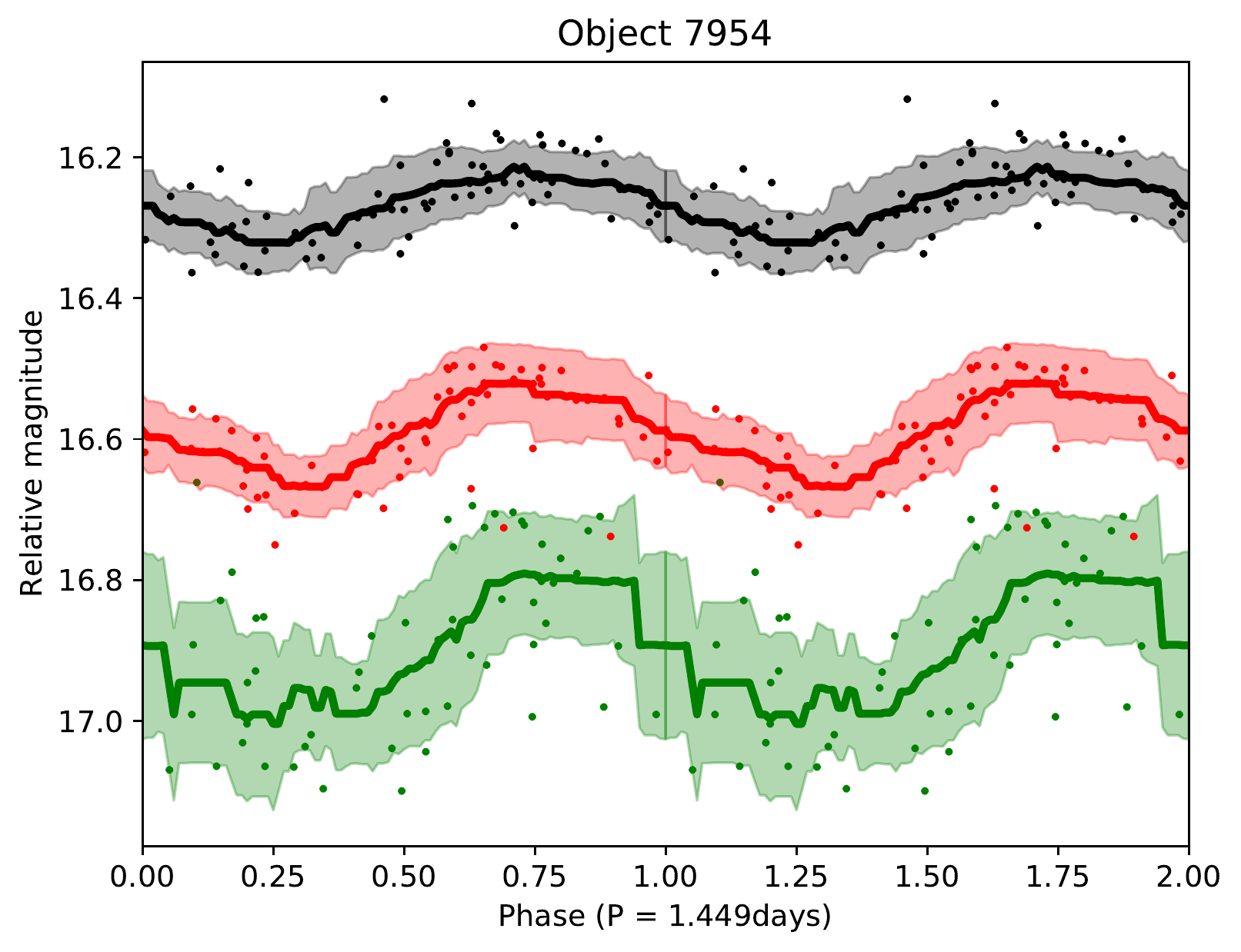} \hfill
\includegraphics[width=\columnwidth]{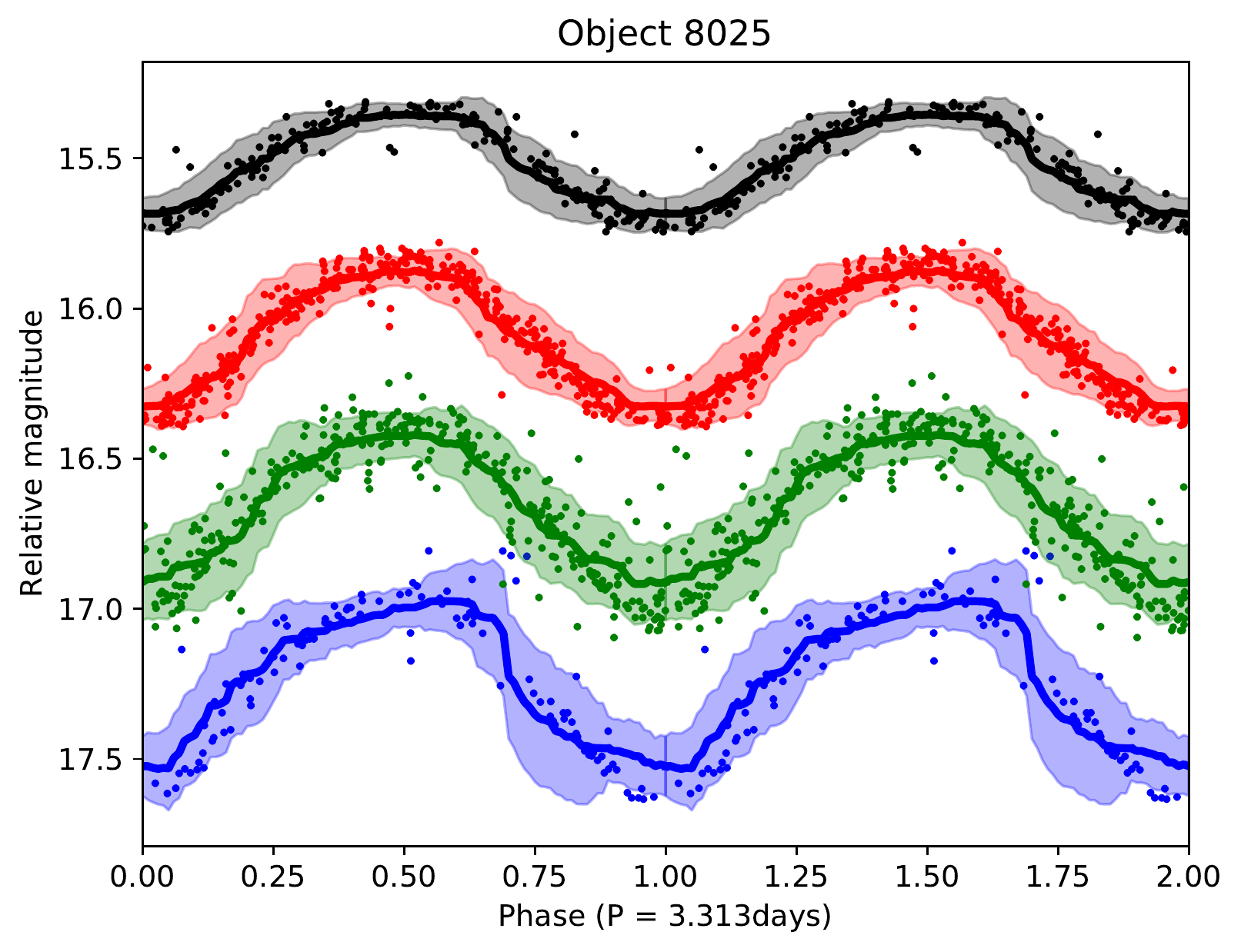} \\
\caption{As Fig.\,\ref{phaseplots} but for objects 7609, 7632, 7639, 7896, 7954, and 8025.}
\end{figure*}

\clearpage\newpage

\begin{figure*}
\centering
\includegraphics[width=\columnwidth]{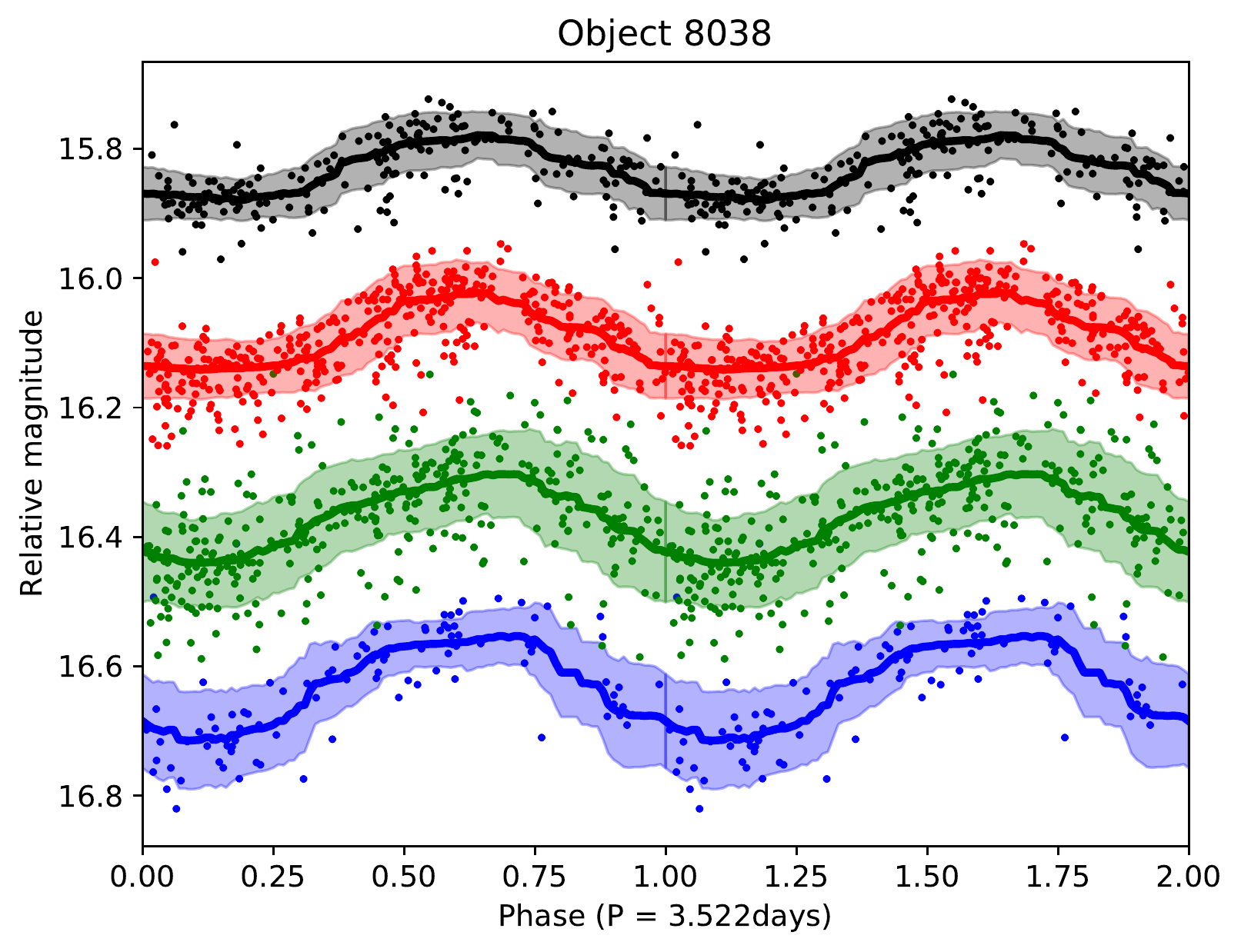} \hfill
\includegraphics[width=\columnwidth]{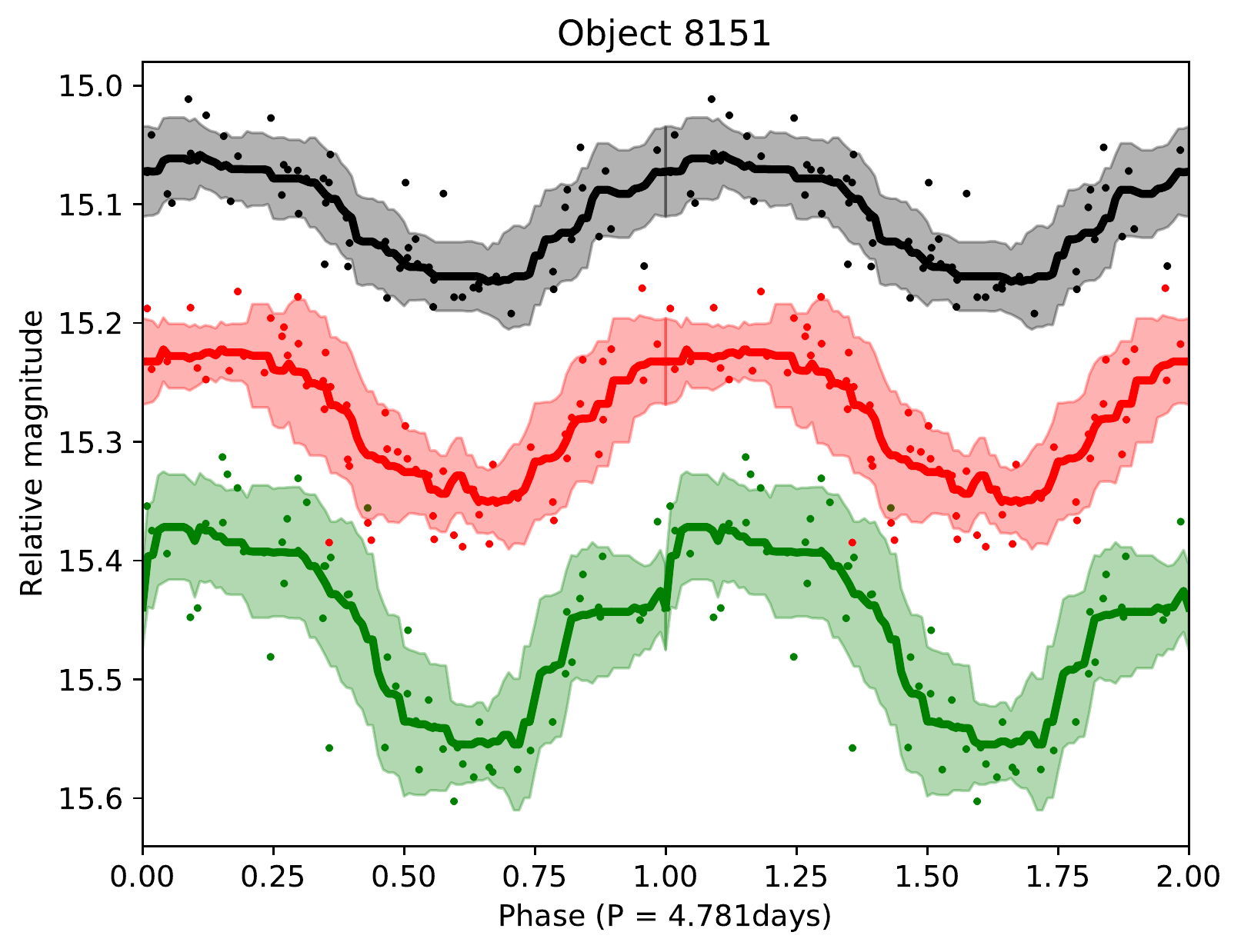} \\
\includegraphics[width=\columnwidth]{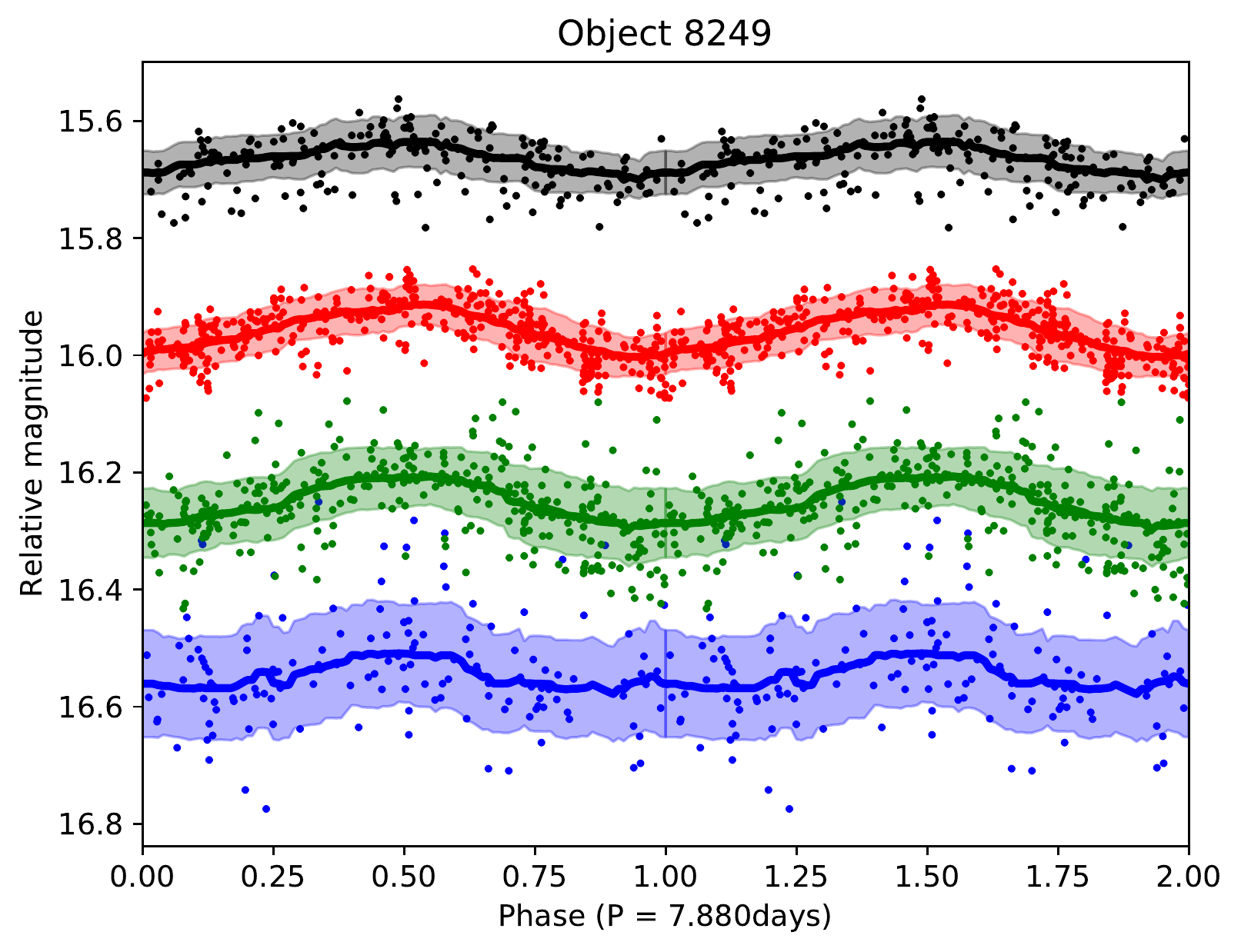} \hfill
\includegraphics[width=\columnwidth]{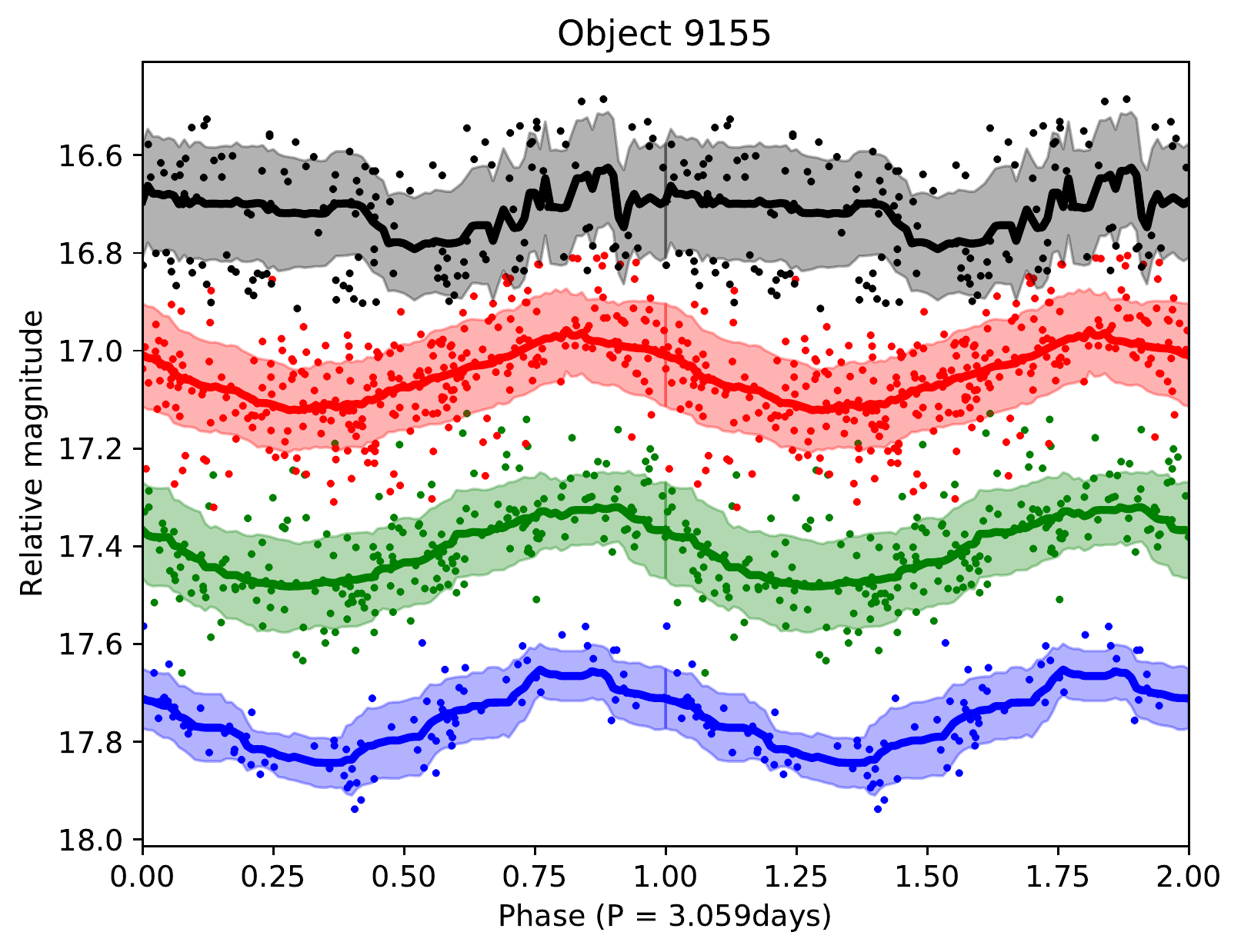} \\
\includegraphics[width=\columnwidth]{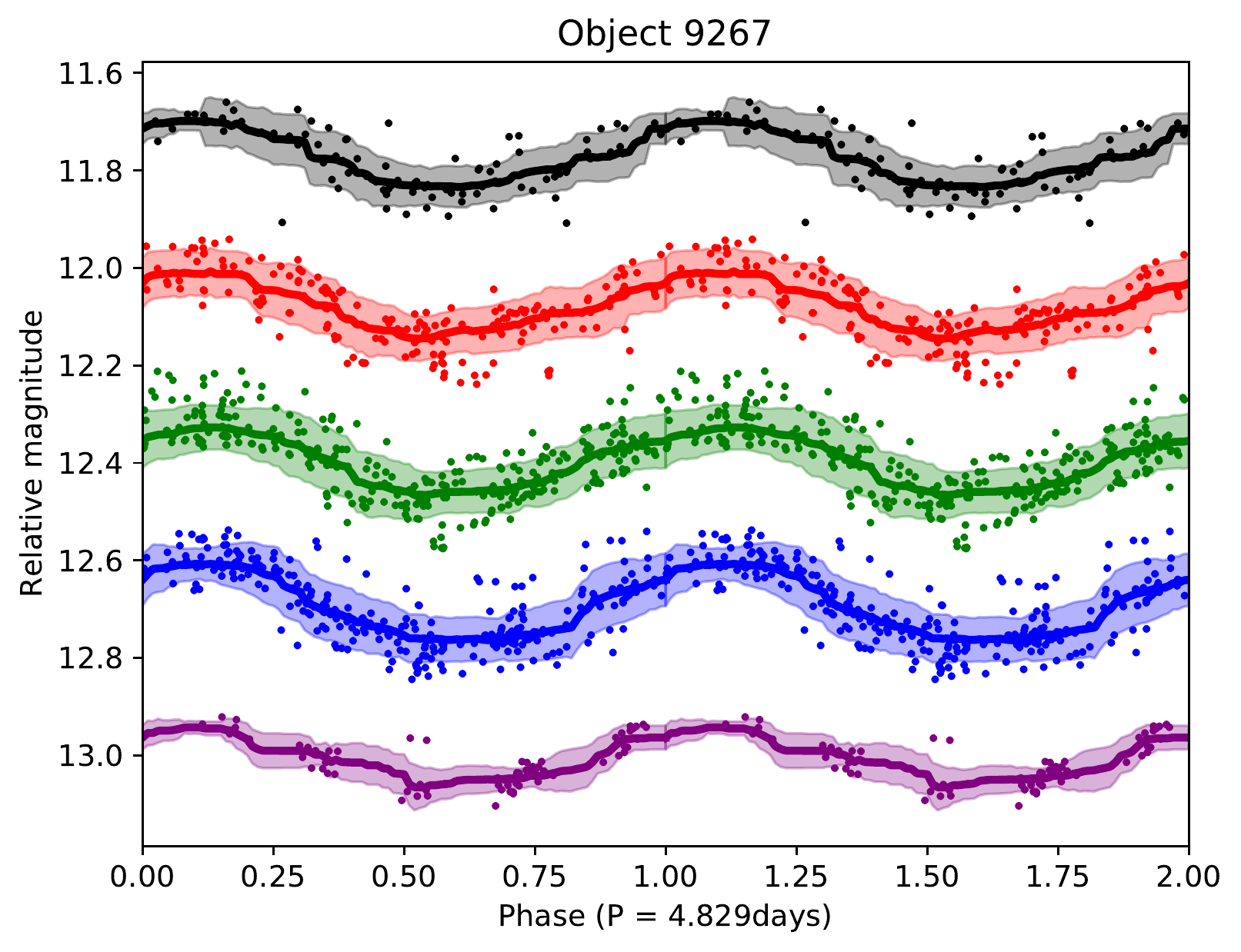} \hfill
\includegraphics[width=\columnwidth]{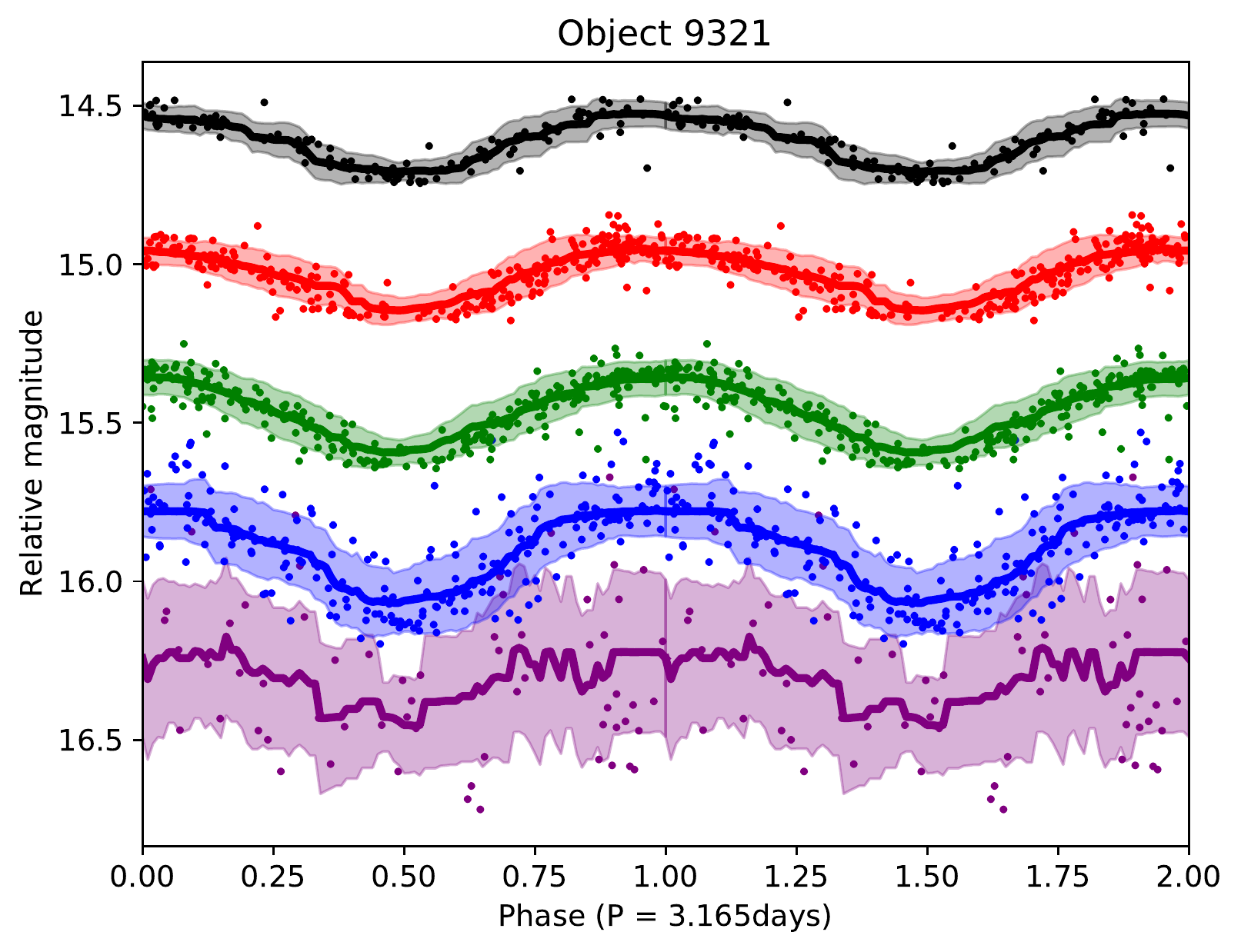} \\
\caption{As Fig.\,\ref{phaseplots} but for objects 8038, 8151, 8249, 9155, 9267, and 9321.}
\end{figure*}

\clearpage\newpage

\begin{figure*}
\centering
\includegraphics[width=\columnwidth]{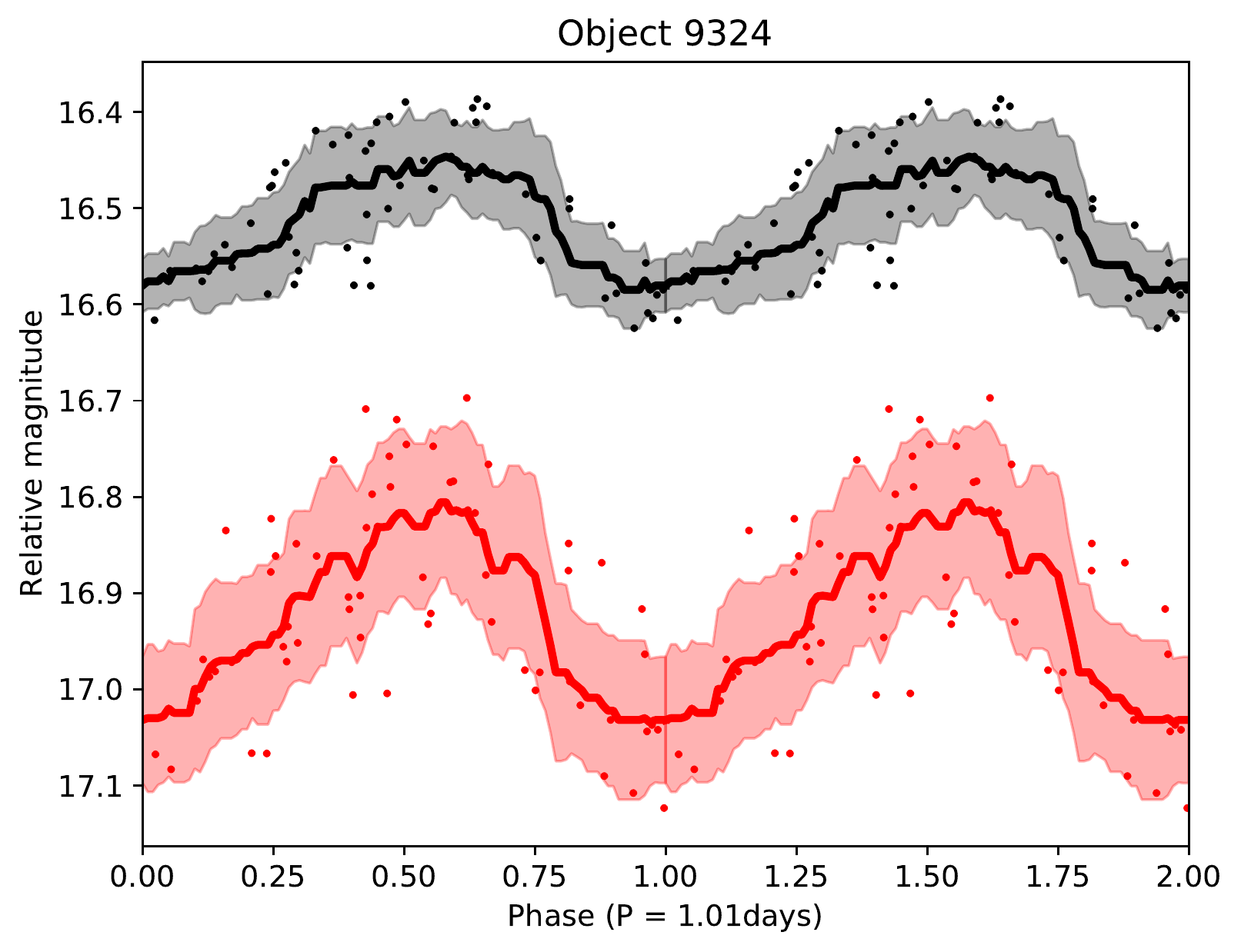} \hfill
\includegraphics[width=\columnwidth]{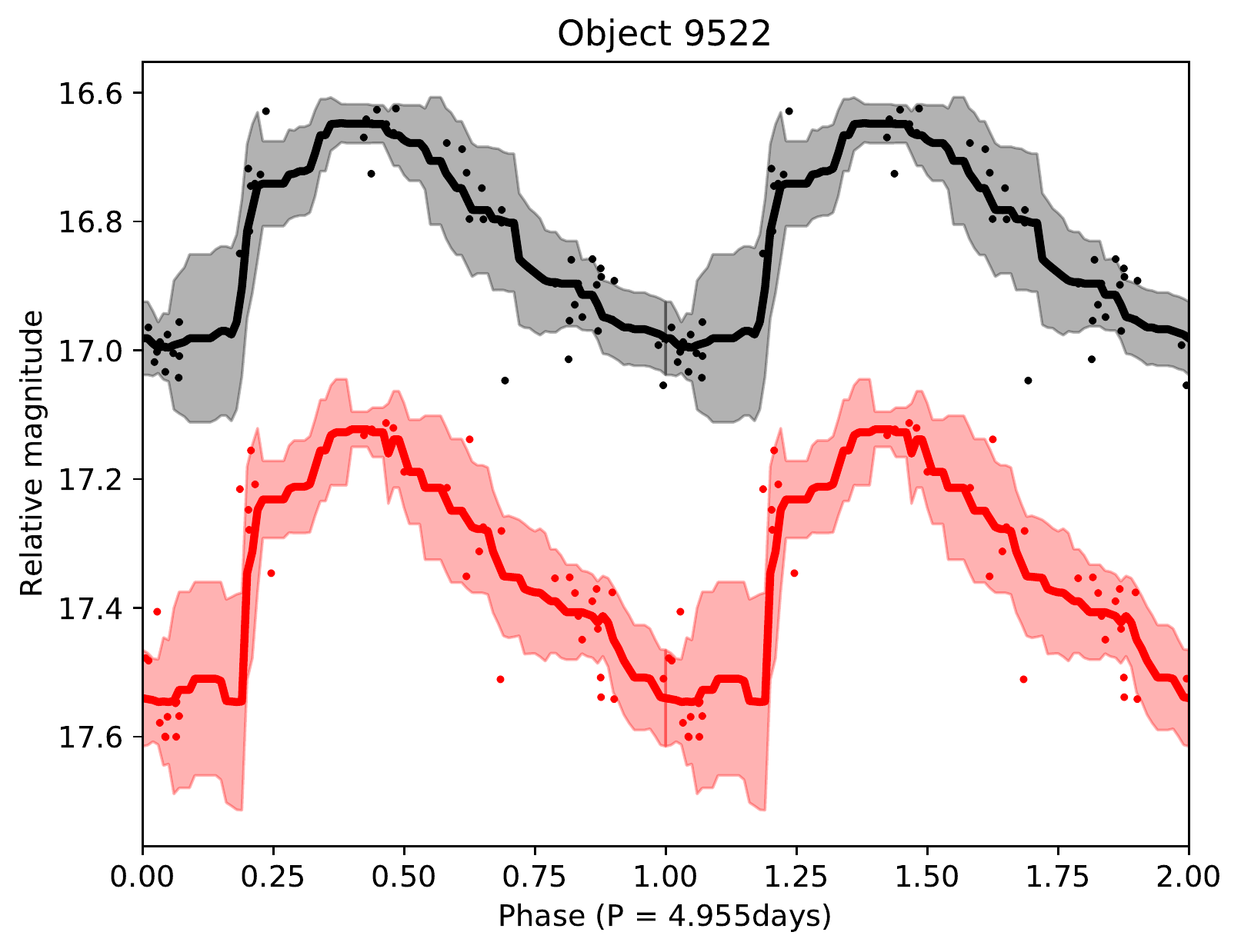} \\
\includegraphics[width=\columnwidth]{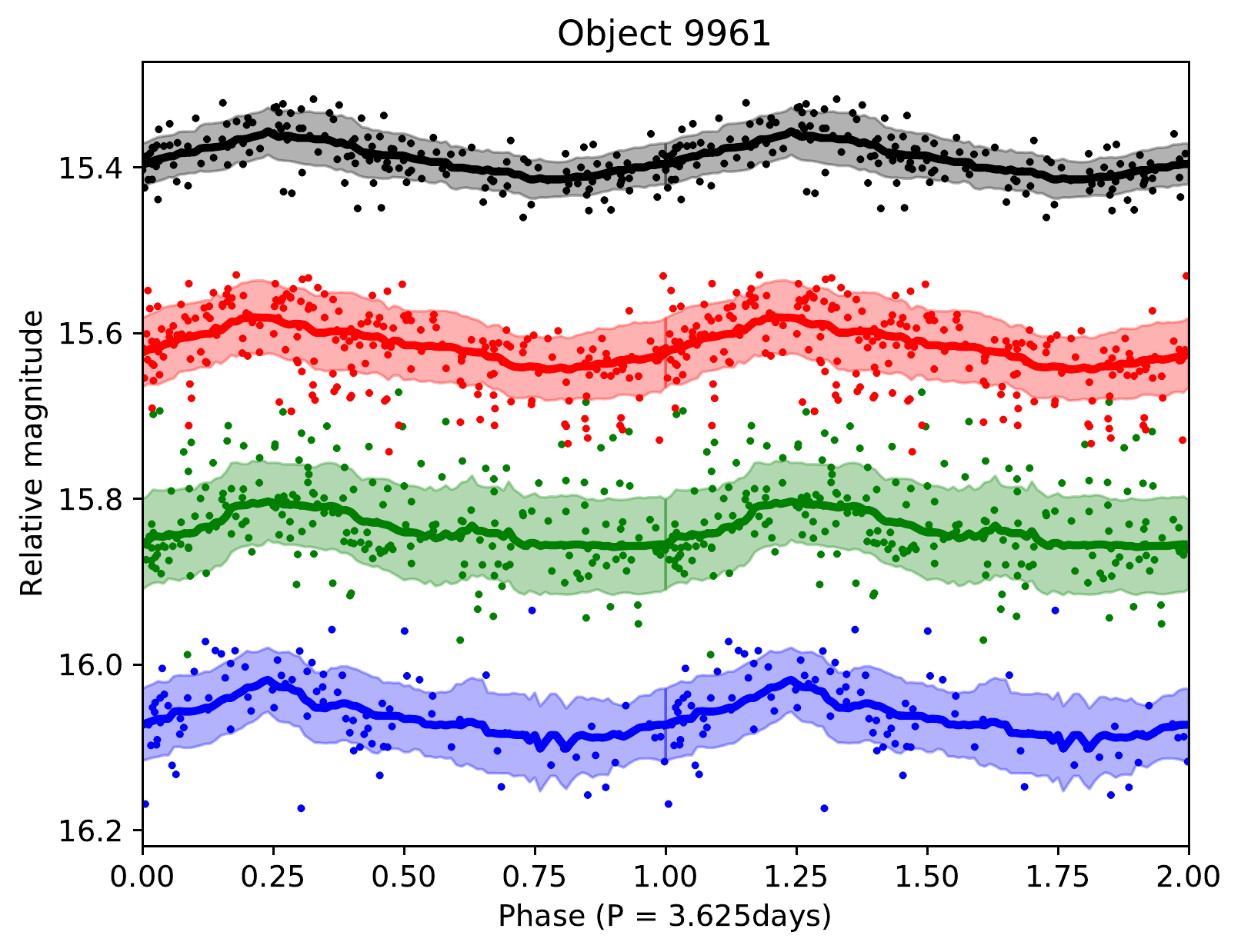} \hfill
\includegraphics[width=\columnwidth]{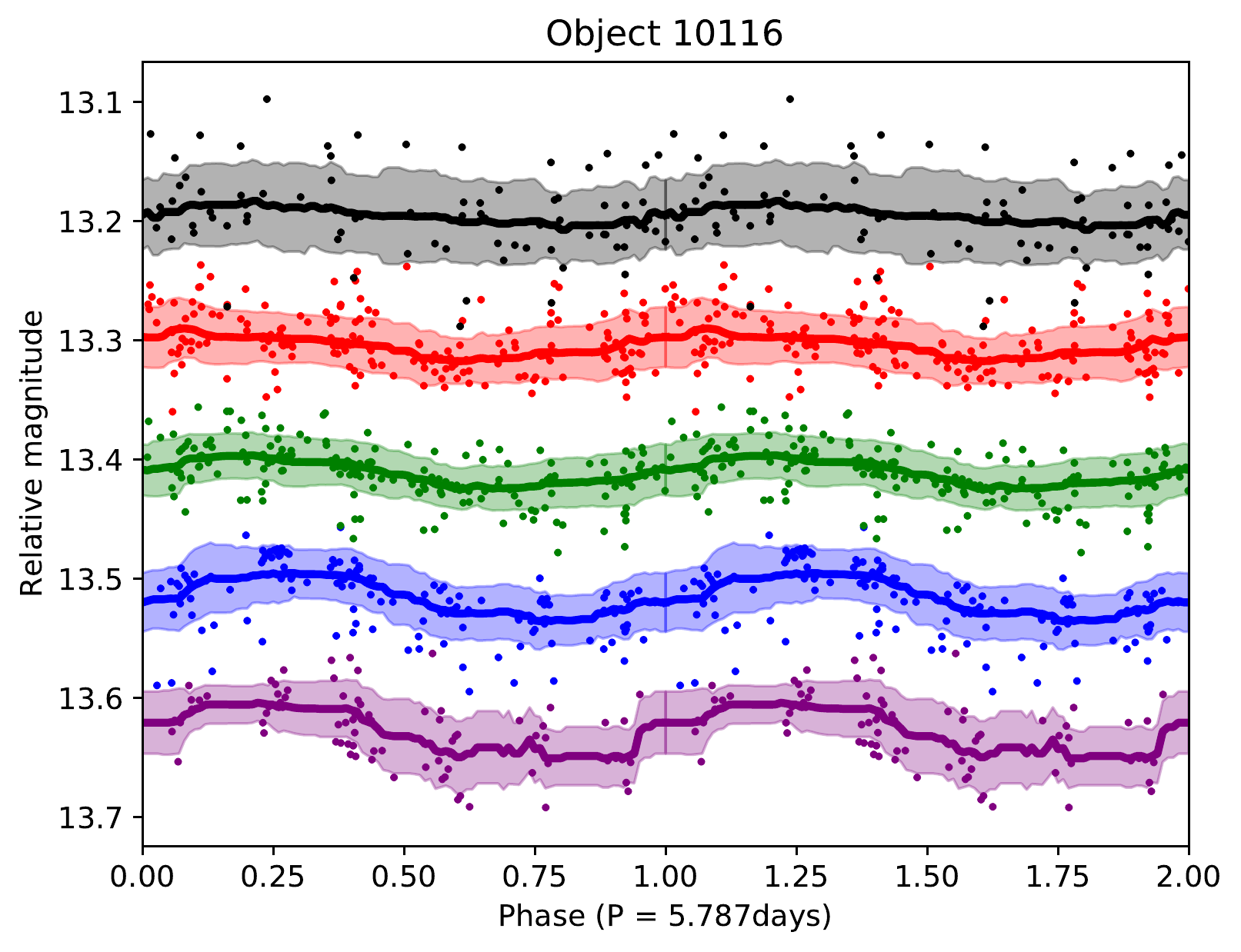} \\
\includegraphics[width=\columnwidth]{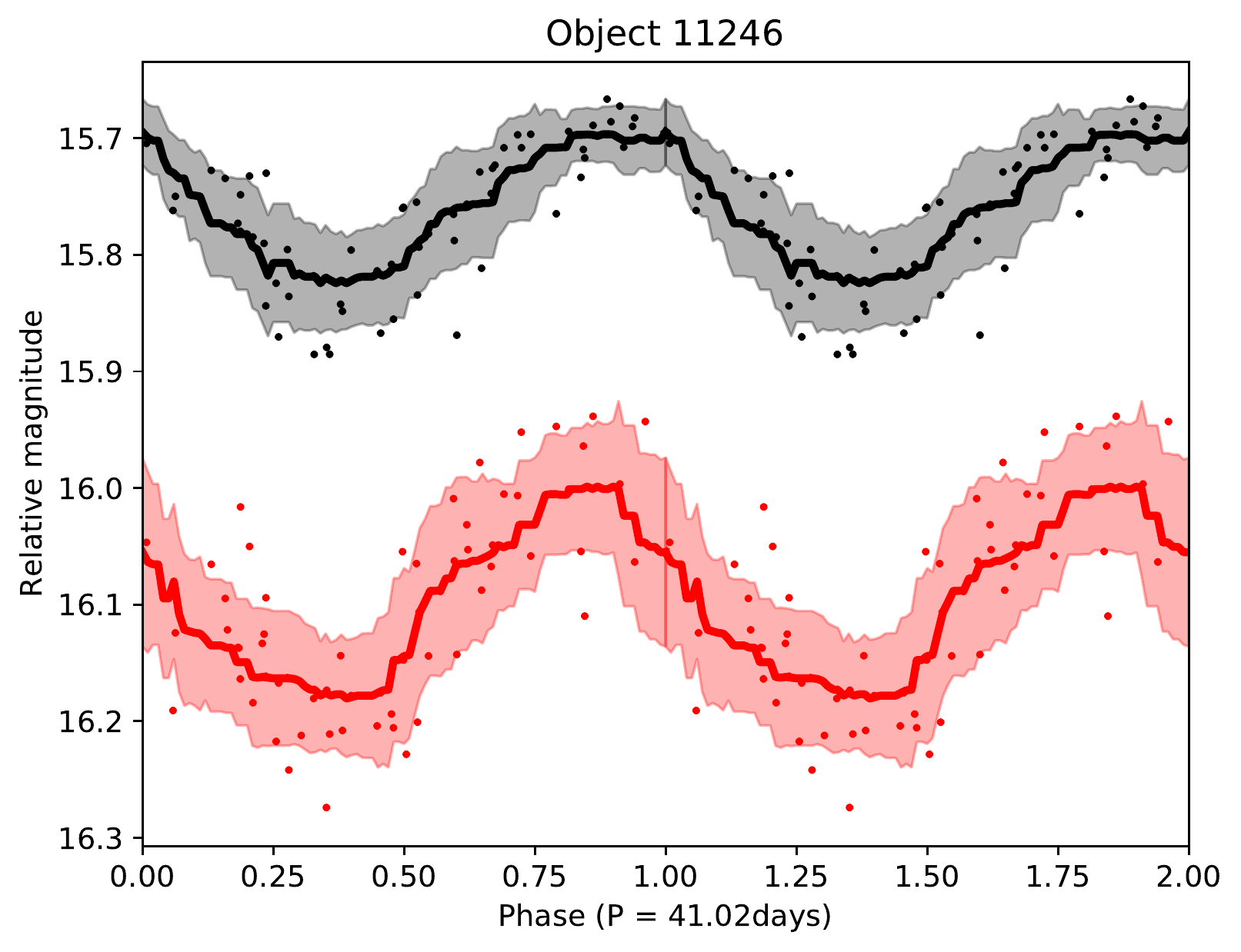} \\
\caption{As Fig.\,\ref{phaseplots} but for objects 9324, 9522, 9961, 10116, and 11246.}
\end{figure*}

\clearpage

\bsp	
\label{lastpage}
\end{document}